\documentclass[aps,groupedaddress,pra,preprint,amsmath,amssymb,nofootinbib,11pt]{revtex4-1}

\usepackage{graphicx}
\usepackage{xcolor}

\usepackage{hyperref}
\hypersetup{
	colorlinks,
	linkcolor={red!50!black},
	citecolor={blue!50!black},
	urlcolor={blue!50!black}
	}

\usepackage{setspace}

\usepackage[english]{babel} 
\usepackage[T1]{fontenc}

\begin{document}
\title{Matter-wave interferometry with composite quantum objects}

\author{M. Arndt}
\author{N. D{\"o}rre}
\author{S. Eibenberger}
\author{P. Haslinger}
\author{J. Rodewald}
\affiliation{University of Vienna, Faculty of Physics, VCQ, Boltzmanngasse 5, 1090 Vienna, Austria}

\author{K. Hornberger}
\author{S. Nimmrichter}
\affiliation{University of Duisburg-Essen, Faculty of Physics, Lotharstra{\ss}e 1, 47048, Germany}

\author{M. Mayor}
\affiliation{University of Basel, Department of Chemistry, St. Johannsring 19, Basel, Switzerland and Forschungszentrum Karlsruhe, Institute for Nanotechnology, P.O.Box 3640, 76021 Karlsruhe, Germany}

\begin{abstract}
\singlespacing
We discuss modern developments in quantum optics with organic molecules, clusters and nanoparticles --
in particular recent realizations of near-field matter-wave interferometry. A unified theoretical description in phase space allows us to describe quantum interferometry in position space and in the time domain on an equal footing. In order to establish matter-wave interferometers as a universal tool, which can accept and address a variety of nanoparticles, we elaborate on new quantum optical elements, such as diffraction gratings made of matter and light, as well as their absorptive and dispersive interaction with complex materials. We present Talbot-Lau interferometry (TLI), the Kapitza-Dirac-Talbot-Lau interferometer (KDTLI) and interferometry with pulsed optical gratings (OTIMA) as the most advanced devices to study the quantum wave nature of composite matter. These experiments define the current mass and complexity record in interferometric explorations of quantum macroscopicity and they open new avenues to quantum assisted metrology with applications in physical chemistry and biomolecular physics.
\end{abstract}
\maketitle
\newpage
\singlespacing
\tableofcontents
\newpage

\singlespacing
\section{Introduction and outline}
\label{sec:intro}

\begin{figure}
\begin{center}
\includegraphics[width=0.9\textwidth]{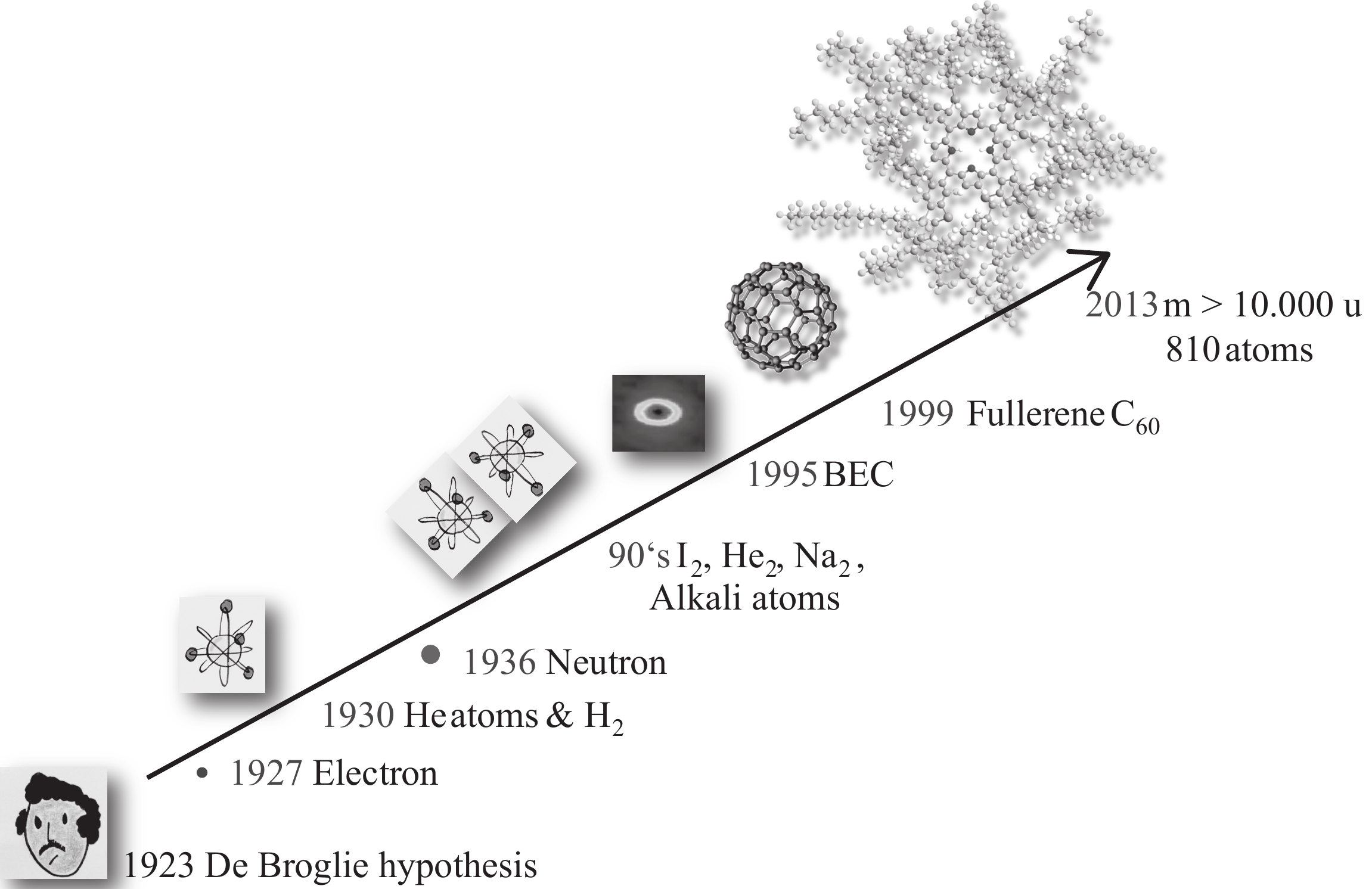}
\caption{Matter-wave interferometry in a historic context. Starting from de Broglie's hypothesis in 1923 \cite{Broglie1923a} the quantum wave nature of matter was demonstrated for electrons \cite{Davisson1927a}, atoms and diatomic molecules \cite{Estermann1930} and neutrons \cite{Halban1936}. Modern breakthroughs in laser physics, nanotechnology, and laser cooling led to the development of atomic matter-wave experiments \cite{Keith1988,Borde1989,Keith1991,Kasevich1991}, and to studies with quantum degenerate coherent ensembles (BEC)~\cite{Anderson1995,Davis1995}. Macromolecule interferometry started with fullerene diffraction in 1999~\cite{Arndt1999} and led to quantum interference with the currently
most complex organic particles~\cite{Eibenberger2013}.}
\label{fig:timeline}
\end{center}
\end{figure}

Macromolecule interferometry builds upon the `shoulders of giants', on many ideas from matter-wave physics with electrons~\cite{Hasselbach2010}, neutrons~\cite{Rauch2000} and atoms~\cite{Cronin2009} which have been developed over the last decades (Fig.\,\ref{fig:timeline}). The birth of quantum coherence experiments with molecules may be dated to the early days of Estermann and Stern in 1930~\cite{Estermann1930},  who diffracted H$_2$ molecules at crystal surfaces.
Pioneering experiments with molecular beams based on high-resolution Ramsey spectroscopy of SF$_6$~\cite{Borde1981} and I$_2$~\cite{Borde1994} which may also be interpreted as de Broglie wave interferometry~\cite{Borde1989}.

Dedicated molecule diffraction experiments at nanofabricated material gratings started in 1994 and enabled proving the existence of the weakly bound He dimer~\cite{Schoellkopf1994}.
Shortly later, a three-grating interferometer was applied to measure the collisional properties of Na$_2$~\cite{Chapman1995}.
Quantum coherence experiments with hot macromolecules were established in 1999 when we studied  diffraction of the fullerenes C$_{60}$ and C$_{70}$~\cite{Arndt1999}.
At an internal temperature of 900\,K or higher these objects share many properties with those of little rocks: With almost 200 internal vibrational and rotational states hot fullerenes can be seen as carrying their own internal heat bath and with increasing temperature they show many phenomena known from macroscopic lumps of condensed matter: they are capable of emitting thermal electrons, they emit a black-body-like electromagnetic spectrum and they even evaporate C$_2$ subunits when heated beyond 1500\,K ~\cite{Kolodney1995}. Yet, on a time scale of several milliseconds, the internal and external degrees of freedom decouple sufficiently not to perturb the quantum evolution of the motional state. Decoherence-free propagation can thus be established, as long as the internal temperature~\cite{Hackermueller2004} and the residual pressure~\cite{Hornberger2003} are small enough.

The fullerene experiments triggered a journey into the world of quantum optics with complex molecules, whose rich internal structure is both a blessing and a curse:
On the one hand, complexity is interesting and it turns out that quantum interferometry provides a well-adapted tool for studying a large variety of internal molecular properties, with the potential for much better accuracy than classical experiments.  On the other hand, many developments are still necessary to prepare ever more complex molecules in pure quantum states of motion. This is particularly challenging since only very few techniques from atomic physics can be carried over to the handling of macromolecules  whose internal states often cannot be addressed individually.

Before diving into the details of quantum manipulation techniques for macromolecules, let us discuss what makes nanoparticle interferometry such an interesting and thriving field. There is a number of answers to this question with two of them providing the key motivation for our current research.
The experiments are driven mainly by the question whether quantum coherence can be maintained at high mass and in the presence of environmental disturbances \cite{Arndt2005}.
Indeed, one might have doubts whether macromolecules can be prepared and kept in a sufficiently coherent state for de Broglie interference. One reason lies in the rich dynamics of internally excited  macromolecules which may lead to internal state changes on the picosecond time scale and to increased interactions with the environment over milliseconds to seconds, depending on the particle structure and temperature.

Meanwhile the original concept of decoherence theory~\cite{Zeh1970, Caldeira1985, Joos1985, Zurek1991} was confirmed experimentally in a number of  settings~\cite{Haroche2013,Hornberger2003, Hackermueller2004}.
In modern matter-wave experiments decoherence can be kept well under control, and we expect that this should still be the case with particles in the mass range of up to $10^7-10^9$\,u using present day technology \cite{Romero-Isart2010,Chang2010,Nimmrichter2011}.

Yet, one may raise the fundamental question whether the linearity of quantum mechanics holds on all mass scales, or whether it breaks down at some point on the way to the macroscopic world, a view denoted as \emph{macrorealism} \cite{Leggett2002a}.
One particular example of a macrorealist theory is the model of continuous spontaneous localization (CSL), which was introduced to solve the quantum measurement problem and to restore objective reality in our everyday world \cite{Ghirardi1990,Bassi2013}.
It is a consistent nonlinear and stochastic modification of the Schr\"odinger equation  compatible with all experimental observations to date. It predicts the wave function of any material particle to spontaneously localize with  a mass dependent rate to an extension of less than 100\,nm.
Other suggestions to modify the Schr\"odinger equation involve gravitational effects, which might become relevant beyond a certain mass scale \cite{Diosi1989, Penrose1996, Giulini2011}.

Such fundamental issues aside, quantum interferometry of complex nanoparticles is developing into a metrological tool.
The ultra-short length scale set by de Broglie interference can be used to measure particle properties by letting them interact with external fields \cite{Nimmrichter2008,Gerlich2008,Gring2010}. Quantum-enhanced molecule and nanoparticle metrology may well become an important method,
in particular with regard to the biomolecular world where increased knowledge of particle properties has an immediate relevance.

Molecule interferometry may also lead to future applications not treated here \cite{Juffmann2013}.
For instance, the molecular interferograms, which are formed during free flight, may be captured on a clean screen, constituting the basis for quantum-assisted molecule lithography. First steps in this direction have been taken in our labs recently~\cite{Juffmann2009}.

In all experiments described below, what matters is the dynamics of the center-of-mass motion. It is the total mass of all atoms in the molecule which determines the de Broglie wavelength and the interference fringe separation. In contrast to that, atomic Bose-Einstein condensates (BEC) are described as a coherent atomic ensemble, whose properties are similar to that of a laser. A laser does not change color when the number of photons is increased; similarly, conventional BEC interference is described by the de Broglie wavelength  of the individual atoms. Only recently, a number of intriguing experiments on BEC squeezing rely on entangled many-particle states~\cite{Eiermann2003,Ockeloen2013}.

The structure of the remaining text is as follows:
 Section~\ref{sec:concepts} presents an overview of different diffraction and interferometer techniques for complex particles. Section~\ref{sec:theory} contains a unified theoretical phase-space description of near-field matter wave physics, which serves as a common framework for all following interferometer implementations.
 The Talbot-Lau (\emph{TL}) interferometer with nano-fabricated gratings is discussed in Section~\ref{sec:tli}. In the following Section~\ref{sec:kdtli} a description of the Kapitza-Dirac-Talbot-Lau (\emph{KDTL}) interferometer with a central optical phase grating is given. An all-optical time-domain matter-wave (\emph{OTIMA}) interferometer with pulsed ionization gratings is then discussed in Section~\ref{sec:otima}.
We conclude in Section~\ref{sec:perspectives} with a perspective on future quantum experiments with high-mass composite materials.

\section{Concepts and tools of coherent nanoparticle manipulation}
\label{sec:concepts}
This section introduces quantum manipulation techniques for nanoparticles. With the term nanoparticle or molecule we will designate objects in the size range between atoms and 100\,nm. We will emphasize the common principles here, leaving the technical details to the following sections.
To make matter-wave interference work, we need to establish coherence in the first place. We need efficient beam splitters and interferometer arrangements which fit existing molecular beam sources, characterized by short de Broglie wavelengths, poor coherence, and low particle fluxes.

\subsection{Coherence preparation}
The degree of motional quantum coherence can be defined as the normalized correlation function between probability amplitudes at different points in space and time. It describes the ability of a particle ensemble to show matter wave interference, i.e.  the extent to which one cannot predict the path individual particles will take in an interferometer arrangement.

The molecules are not required to be coherent with respect to their internal states. This is fortunate since it is almost impossible in practice  to prepare two molecules in the same internal state. This even holds for a few-atom molecule where the density of states grows rapidly with the number of atoms.
Quantum interference can always be established as long as the internal state dynamics remains uncorrelated with the state of motion at all times.

Moreover, also the motional state of the ensemble can be mixed. It is often  tolerable to have a distribution of velocities $v$ and masses $m$ in the ensemble, if the associated de Broglie wavelengths  $\lambda_{dB} = h/mv$ agree to within around 10\%.

As regards the coherence properties, similar rules hold for matter waves as for light fields:  The extension of the beam source,  as well as its distance to the diffraction elements,  determines the transverse spatial coherence.
The width of the wavelength distribution on the other hand, governs the longitudinal (or temporal) coherence of the matter waves. Mathematically, these relations are engraved in the van Cittert-Zernike theorem and the Wiener-Khinchin theorem \cite{Born1993}. The spatial coherence function is determined by the Fourier transform of the intensity transmission function of the source aperture, whereas the longitudinal coherence length is inversely proportional to the wavelength or velocity spread.

Spatial coherence can also be related to the quantum uncertainty principle, providing an intuitive picture. The smaller the source size the larger is the momentum uncertainty of the particles. As a rule of thumb, we may take the coherence width behind a spatially incoherent but monochromatic source of aperture diameter $D$ to be determined by the distance between the first minima of the diffraction lobe under plane wave illumination.
We can therefore estimate the spatial coherence width $W_c$ to grow with the distance $L$ behind the source as: $W_c \simeq L \lambda_{dB} /D$.

The longitudinal (temporal) coherence length $L_c$ is a measure for the spectral purity of the beam, given by $L_c \simeq \lambda^2/\Delta \lambda$  \cite{Born1993}.
Different prefactors of the order of $2\pi$ can be found in the literature, depending on the assumption made about the shape of the wave packet, as well as the definition of its width. Since most macromolecular interference experiments so far were performed with thermal beams, the Maxwell-Boltzmann velocity distribution sets a typical limit to the initial longitudinal coherence of about $L_c \simeq 2\lambda_{dB}$.
The observation of N$^{th}$-order interference requires a coherence length of at least $N \cdot \lambda$.
This can be achieved by using slotted disk~\cite{Scoles1989} or helical velocity selectors~\cite{Szewc2010}, time-of-flight measurements, or by exploiting gravitational free-fall and a selection of ballistic parabolas~\cite{Nairz2000,Brezger2002}.

Since the de Broglie wavelength in nanoparticle interferometry is usually smaller than 10 pm, it is desirable to improve the coherence parameters by preparing a very small and very cold beam source.
The loss-less preparation of micron-sized sources can be realized by covering the thermal source with a tiny slit or by evaporating the desired material only locally using a laser-micro-focus evaporator~\cite{Juffmann2012}. Motional cooling,  however, is still a big challenge.
Over the last decades, laser manipulation and trapping techniques for atoms have progressed to a level that it is nowadays possible to routinely generate ultra-cold coherent atom ensembles.
In contrast to that, cooling of free nanoparticles to less than the temperature of a cryogenic buffer gas remains a true challenge. First experiments on optical feedback cooling~\cite{Gieseler2012} and cavity cooling of nanoparticles~\cite{Kiesel2013,Asenbaum2013} have recently emerged for particles in the size range between 70\,nm and several micrometers. It is however still an open question how to extrapolate these achievements to objects in the mass range of $10^3$ to $10^7$\,u and in an ultra-high vacuum environment.
All molecule diffraction and interference experiments presented here have been performed with conventional beam sources, based on thermal evaporation or sublimation of the material, in some cases followed by adiabatic expansion in a dense seed gas to favor the formation of clusters and to reduce the longitudinal velocity distribution.

\subsection{Far-field diffraction at a nanomechanical grating}
The clearest demonstration of the quantum wave nature of matter is provided by diffraction of particles at a single mechanical grating.
We will therefore start by analyzing the requirements for grating diffraction of massive matter beams~\cite{Arndt1999,Juffmann2012}.
The angular separation of the diffraction peaks
behind a grating of period $d$ is approximately given by $\sin \theta_{\rm diff} = n\cdot \lambda_{dB} / d$, in full analogy to classical wave optics.
To resolve the diffraction fringes in the far-field
the diffraction angle must be greater than the collimation of the molecular beam, $\theta_{\rm diff}  \ge \theta_{\rm coll} $.
Signal constraints, as well as the onset of van der Waals interactions between the molecules and the collimator slits, suggest that the particle beam collimation should not be reduced below $\theta_{\rm coll} \simeq 5\,\mu$rad.
A thermal molecular beam of particles in the mass range of 1000\,u requires high temperatures (500-1000\,K). It therefore operates typically at a most probable velocity of $v=\sqrt{2 k_B T/m}\simeq 200$\,m/s with de Broglie wavelengths around $\lambda = h/mv\simeq 2-5$\,pm.
A tight collimation is needed to fulfill the coherence requirements and for a de Broglie wavelength of 2\,pm the source width has to be as small as 5\,$\mu$m to obtain transverse coherence of 200\,nm in 50\,cm distance behind the source.
These tight constraints imply already the use of diffraction structures with a period around $d=100$\,nm.
Nanofabricated masks this small became available about 25 years ago~\cite{Keith1988}. First realized by electron beam lithography they can nowadays also been inscribed using photo lithography~\cite{Savas1995} as well as focused ion beam lithography~\cite{Juffmann2012}.  Diffraction at mechanical gratings has led to numerous experiments with atoms \cite{Keith1988,Carnal1991}, molecular dimers \cite{Schoellkopf1994,Chapman1995} and complex hot molecules \cite{Arndt1999,Juffmann2012}.
Only recently we were able to demonstrate that even Nature provides nanomasks suitable for molecular diffraction in form of the silified frustule (skeleton) of algae, such as Amphipleura pellucida~\cite{Sclafani2013}.

\subsection{Optical gratings}
The use of mechanical nanomasks with high-mass molecules is challenged by two facts: the attractive van der Waals interaction between the polarizable molecules and the dielectric or  metallic grating wall leads to a strongly velocity-dependent phase modulation of the molecular matter wave~\cite{Grisenti1999,Arndt1999,Nairz2003}. It may even cause molecular loss for large particles that stick to the grating walls~\cite{Juffmann2012a}.
In this case, it seems appealing to replace mechanical structures by elements made of light.

\subsubsection{Measurement induced absorptive gratings}

Every absorptive grating may be viewed as realizing a projective measurement: the periodic arrangement of the slits in a membrane defines a transmission function for the particles to pass. If the particle does not hit the grating bars, its wave function is projected onto the comb of slit openings.

The importance of measurement-induced gratings was described by Storey et al. \cite{Storey1992}, who suggested that optical interactions may serve a similar purpose. For atoms, an optical mask can be realized in the form of a standing light wave which pumps metastable noble gas atoms into undetected states \cite{Abfalterer1997}. Only in the dark nodes of the optical lattice, atoms pass the grating without excitation and can be detected. This idea was used in a full three-grating interferometer with absorptive light gratings, where the atoms were optically shelved in undetected magnetic states~\cite{Fray2004}. The idea can be generalized to a much larger class of particles if we replace the resonant atom-light coupling by single-photon ionization in an ultraviolet standing light wave grating. This proposal \cite{Reiger2006,Nimmrichter2011} was only recently realized for the first time in our lab~\cite{Haslinger2013} and applied to clusters of molecules, as discussed in Section 6.

\subsubsection{Optical phase gratings}
Nanomechanical and optical absorption gratings are typically accompanied by an additional phase contribution. In mechanical structures this is caused by the van der Waals interaction between the molecule and the grating wall. In optical diffraction elements it arises due to the interaction between the optical polarizability of the particle  and the electric field of the laser grating.
The concept of phase gratings goes back to the idea of Kapitza and Dirac who suggested in 1933 that electrons might be reflected at a standing light wave due to the ponderomotive potential~\cite{Kapitza1933}. Diffraction of particles at light was first observed with neutral atoms~\cite{Moskowitz1983} via the dipole interaction.
The dipole force accounts for the fact that the electric laser field induces an electric dipole moment in the particle which in turn interacts with the electric field of the laser beam. The dipole interaction potential modulates the phase of the matter wave.

As compared to nanoparticles, atomic resonance lines are often narrow and the atom-field interaction can be strongly enhanced when the detuning $\delta= \omega_L-\omega_A$ between the laser frequency and the atomic transition frequency is small, i.e. comparable to the transition line width $\Gamma$.
The atomic response to an external light field may vary by to about five orders of magnitude across the excitation spectrum with resonance widths of the order of  $\Gamma \simeq 10-100$\,MHz.
In contrast to that, electronic transitions in complex molecules can be as broad as 1-50\,THz, including a range of vibrational and rotational states.
Even far-off resonance laser beams may still exert a force on the particles but their magnitude is substantially smaller than in the resonant atomic case.

Pure phase gratings with little absorption were first realized with atoms~\cite{Gould1986,Martin1988}.
In complex molecules there is often also a chance for a photon to be swallowed~\cite{Nairz2001}, a process not always followed by the emission of light.
In the case of the fullerenes C$_{60}$ and C$_{70}$, for example, the optical excitation of a singlet state $S_0 \rightarrow S_1$ is followed by a non-radiative intersystem crossing $S_1 \rightarrow T_1$ with a probability greater than $99\%$.
The triplet is supposed to live at least a dozen microseconds and if it decays to the ground state it does so non-radiatively \cite{Dresselhaus1998}. Generally, in large molecules a great number of dissipative processes compete with electro-magnetic emission. Such internal energy conversion does not affect the coherence of the center-of-mass evolution.

\subsection{Matter-wave interferometry in the time domain}
Early matter-wave experiments were operated with continuous beams of electrons and neutrons.
A first time-dependent aspect of diffraction was emphasized when Moshinski \cite{Moshinski1952} proposed the possibility of neutron diffraction at a rapidly opening shutter.
This idea was realized in 1998~\cite{Hils1998}, two years earlier in different variants with cold atoms, which could be diffractively reflected at time-modulated potentials of light~\cite{Steane1995,Szriftgiser1996}. In these experiments the emphasis was put on exploiting energy-time uncertainty in analogy to the position-momentum uncertainty relation in the more common diffraction in position space.


Here we focus on time-dependent (pulsed) gratings and the momentum redistribution during the diffraction process. In the paraxial approximation, we may restrict our considerations to the dynamics of the one-dimensional transverse wave function, regardless of its longitudinal position.
\begin{figure}
\begin{center}
\includegraphics[width=.45\textwidth]{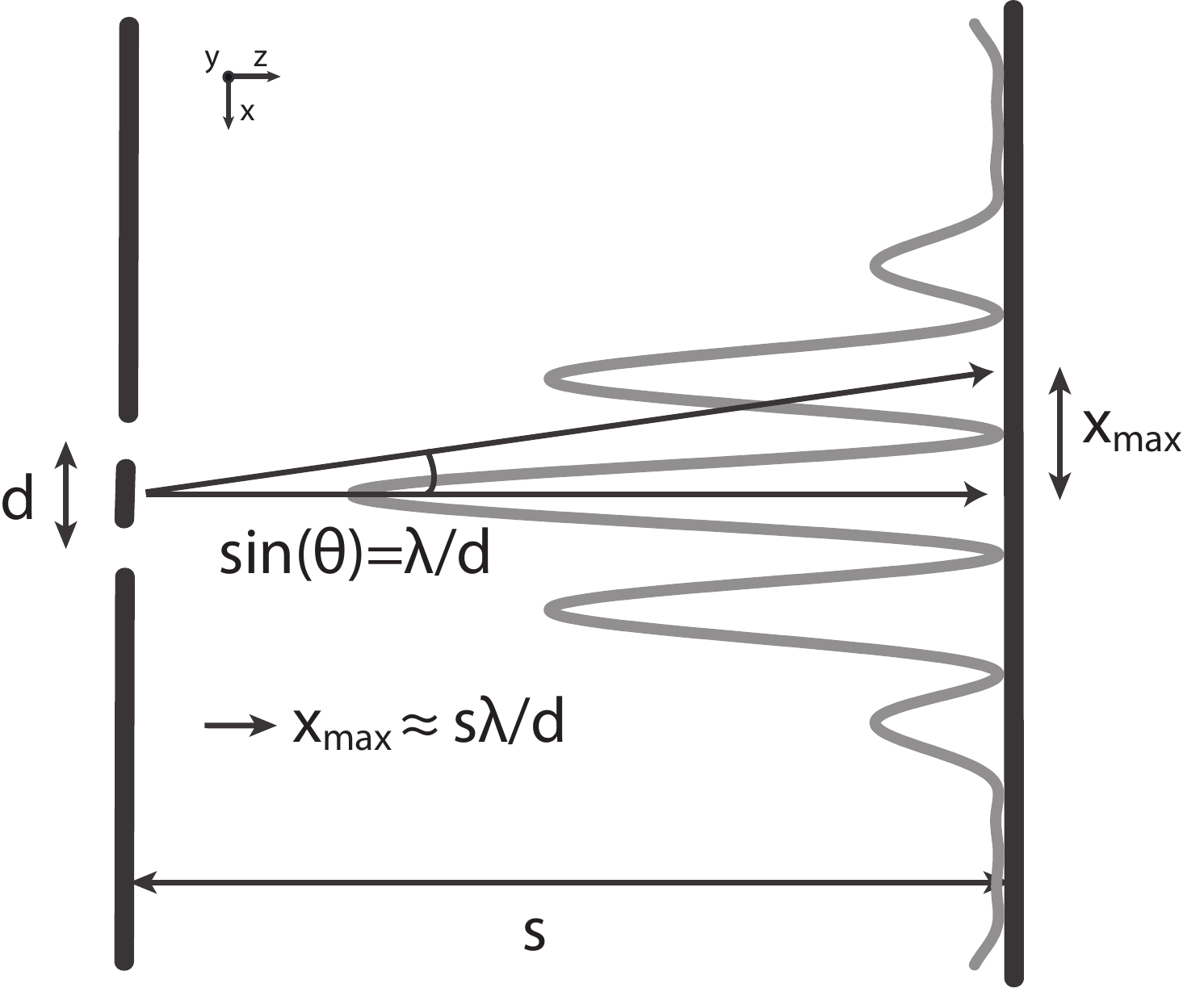}\quad
\includegraphics[width=.48\textwidth]{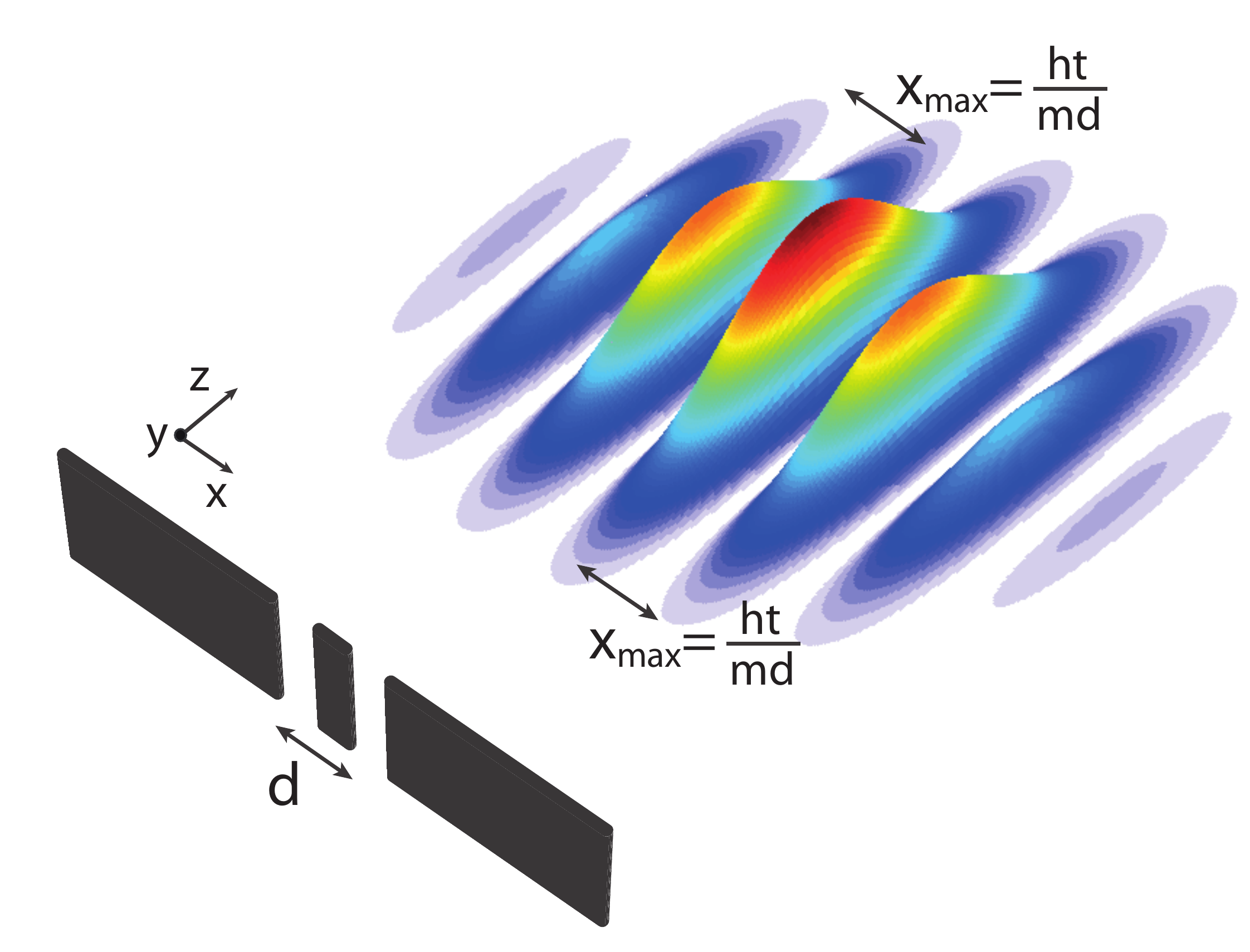}
\caption{Far-field diffraction behind a double slit leads to the sketched interference fringes. Adjacent fringe maxima are separated in momentum by multiples of the grating momentum $\Delta p=h/d$, which depends on the grating period $d$. This alone defines the distance between the fringes after a given flight time $t$.
Molecules of different velocity will travel different distances, but if they are observed at the same time, they will show the same fringe pattern.}
\label{fig:timedomain}
\end{center}
\end{figure}
As a function of time, the interference maxima behind a grating of period $d$ expand like $x_{max}=ht/md $, independently of the velocity. The latter comes into play only in stationary setups, where the detection plane is placed at a fixed distance $L=vt$ behind the diffraction element. In terms of the de Broglie wavelength, the separation then reads as $x_{max}=\lambda_{dB} L/d$.
The momentum of the particle is redistributed in units of the grating momentum $p_{d}=h/d$, irrespectively of velocity or mass (see Figure~\ref{fig:timedomain}).

Since 1991 several implementations of precision atom interferometry have followed the example of Kasevich and Chu~\cite{Kasevich1991a} in implementing atom interferometry with pulsed Raman beam splitters. Time-domain Talbot-Lau interferometry with laser phase gratings  was later demonstrated for cold atoms \cite{Cahn1997,Turlapov2005} and ultra-cold Bose Einstein condensates~\cite{Deng1999}. Most recently, we extended this concept to interferometry with clusters of molecules using pulsed ionization gratings~\cite{Haslinger2013}.

Pulsed optical beam splitters offer many advantages: Depending on the beam splitter mechanism, the interference fringe shifts can be largely independent of the particle velocity. A certain degree of selection is needed only to ensure that every detected molecule has interacted sequentially with all three gratings. Optical gratings are widely adaptable in situ and over time.
Ionization gratings and, more generally photo-depletion gratings are rather insensitive to frequency fluctuations;
they are rugged and indestructible. Modern lasers define gratings with a precise periodicity and allow one to time their separation in time with nanosecond accuracy.

\subsection{From far-field to near-field diffraction and near-field interferometry}
In textbooks on classical optics the effect of diffraction is usually described in the Fraunhofer limit of long distances behind the grating. The interference pattern can then be understood as the Fourier transform of the aperture transmission function. In the near field~\cite{Born1993} the theoretical description involves Kirchhoff-Fresnel integrals, which account for the wave front curvature.

Far-field interference is easily understood as a wave phenomenon since the separated diffraction orders constitute a clear signature of wave behavior. However, this requires a tight collimation which dramatically reduces the transmitted signal. In experiments on molecule diffraction less than one billionth of all emitted molecules contribute to the final interferogram~\cite{Juffmann2012a}.
\begin{figure}
\begin{center}
\includegraphics[width=.5\textwidth]{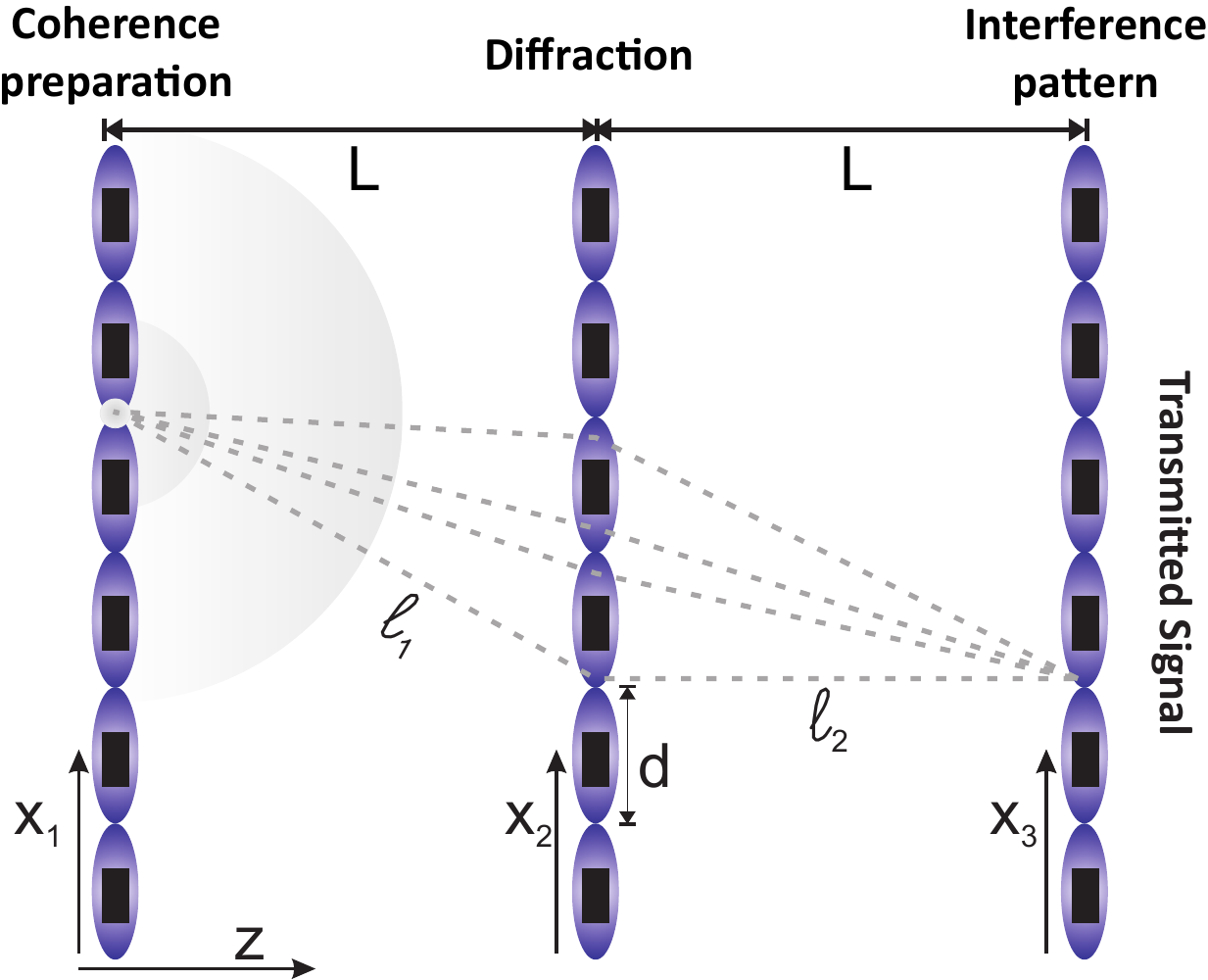}
\caption{Concept of Talbot-Lau interferometry: Spatially incoherent molecular beams can be brought to interference in a three-grating setup where the first grating serves as an absorptive filter that acts as a periodic array of coherent sources. This grating can be realized by a nanomechanical mask or by a standing light wave that depletes the molecular beam via ionization or fragmentation.
Diffraction at a second grating with the same period leads to interference at the position of the third grating. The resulting molecular density pattern has the same periodicity as the gratings; it is resolved by scanning the third grating over the fringes. This setup was used in optics \cite{Patorski1989}, x-ray imaging \cite{Pfeiffer2006} and with atoms \cite{Clauser1994}. Here we use it with complex molecules in three different grating configurations: with material gratings (Talbot-Lau) \cite{Brezger2002}, with two material and one phase grating (KDTLI) \cite{Gerlich2007}, and with three pulsed ionization gratings (OTIMA)\cite{Haslinger2013}.} \label{fig:TLI_Idea}
\end{center}
\end{figure}

Talbot-Lau (TL) near-field interferometers circumvent this problem (see Figure \ref{fig:TLI_Idea}).
They use at least two, in most cases three gratings of the same grating period.
The TL setting is based on the observation by H.F. Talbot in 1836 \cite{Talbot1836} that under coherent illumination wavelength-dependent self-images will form behind a periodic structure, at multiples of the Talbot time (or Talbot length)
\begin{equation}
T_{T}=\frac{md^{2}}{h}\quad\text{or}\quad   L_{T}=\frac{mvd^{2}}{h}=\frac{d^{2}}{\lambda_{dB}}.\label{eq:TalbotTimeLength}
\end{equation}
This is known as the Talbot effect and was observed with collimated beams of atoms \cite{Chapman1995,Nowak1997}. The strong collimation requirement is alleviated in the TL setup by adding two further gratings of equal period. The first grating acts as a mask and prepares matter-wave coherence at a distance $L$, which must be at least of the order of the Talbot length. There, the second grating diffracts the molecular center-of-mass wave function.
An interference pattern with the periodicity $d$ is formed at the same distance $L$ behind the second grating, and a third grating can be used to spatially probe the molecular density pattern.
The detected signal in Talbot-Lau experiments can exceed the throughput of far-field experiments by more than a factor of ten thousand: The first grating effectively realizes thousands of tiny molecular beam sources, and the detection can be extended over large areas, either by direct surface imaging or by a highly multiplexing detector, part of which is the third grating in the TL-interferometer.

The TL concept has proven fruitful in optical interferometry for several decades~\cite{Patorski1989}. Its generalization to atom interferometry is due to John F. Clauser~\cite{Clauser1994} who demonstrated the effect for potassium atoms and also proposed to extrapolate it to 'small rocks and little viruses'~\cite{Clauser1997}.
Since the year 2002, a series of experiments in our group at the University of Vienna has then explored increasingly massive and complex particles
in a number of different interferometers that built on this concept~\cite{Hornberger2012,Juffmann2013} with important adaptations.
The first Talbot-Lau interferometer with three mechanical gratings was realized in 2002~\cite{Brezger2002}. It was used for collisional~\cite{Hornberger2003} and thermal decoherence~\cite{Hackermueller2004} studies, for our first steps into biomolecule interferometry~\cite{Hackermueller2003} and molecule metrology~\cite{Berninger2007}. The original three-grating interferometer was adapted to an interferometer with two mechanical gratings and one atomically clean surface detector in 2009~\cite{Juffmann2009}.
The strong dephasing effects caused by van der Waals forces between the molecules and the nanogratings motivated us to propose~\cite{Brezger2003} and realize~\cite{Gerlich2007} Kapitza-Dirac-Talbot-Lau interferometry (KDTLI), as described in Section\,5. In order to generalize the concept of light gratings,  OTIMA interferometry was established as described in Section\,6~\cite{Reiger2006,Nimmrichter2011,Haslinger2013}.

Care has, however, to be taken: In all these cases the detection of a periodic fringe pattern might in principle also be mimicked by a moir{\'e} shadow effect~\cite{Oberthaler1996,Berman1997}, under certain circumstances.
A detailed quantitative analysis, provided in the following section, is therefore needed to distinguish genuine quantum interference effects from classical ballistic trajectories~\cite{Brezger2002,Brezger2003,Hornberger2009}.
\section{A unified phase-space description of three-grating matter wave interferometry}
\label{sec:theory}

%

\newcommand{\Sp}{\text{tr}}
\newcommand{\tr}[1]{\text{tr}\left( #1 \right)}
\newcommand{\trA}[1]{\text{tr}_A \left( #1 \right)}
\newcommand{\trB}[1]{\text{tr}_B \left( #1 \right)}
\newcommand{\trS}[1]{\text{tr}_S \left( #1 \right)}
\newcommand{\trC}[1]{\text{tr}_C \left( #1 \right)}
\newcommand{\trM}[1]{\text{tr}_M \left( #1 \right)}
\newcommand{\qerw}[1]{\left\langle #1 \right\rangle}
\newcommand{\erw}[1]{ {\cal E} \left[ #1 \right]}
\newcommand{\req}[1]{\stackrel{\text{(\ref{#1})}}{=}}
\newcommand{\teq}[1]{\stackrel{\text{#1}}{=}}
\newcommand{\ito}{(\text{\textbf{I}})\,}
\newcommand{\strat}{(\text{\textbf{S}})\,}

\newcommand{\vac}{\text{vac}}

\newcommand{\eps}{\varepsilon}
\newcommand{\vareps}{\varepsilon}
\newcommand{\id}{\text{id}}

\newcommand{\nn}{\nonumber \\}

\newcommand{\la}{\langle}
\newcommand{\ra}{\rangle}
\newcommand{\lla}{\left\langle}
\newcommand{\rra}{\right\rangle}
\newcommand{\ida}{{\dagger}}

\newcommand{\da}{ \mathrel{\reflectbox{\rotatebox[origin=c]{180}{\smaller$\dagger$}}} }

\newcommand{\diff}{\text{\rm d}}
\newcommand{\vnabla}{\overrightarrow{\nabla}}
\newcommand{\grad}{\text{\rm grad\,}}
\newcommand{\rot}{\text{\rm rot}}
\newcommand{\divergenz}{\text{\rm div}}
\newcommand{\difft}{\frac{\diff}{\diff t}}
\newcommand{\diffx}{\frac{\diff}{\diff x}}
\newcommand{\pdx}{\partial_x}
\newcommand{\pdy}{\partial_y}
\newcommand{\pdz}{\partial_z}
\newcommand{\pdt}{\partial_t}

\newcommand{\ind}[1]{\int \text{\rm d}^{#1}} 
\newcommand{\indd}{\int \text{\rm d}} 
\newcommand{\iindp}[3]{\iint \frac{\text{\rm d}^{#1} #2 \, \text{\rm d}^{#1} #3 }{\left(2 \pi \hbar \right)^{#1}}}
\newcommand{\iinddp}[2]{\iint \frac{\text{\rm d} #1 \, \text{\rm d} #2 }{2 \pi \hbar}}
\newcommand{\iind}[3]{\iint \text{\rm d}^{#1} #2 \, \text{\rm d}^{#1} #3 }
\newcommand{\iindd}[2]{\iint \text{\rm d} #1 \, \text{\rm d} #2 }
\newcommand{\suminf}[1]{\sum_{#1 = -\infty}^{\infty}} 

\newcommand{\casedis}[1]{\left\{ \begin{array}{ll} #1 \end{array} \right.}

\newcommand{\svekt}[1]{\left(#1\right)}
\newcommand{\ex}{\mathbf{e}_x}
\newcommand{\ey}{\mathbf{e}_y}
\newcommand{\ez}{\mathbf{e}_z}
\newcommand{\ve}{\boldsymbol{e}}
\newcommand{\vpi}{\boldsymbol{\pi}}
\newcommand{\vk}{\boldsymbol{k}}
\newcommand{\vl}{\boldsymbol{\ell}}
\newcommand{\vx}{\boldsymbol{r}}
\newcommand{\vy}{\boldsymbol{y}}
\newcommand{\va}{\boldsymbol{a}}
\newcommand{\vb}{\boldsymbol{b}}
\newcommand{\vc}{\boldsymbol{c}}
\newcommand{\vf}{\boldsymbol{f}}
\newcommand{\vn}{\boldsymbol{n}}
\newcommand{\vX}{\boldsymbol{X}}
\newcommand{\vA}{\boldsymbol{A}}
\newcommand{\vT}{\boldsymbol{T}}
\newcommand{\vD}{\boldsymbol{D}}
\newcommand{\vd}{\boldsymbol{d}}
\newcommand{\vE}{\boldsymbol{E}}
\newcommand{\vY}{\boldsymbol{Y}}
\newcommand{\vp}{\boldsymbol{p}}
\newcommand{\vs}{\boldsymbol{s}}
\newcommand{\vS}{\boldsymbol{S}}
\newcommand{\vsigma}{\boldsymbol{\sigma}}
\newcommand{\vGamma}{\boldsymbol{\Gamma}}
\newcommand{\vq}{\boldsymbol{q}}
\newcommand{\vQ}{\boldsymbol{Q}}
\newcommand{\vB}{\boldsymbol{B}}
\newcommand{\vH}{\boldsymbol{H}}
\newcommand{\vP}{\boldsymbol{P}}
\newcommand{\vu}{\boldsymbol{u}}
\newcommand{\vv}{\boldsymbol{v}}
\newcommand{\vw}{\boldsymbol{w}}
\newcommand{\vj}{\boldsymbol{j}}
\newcommand{\vR}{\boldsymbol{R}}
\newcommand{\vW}{\boldsymbol{W}}
\newcommand{\vxi}{\boldsymbol{\xi}}
\newcommand{\veta}{\boldsymbol{\eta}}
\newcommand{\vchi}{\boldsymbol{\chi}}
\newcommand{\vF}{\boldsymbol{F}}
\newcommand{\veps}{\boldsymbol{\epsilon}}

\newcommand{\C}{\mathbb {C}}
\newcommand{\PP}{\mathbb {P}}
\newcommand{\N}{\mathbb {N}}
\newcommand{\Z}{\mathbb {Z}}
\newcommand{\Q}{\mathbb {Q}}
\newcommand{\R}{\mathbb {R}}
\newcommand{\K}{\mathbb {K}}

\newcommand{\HR}{ {\cal H} } 
\newcommand{\A}{ {\cal A} }
\newcommand{\B}{ {\cal B} }
\newcommand{\G}{ {\cal G} }
\newcommand{\X}{ {\cal X} }
\newcommand{\cC}{ {\cal C} }
\newcommand{\Lag}{ {\cal L} }
\newcommand{\cL}{ {\cal L} }
\newcommand{\F}{ {\cal F} } 
\newcommand{\FT}[2]{ {\cal F}_{#1} \left[ #2 \right] } 
\newcommand{\cS}{ {\cal S} }
\newcommand{\cR}{ {\cal R} }
\newcommand{\cP}{ {\cal P} }
\newcommand{\cQ}{ {\cal Q} }
\newcommand{\T}{ {\cal T} }
\newcommand{\V}{ {\cal V} }
\newcommand{\D}{ {\cal D} }
\newcommand{\J}{ {\cal J} }
\newcommand{\cI}{ {\cal I} }
\newcommand{\Order}[1]{ {\cal O} \left( #1 \right) } 
\newcommand{\I}{\mathbb {I}} 

\newcommand{\Op}[1]{\mathsf #1} 
\newcommand{\oH}{ \Op{H} } 
\newcommand{\oU}{ \Op{U} } 
\newcommand{\oA}{ \Op{A} } 
\newcommand{\oE}{ \Op{E} } 
\newcommand{\oF}{ \Op{F} } 
\newcommand{\oB}{ \Op{B} } 
\newcommand{\oC}{ \Op{C} } 
\newcommand{\oD}{ \Op{D} } 
\newcommand{\oS}{ \Op{S} } 
\newcommand{\oG}{ \Op{G} } 
\newcommand{\oT}{ \Op{T} } 
\newcommand{\oL}{ \Op{L} } 
\newcommand{\ovL}{ \text{\sf \textbf{L}}} 
\newcommand{\oV}{ \Op{V} } 
\newcommand{\oW}{ \Op{W} }
\newcommand{\oX}{ \Op{X} }
\newcommand{\oY}{ \Op{Y} }
\newcommand{\oZ}{ \Op{Z} }
\newcommand{\oz}{ \Op{z} }
\newcommand{\oO}{ \Op{O} }
\newcommand{\oM}{ \Op{M} }
\newcommand{\oN}{ \Op{N} }
\newcommand{\on}{ \Op{n} }
\newcommand{\om}{ \Op{m} }
\newcommand{\oR}{ \Op{R} } 
\newcommand{\ovR}{ \text{\sf \textbf{R}}} 
\newcommand{\ovP}{ \text{\sf \textbf{P}}} 
\newcommand{\ovr}{ \text{\sf \textbf{r}}} 
\newcommand{\ovp}{ \text{\sf \textbf{p}}} 
\newcommand{\ovQ}{ \text{\sf \textbf{Q}}} 
\newcommand{\oa}{ \Op{a} } 
\newcommand{\ob}{ \Op{b} } 
\newcommand{\ok}{ \Op{k} } 
\newcommand{\oK}{ \Op{K} } 
\newcommand{\og}{ \Op{g} } 
\newcommand{\oh}{ \Op{h} } 
\newcommand{\op}{ \Op{p} } 
\newcommand{\oq}{ \Op{q} } 
\newcommand{\oP}{ \Op{P} } 
\newcommand{\oQ}{ \Op{Q} } 
\newcommand{\ox}{ \Op{x} } 
\newcommand{\OP}{ \text{\sf \textbf{p}}} 
\newcommand{\OX}{ \text{\sf \textbf{r}}} 

\newcommand{\opsi}{ \hat{\psi} }
\newcommand{\ochi}{ \hat{\chi} }

\newcommand{\euler}{\text{\rm e}}
\newcommand{\imu}{\text{\rm i}}
\newcommand{\field}[1]{\mathds{#1}}
\newcommand{\re}{\text{Re}}
\newcommand{\im}{\text{Im}}
\newcommand{\sgn}{\text{\rm sgn}}
\newcommand{\sinc}{\text{\rm sinc}}
\newcommand{\Si}{\text{\rm Si}}
\newcommand{\erf}{\text{\rm erf}}
\newcommand{\sincc}[1]{\text{\rm sinc}\, \left( #1 \right)}
\newcommand{\const}{\text{const}}

\newcommand{\dd}{\frac{d}{d_1}}
\newcommand{\eik}{\text{\rm eik}}
\newcommand{\LLlambda}{\dfrac{L}{L_{\lambda}}}
\newcommand{\Llambda}{L_{\lambda}}
\newcommand{\ih}{\frac{i}{\hbar}}
\newcommand{\pihh}{\frac{1}{2\pi \hbar}} 
\newcommand{\pih}[1]{\frac{1}{\left(2\pi \hbar \right)^{#1}}} 

\newcommand{\phii}[1]{\phi \left(#1 \right)}
\newcommand{\chii}[1]{\chi \left(#1 \right)}
\newcommand{\psii}[1]{\psi \left(#1 \right)}
\newcommand{\varthetaa}[1]{\vartheta \left(#1 \right)}
\newcommand{\varthetas}[1]{\vartheta_S \left(#1 \right)}
\newcommand{\phic}[1]{\phi^{*} \left(#1 \right)}
\newcommand{\chic}[1]{\chi^{*} \left(#1 \right)}
\newcommand{\psic}[1]{\psi^{*} \left(#1 \right)}
\newcommand{\rhoo}[1]{\rhoo \left(#1 \right)}
\newcommand{\expp}[1]{\exp \left\{#1 \right\}}

The conventional description of matter-wave interference and diffraction
at fixed apertures involves solving the Kirchhoff-Fresnel diffraction
integral for the wave function. This is justified in the presence of stationary potentials and as long as we can approximate the source as emitting pure quantum states.
In most molecular quantum delocalization experiments so far, the longitudinal distance between the diffractive mask and the detection
screen was also much larger than the size of both the aperture
and the interference pattern. This permits us to change to a time-domain
treatment by eliminating the longitudinal $z$-coordinate. The paraxial
approximation allows us to describe the propagation of the transverse
single-particle wave function in a co-moving frame. Diffraction at
a grating structure, which varies only in the $x$-direction, transforms only the one-dimensional wave function $\psi\left(x\right)$.
In a stationary setup the propagation time $t=L/v_{z}$ is
related to the screen distance $L$ via the longitudinal velocity
$v_{z}$. One thus evaluates the interference pattern for fixed $v_z$ and then averages over the velocity distribution.

\subsection{The Wigner function representation}

In the following we will sketch the theory behind the Talbot-Lau interferometer
(TLI) scheme using a one-dimensional phase-space model. The latter
applies to both the stationary KDTLI setup and the pulsed OTIMA setup
(as well as to the conventional TLI schemes with material gratings
not discussed here). The description is based on the Wigner function
formalism \cite{Schleich2001}, a most suitable and elegant representation for
near-field interference phenomena. Given the density operator $\rho$
that describes the reduced one-dimensional quantum state of motion in the transverse direction,
the Wigner function reads as\begin{equation}
w\left(x,p\right)=\frac{1}{2\pi\hbar}\int\diff s\, e^{ips/\hbar}\la x-\frac{s}{2}|\rho|x+\frac{s}{2}\ra.\label{eq:wignerfunction}\end{equation}
It is normalized to $\int\diff x\diff p\, w\left(x,p\right)=1$, a
real-valued function, but not necessarily positive. In fact, its negativity
implies the presence of quantum coherence. The main advantage for
our purposes lies in its natural implementation of the quantum-classical
correspondence: Not only does it reproduce the same marginal (position
and momentum) distributions as its classical counterpart, e.g. $\int\diff p\, w\left(x,p\right)=\la x|\rho|x\ra$.
The Wigner function is actually positive and identical to the corresponding
classical phase-space distribution $f\left(x,p\right)$ for mixed
states that do not exhibit quantum coherence. Moreover, $w\left(x,p\right)$
and $f\left(x,p\right)$ share the same time evolution along classical
trajectories when they are subject to at most harmonic potentials \cite{Schleich2001}.
Given a constant acceleration $a$, for instance, the Wigner function
evolves in time by means of the shearing-displacement transformation\begin{equation}
w_{t}\left(x,p\right)=w_{0}\left(x-\frac{pt}{m}+a\frac{t^{2}}{2},p-mat\right),\label{eq:wigner_freeprop}\end{equation}
and so does $f_{t}\left(x,p\right)$.

Putting everything together, the Wigner function is a convenient tool
to assess near-field quantum interference and to compare it with the
moir\'{e} shadow effect arising from a hypothetical classical description.
Moreover, we will see later that standard environmental decoherence
and phase averaging effects are also easily incorporated into the
description; they render the quantum and the classical description indistinguishable.

\subsection{Grating diffraction in phase space\label{sec:gratingtrafo}}

Essentially, two ingredients are required for the coherent description
of Talbot-Lau interferometry in phase space: the transformation \eqref{eq:wigner_freeprop}
corresponding to the free propagation of the particles between the
gratings and the grating transformation, which is to be discussed
now. As already noted above, we can resort to an effective one-dimensional
treatment of the interferometer, because the relevant diffraction
effects are restricted to the transverse $x$-coordinate of the grating
structure. The grating transformation shall
not influence notably the motion of the incident particles along the
$y$-axis and the longitudinal $z$-axis. This is guaranteed in a
short-time high-energy scattering limit, where the transverse wave
function of an incident particle is subject to a local scattering
transformation of the form \cite{Nimmrichter2008a}
\begin{equation}
\psi\left(x\right)\mapsto t\left(x\right)\psi\left(x\right)=\left|t\left(x\right)\right|\exp\left[i\phi\left(x\right)\right]\psi\left(x\right).\label{eq:gratingtrafo_psi}
\end{equation}
We denote by $t\left(x\right)$ the complex transmission function
of the $d$-periodic grating. Disregarding the finite grating size,
the function can be expanded into the Fourier sum $t\left(x\right)=\sum_{n}b_{n}\exp\left(2\pi inx/d\right)$.
The standard textbook example of an incident plane wave transforms
into a superposition of plane waves,
\begin{equation}
\exp\left(\frac{ipx}{\hbar}\right)\mapsto\sum_{n=-\infty}^{\infty}b_{n}\exp\left[\ih\left(p+n\frac{h}{d}\right)x\right],
\end{equation}
which illustrates the emergence of discrete diffraction orders shifted
by multiples of the grating momentum $h/d$. The basic effect,
i.e.~the redistribution to different momenta, is the same both if the grating modulates the amplitude or the phase of an incident matter wave. An absorptive mask differs from a pure phase grating merely in terms of the transmission
probability, $\left|t\left(x\right)\right|^{2}=\sum_{n}A_{n}\exp\left(2\pi inx/d\right)$,
which equals unity in the latter case.

The redistribution of momenta in steps of $h/d$ due to diffraction
also shows up explicitly in the phase-space picture. The grating transformation
\eqref{eq:gratingtrafo_psi} translates into an integral transformation
of the Wigner function, $w\left(x,p\right)\mapsto\int\diff q\, T\left(x,p-q\right)w\left(x,q\right)$,
with the convolution kernel \cite{Hornberger2004,Nimmrichter2008a} \begin{align}
T\left(x,p\right) & =\frac{1}{2\pi\hbar}\int\diff s\, e^{ips/\hbar}t\left(x-\frac{s}{2}\right)t^{*}\left(x+\frac{s}{2}\right)\nonumber \\
 & =\sum_{n}\exp\left(\frac{2\pi inx}{d}\right)\sum_{j}b_{j}b_{j-n}^{*}\delta\left(p-\frac{2j-n}{2}\frac{h}{d}\right).\label{eq:Tq}\end{align}
It will be convenient to introduce the so-called Talbot coefficients
at this point, \begin{equation}
B_{n}\left(\xi\right)=\sum_{j}b_{j}b_{j-n}^{*}\exp\left[i\pi\left(n-2j\right)\xi\right],\label{eq:Bn_q}\end{equation}
and write\begin{equation}
T\left(x,p\right)=\frac{1}{2\pi\hbar}\sum_{n}\exp\left(\frac{2\pi inx}{d}\right)\int\diff s\, e^{ips/\hbar}B_{n}\left(\frac{s}{d}\right).\label{eq:Tq_TalbotCoeff}\end{equation}

\subsubsection{Thin stationary gratings for fast particles}
For stationary grating arrangements, the above local scattering transformation
\eqref{eq:gratingtrafo_psi} is valid for paraxial matter-wave beams
and thin gratings. To be specific, we ignore the transverse $x$-motion
of the particles while they are passing through the grating. For the
case of, say, a material grating mask of thickness $b$ and slit opening
$s=fd$, and given a matter-wave beam divergence angle $\alpha$,
we require that $b\tan\alpha/fd\ll1$. The Fourier components of the
aperture function $\left|t\left(x\right)\right|^{2}$ then read as
$A_{n}=B_{n}\left(0\right)=f\sinc\left(\pi nf\right)$, with $f=s/d$
the slit opening fraction.

In addition, the phase of the incident matter waves is modulated in
the presence of an interaction potential $V\left(x,z\right)$ between
the particles and the grating. In the limit of high longitudinal velocities
$v_{z}$, where the kinetic energy of the particles is large compared
to the interaction strength, the modulation is approximately given
by the so-called eikonal phase \cite{Glauber1959,Nimmrichter2008a},
\begin{equation}
\phi\left(x\right)=-\frac{1}{\hbar v_{z}}\int\diff z\, V\left(x,z\right).\label{eq:eikophase}
\end{equation}
Note that this adds to the velocity dependence of stationary Talbot-Lau
interference, where the propagation time between the gratings is determined
by $L/v_{z}$. This additional dependence can, in fact, get prohibitively
severe for heavy molecules in material gratings due to the strong
van der Waals attraction between the particles and the grating walls.
When averaged over realistic velocity distributions this could kill
the interference effect entirely \cite{Gerlich2007}.

The KDTLI scheme discussed here employs a standing-wave laser of wavelength
$\lambda$ and power $P_{L}$ as a pure phase grating for polarizable
nanoparticles. Given the polarizability $\alpha$, the particle interacts
with the standing-wave field $\vE\propto\sqrt{f\left(y,z\right)}\cos\left(2\pi x/\lambda\right)$
via the optical potential $V=-\alpha\left|\vE\right|^{2}/4$. This
yields the phase\begin{equation}
\phi\left(x\right)=\phi_{0}\cos^{2}\left(\frac{2\pi x}{\lambda}\right),\qquad\text{with }\,\,\phi_{0}=\frac{8\alpha P_{L}}{hc\eps_{0}w_{y}w_{z}}\int\frac{\diff z}{v_{z}}\, f\left(0,z\right)=\frac{4\sqrt{2\pi}\alpha P_{L}}{hc\eps_{0}w_{y}v_{z}}.\label{eq:phi0_KDTLI}\end{equation}
We assume a Gaussian laser profile, $f\left(y,z\right)=2\exp\left[-2\left(y/w_{y}\right)^{2}-2\left(z/w_{z}\right)^{2}\right]/\pi w_{y}w_{z}$,
with a waist $w_{y}$ that is large compared to the width of the matter-wave
beam, whereas the longitudinal waist $w_{z}$ shall be small (thin
grating). The standing-wave grating period amounts to $d=\lambda/2$.

We find that the Fourier coefficients of the transmission function
of the standing-wave phase grating, $t\left(x\right)=\exp\left[i\phi\left(x\right)\right]$,
are given by Bessel functions, $b_{n}=i^{n}\exp\left(-i\phi_{0}/2\right)J_{n}\left(\phi_{0}/2\right)$.
The explicit form of the corresponding Talbot coefficients \eqref{eq:Bn_q}
follows from an addition theorem for Bessel functions \cite{Abramowitz1965},\begin{equation}
B_{n}\left(\xi\right)=J_{n}\left[\zeta_{{\rm coh}}\left(\xi\right)\right],\qquad{\rm with}\,\,\zeta_{{\rm coh}}\left(\xi\right)=\phi_{0}\sin\pi\xi.\label{eq:Bn_q_KDTLI}\end{equation}
In practice, one may need to include momentum averaging effects due
to the absorption or Rayleigh scattering of grating photons \cite{Hornberger2009,Nimmrichter2011a};
they result in modified Talbot coefficients depending on the absorption
and the scattering cross-section of the particles. However, these
are mostly minor, negligible modifications in all the cases presented
here.

\subsubsection{Short ionizing grating pulses}
The gratings in the OTIMA scheme are short standing-wave laser pulses,
and the local scattering transformation \eqref{eq:gratingtrafo_psi}
holds when the particles are approximately at standstill during the
pulse length $\tau$. Given the integrated pulse energy $E_{L}=\int_{\tau}\diff t\, P_{L}\left(t\right)$,
and assuming that the laser spot profile is much wider than the illuminated
particle ensemble \cite{Nimmrichter2011a}, the eikonal phase shift
can be written as\begin{equation}
\phi_{0}=\frac{4\pi\alpha E_{L}}{hc\eps_{0}}f\left(0,0\right).\end{equation}
The ionizing UV standing-wave pulses, however, are not just pure phase
gratings. It is, in fact, required for the Talbot-Lau scheme to work
that the first and the third grating are amplitude-modulating masks
to establish initial coherence and to
implement a position-resolving detection of the interference pattern.
Here, single-photon ionization is employed to achieve this task, as
many nanoparticles, e.g. metal clusters, ionize upon the absorption
of a single UV energy quantum.

We model the photon absorption as a Poisson process, where the probability
of absorbing $k$ photons from the pulse, $P_{k} (x)=\exp\left[-n\left(x\right)\right]n^{k}\left(x\right)/k!$,
is determined by \begin{equation}
n\left(x\right)=n_{0}\cos^{2}\left(\frac{2\pi x}{\lambda}\right),\qquad\text{with }\,\, n_{0}=\frac{4\sigma_{{\rm abs}}E_{L}\lambda}{hc\eps_{0}}f\left(0,0\right).\label{eq:n0_OTIMA}\end{equation}
The term $n_{0}$ denotes the mean number of absorbed photons at the
antinodes of the standing wave, with $\sigma_{{\rm abs}}$ the absorption
cross-section of the particles. The latter is typically written as
the imaginary part of the complex dipole polarizability, $\chi=\alpha+i\lambda\eps_{0}\sigma_{{\rm abs}}/2\pi$.
For example, the complex polarizability of a nanosphere of radius
$R$ and dielectric permittivity $\eps$ of its material reads as
\cite{Kreibig1995} $\chi=4\pi\eps_{0}R^{3}\left(\eps-1\right)/\left(\eps+2\right)$.

Assuming that the first absorbed photon ionizes the particle, and
that the ions are removed from the ensemble, the transmission probability
is given by $\left|t\left(x\right)\right|^{2}=P_{0}=\exp\left[-n\left(x\right)\right]$.
This leaves us with the following transmission function and its Fourier
components;\begin{equation}
t\left(x\right)=\exp\left[\left(i\phi_{0}-\frac{n_{0}}{2}\right)\cos^{2}\frac{\pi x}{d}\right],\qquad b_{n}=\exp\left(i\frac{\phi_{0}}{2}-\frac{n_{0}}{4}\right)I_{n}\left(i\frac{\phi_{0}}{2}-\frac{n_{0}}{4}\right),\end{equation}
with $I_{n}$ a modified Bessel function. Once again, the Talbot coefficients
\eqref{eq:Bn_q} are obtained from an addition theorem \cite{Nimmrichter2011a},\begin{equation}
B_{n}\left(\xi\right)=e^{-n_{0}/2}\left[\frac{\zeta_{{\rm coh}}\left(\xi\right)-\zeta_{{\rm ion}}\left(\xi\right)}{\zeta_{{\rm coh}}\left(\xi\right)+\zeta_{{\rm ion}}\left(\xi\right)}\right]^{n/2} \\ J_{n}\left[\sgn\left\{ \zeta_{{\rm coh}}\left(\xi\right)+\zeta_{{\rm ion}}\left(\xi\right)\right\} \sqrt{\zeta_{{\rm coh}}^{2}\left(\xi\right)-\zeta_{{\rm ion}}^{2}\left(\xi\right)}\right],\label{eq:Bn_q_OTIMA}\end{equation}
with $\zeta_{{\rm ion}}\left(\xi\right)=\left(n_{0}/2\right)\cos\pi\xi$
and the same phase modulation term $\zeta_{{\rm coh}}\left(\xi\right)$
as in \eqref{eq:Bn_q_KDTLI}. The ratio between the amplitude and
the phase modulation in the grating is conveniently described by the
material parameter\begin{equation}
\beta=\frac{n_{0}}{2\phi_{0}}=\frac{\im\chi}{\re\chi}=\frac{\sigma_{{\rm abs}}\eps_{0}\lambda}{2\pi\alpha},\end{equation}
which does not depend on the grating intensity. The OTIMA scheme is
intended for strongly absorbing particles with $\beta\gtrsim1$.

\subsubsection{Classical pendant of the grating transformation\label{sec:classGrating}}
A fringe modulation that occurs in the particle density in the near
field behind the grating does not necessarily indicate quantum interference.
Hypothetically speaking, shadow fringes could emerge just as well
if classical particles traversed the grating on ballistic trajectories.
A quantitative analysis is therefore necessary to clearly distinguish
true wave interference from the predictions of a classical ballistic
model.

A classical density modulation behind the grating could arise due
to two effects: (i) a modulation of the incident particle ensemble
by absorptive grating masks blocking a fraction of the trajectories
(e.g. by the walls between adjacent slits in a material grating),
and (ii) a classical lensing effect due to interaction forces exerted
on the particles by the grating.

Substituting the above Wigner function \eqref{eq:wignerfunction}
with the positive phase-space distribution $f\left(x,p\right)$ of
a classical particle ensemble, we can discuss the classical counterpart
of the grating transformation \eqref{eq:Tq}. The masking effect (i)
is simply achieved by a multiplication of the incident particle state
with the transmission probability of the grating, $f\left(x,p\right)\mapsto\left|t\left(x\right)\right|^{2}f\left(x,p\right)$.
For the lensing effect (ii), on the other hand, one must in principle
know the precise deflection of the trajectories through the grating.
In full analogy to the above eikonal approximation \eqref{eq:eikophase}
for the quantum case, we can approximate the particle deflection by
the momentum kick $q\left(x\right)=\hbar\partial_{x}\phi\left(x\right)$,
which represents the integrated transverse force acting on the particle
at position $x$ in the grating (neglecting its motion). The momentum
kick is again $d$-periodic and it transforms $f\left(x,p\right)\mapsto f\left(x,p-q\left(x\right)\right)$.

Putting the two effects together, the classical grating transformation
can be expressed in terms of the convolution kernel \begin{equation}
T_{{\rm cl}}\left(x,p\right)=\left|t\left(x\right)\right|^{2}\delta\left[p-\hbar\partial_{x}\phi\left(x\right)\right]=\frac{1}{2\pi\hbar}\sum_{n}\exp\left(\frac{2\pi inx}{d}\right)\int\diff s\, e^{ips/\hbar}C_{n}\left(\frac{s}{d}\right).\label{eq:Tcl}\end{equation}
We see that it can be brought to the same form as the quantum kernel \eqref{eq:Tq_TalbotCoeff},
but with different, classical Talbot coefficients\begin{equation}
C_{n}\left(\xi\right)=\frac{1}{d}\int_{-d/2}^{d/2}\diff x\,\left|t\left(x\right)\right|^{2}\exp\left[-\frac{2\pi inx}{d}-i\xi d\partial_{x}\phi\left(x\right)\right].\label{eq:Bn_cl}\end{equation}
It turns out that the classical terms differ from the quantum expressions
\eqref{eq:Bn_q_KDTLI} and \eqref{eq:Bn_q_OTIMA} for the laser gratings
in the KDTLI and the OTIMA scheme by the substitution \cite{Hornberger2009,Nimmrichter2011a}
$\zeta_{{\rm coh}}\left(\xi\right)\mapsto\phi_{0}\pi\xi$ and $\zeta_{{\rm ion}}\left(\xi\right)\mapsto n_{0}/2$.

\subsection{The Talbot self-imaging effect}
With the phase-space grating transformation at hand, we can discuss
the elementary near-field diffraction effect that forms the basis
of the Talbot-Lau experiments studied here: the Talbot effect \cite{Talbot1836,Patorski1989}.
It states that a periodic grating structure illuminated by a perfect
plane wave is reconstructed in the density distribution at integer
multiples of the Talbot time behind the grating. This recurring Talbot
image is the result of constructive near-field interference between
all the outgoing wavelets diffracted at the grating.

This effect is easily assessed in phase space. Suppose that an eigenstate
of zero transverse momentum, $w_{0}\left(x,p\right)\propto\delta\left(p\right)$,
illuminates a periodic grating $t\left(x\right)$. The diffracted
Wigner function, $w_{1}\left(x,p\right)=T\left(x,p\right)$, shall
then propagate freely (and without acceleration), according to Equation
\eqref{eq:wigner_freeprop}, and the matter-wave density distribution
$w_{t}\left(x\right)=\la x|\rho|x\ra=\int\diff p\, w_{t}\left(x,p\right)$
is subsequently recorded after a time $t$. Using the explicit form
of the grating transformation kernel \eqref{eq:Tq_TalbotCoeff}, we
find\begin{equation}
w_{t}\left(x\right)=\int\diff p\, T\left(x-\frac{pt}{m},p\right)=\sum_{n}B_{n}\left(n\frac{t}{T_{T}}\right)\exp\left(\frac{2\pi inx}{d}\right),\label{eq:w_TalbotEffect}\end{equation}
with $T_{T}$ the Talbot time as defined in \eqref{eq:TalbotTimeLength}.
This density distribution describes a periodic fringe pattern, where
the time-dependent Fourier amplitudes are given by the Talbot coefficients
\eqref{eq:Bn_q} of the grating.

\begin{figure}
\begin{center}
\includegraphics[width=\textwidth]{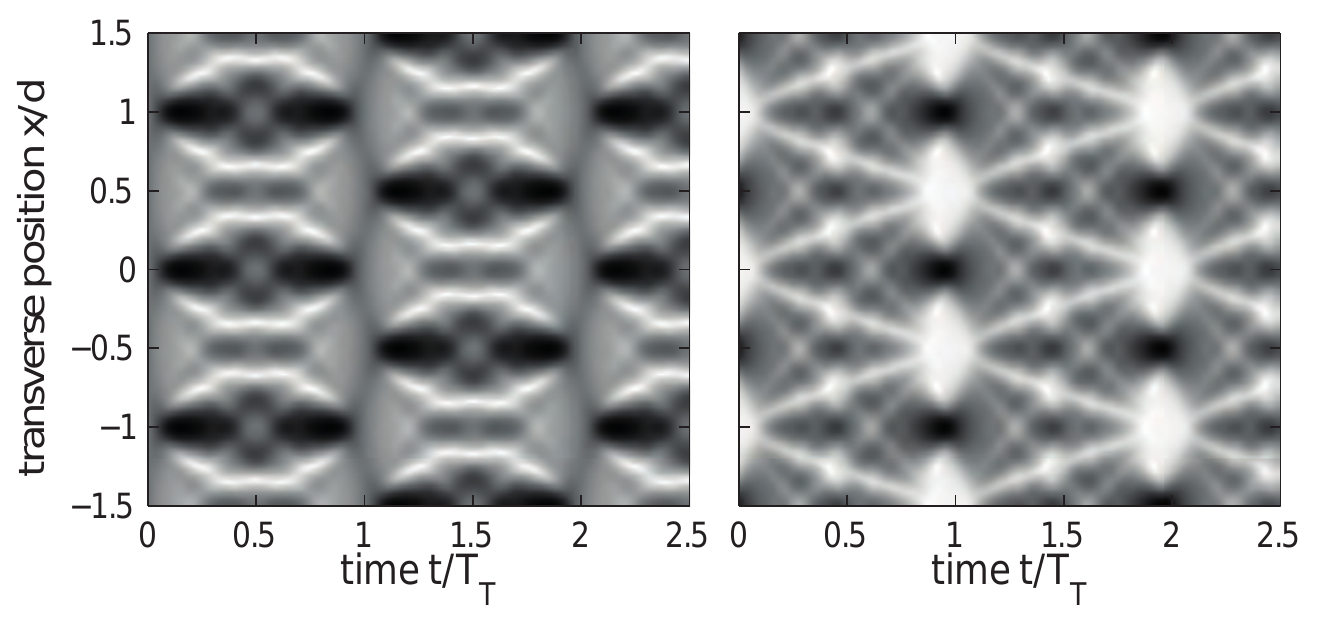}
\caption{\label{fig:TalbotCarpet}Simulated density distribution as function of time
$t$ (in units of the Talbot time) and transverse position $x$ (in
units of the grating period) in the near field behind a standing-wave
phase grating (left) and behind an ionizing laser grating (right). High density regions are dark. The gratings are illuminated by a perfectly collimated matter-wave state $|p=0\ra$. The phase parameter is set to $\phi_{0}=\pi$ in
both cases, and $\beta=1.0$ is assumed for the ionizing grating.
Talbot images of the grating profile occur at integer multiples of
the Talbot time.}
\end{center}
\end{figure}

Figure \ref{fig:TalbotCarpet} depicts two exemplary Talbot carpets,
i.e. matter-wave density distributions behind a standing-wave phase
grating, as used in the KDTLI scheme (left panel), and behind an ionizing
grating, as used in the OTIMA scheme (right panel). The Talbot effect
can be observed at distinct times behind the grating, which are given
by integer multiples of the Talbot time, $t=NT_{T}$. Here, the Talbot
coefficients reduce to $B_{n}\left(N\right)=\left(-\right)^{nN}B_{n}\left(0\right)$
and the interference pattern \eqref{eq:w_TalbotEffect} mimics the
grating mask profile (shifted by half periods),\begin{equation}
w_{NT_{T}}\left(x\right)=\left|t\left(x+N\frac{d}{2}\right)\right|^{2}.\end{equation}
In the case of a pure phase grating, where $\left|t\left(x\right)\right|^{2}=1$,
the Talbot images exhibit no amplitude modulation (see left panel),
while strong fringe oscillations appear in between adjacent Talbot
times.

\begin{figure}
\begin{center}
\includegraphics[width=\textwidth]{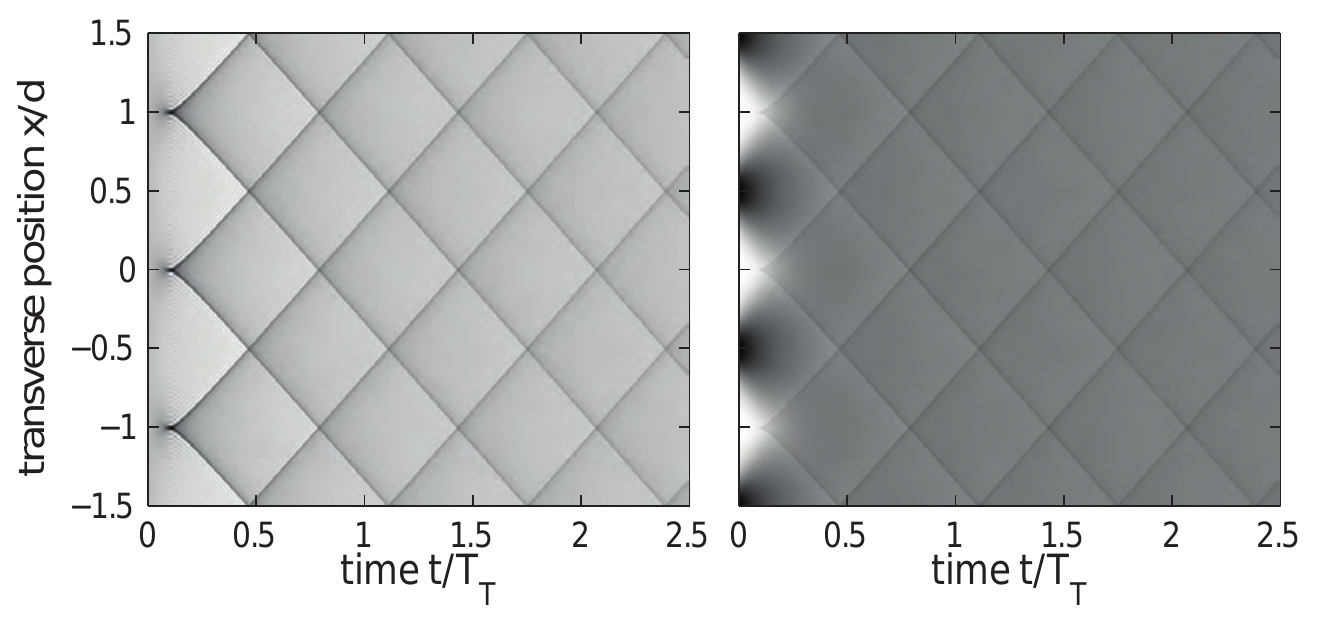}
\caption{\label{fig:classTalbotCarpet}Hypothetical density distribution in
the near field behind a standing-wave phase grating (left) and behind
an ionizing laser grating (right) illuminated by classical ballistic
particles. The same parameters are used as in Figure \ref{fig:TalbotCarpet}.}
\end{center}
\end{figure}

The recurring Talbot image of the grating profile is a characteristic
feature of quantum wave interference, whereas the classical ballistic
description given in sec.~\ref{sec:classGrating} leads to significantly
different results. If we replace the Talbot coefficients in \eqref{eq:w_TalbotEffect}
by their classical counterparts \eqref{eq:Bn_cl} for the standing-wave
phase grating and for the ionizing grating, we obtain the classical
Talbot carpets depicted in fig.~\ref{fig:classTalbotCarpet}. Talbot
images of the grating profile do not occur here, whereas a classical
lensing effect is clearly visible.

\subsection{Talbot-Lau interference in phase space}
The observation of recurring Talbot images and Talbot carpets like
the ones depicted in fig.~\ref{fig:TalbotCarpet} requires strictly
coherent illumination. As in the case of far-field diffraction, matter
waves must be collimated to less than the grating momentum $h/d$,
which is impractical for most high-mass interference experiments.
A single narrow collimation slit would simply throw away too much
of the matter-wave signal.

This problem can be circumvented by placing another grating of the
same period before the actual interference grating%
\footnote{Different grating periods are also possible \cite{Nimmrichter2008a},
but their ratio must be chosen carefully.%
}. The first grating, which must be an absorptive mask, then generates
sufficient matter-wave coherence for the second one while transmitting
a reasonably large fraction of the particles. In principle, the initial
particle ensemble before the first grating need not have any transverse
coherence at all; still, high-contrast interference fringes can emerge
at specific distances (times) behind the second grating, as will be
described in the following before we move on to a realistic model
of the KDTLI and the OTIMA scheme.

\subsubsection{Coherent description}
Let us start with a poorly collimated particle ensemble represented
by an incoherent mixture of momenta, $\rho_{0}=\int\diff p\, D\left(p\right)|p\ra\la p|$,
with $D\left(p\right)\geq0$ a normalized momentum distribution. Its
characteristic width $\Delta p$ shall extend over many grating momenta,
$\Delta p\gg h/d$, which gives a coherence length of the order of
$\hbar/\Delta p\ll d$ and means that the state is basically unsuitable
for diffraction at a grating structure of period $d$. The effect
of the grating is merely that of a classical transmission mask.

The Wigner function of the initial matter-wave state, $w_{0}\left(x,p\right)=D\left(p\right)/\Delta x$,
is indistinguishable from a classical particle ensemble with the same
momentum distribution. For simplicity, we neglect any fringe effects
related to the finite spatial extension of the state assuming that
it uniformly covers many grating periods, $\Delta x\gg d$. It allows
us to work with strictly periodic functions, and the grating transformation
given in Section \ref{sec:gratingtrafo} maps the state into\begin{equation}
w_{1}\left(x,p\right)=\frac{1}{\Delta x}\int\diff q\, T\left(x,p-q\right)D\left(q\right)=\frac{1}{2\pi\hbar\Delta x}\sum_{n}e^{2\pi inx/d}\int\diff s\, B_{n}\left(\frac{s}{d}\right)\widetilde{D}\left(s\right)e^{ips/\hbar}.\end{equation}
Here, the Fourier transform of the momentum distribution, $\widetilde{D}\left(s\right)=\int\diff p\, D\left(p\right)e^{-ips/\hbar}$,
is restricted to arguments $s\sim\hbar/\Delta p\ll d$, which allows
us to approximate the Talbot coefficients by the Fourier components
of the grating mask, $B_{n}\left(s/d\right)\approx B_{n}\left(0\right)=A_{n}$.
In other words, the broad initial momentum spread is essentially unaltered
by the grating and the transmitted state is simply given by the masked
Wigner function\begin{equation}
w_{1}\left(x,p\right)\approx\frac{1}{\Delta x}\sum_{n}e^{2\pi inx/d}A_{n}D\left(p\right)=\frac{1}{\Delta x}D\left(p\right)\left|t\left(x\right)\right|^{2}.\label{eq:w1}\end{equation}
The same expression holds in the classical case, but this does not
mean that no quantum coherence is present. In fact, the matter waves
emerging from the grating slits begin to delocalize as time evolves.
For the moment, let us omit acceleration in the free state evolution
\eqref{eq:wigner_freeprop}. After a time $T_{1}$, nondiagonal elements
of the density matrix (i.e. spatial coherences) appear,\begin{equation}
\la x-\frac{s}{2}|\rho|x+\frac{s}{2}\ra=\int\diff p\, w_{1}\left(x-\frac{pT_{1}}{m},p\right)e^{-ips/\hbar}=\sum_{n}A_{n}\widetilde{D}\left(s+nd\frac{T_{1}}{T_{T}}\right)e^{2\pi inx/d}.\end{equation}
This coherence function exhibits a dominant peak of magnitude $\left|A_{1}\right|$
at the values $s=\pm dT_{1}/T_{T}$. Hence, adjacent slits of a second
grating placed at $T_{1}=T_{T}$ would be coherently illuminated by
the matter waves---the basis of the Talbot-Lau effect. Clearly, the
coherence is sharply limited to those $s$-values where the contributions
of matter waves emerging from adjacent source slits of the first grating
add up in phase.

In a sense, each slit of the first grating acts as a single partly
coherent matter-wave source for a Talbot image of the second grating
\cite{Brezger2003}, and it overlaps with the offset images of all other
source slits. A visible interference pattern can be expected only
at specific resonance times behind the second grating where the offsets
match the fringe oscillations. Let us work out how this is delivered
by our theoretical model.

The state \eqref{eq:w1} emerging from the first grating G$_{1}$
shall evolve for the time $T_{1}$ according to \eqref{eq:wigner_freeprop},
now including the possibility for an external acceleration $a\neq0$.
It then transforms as \eqref{eq:Tq_TalbotCoeff} at a second grating
G$_{2}$ of the same period, which results in the Wigner function\begin{align}
w_{2}\left(x,p\right) & =\int\diff q\, T^{(2)}\left(x,p-q\right)w_{1}\left(x-\frac{qT_{1}}{m}+\frac{aT_{1}^{2}}{2},q-maT_{1}\right)\nonumber \\
 & =\frac{1}{2\pi\hbar\Delta x}\sum_{n,\ell}A_{n}^{(1)}\exp\left\{ \frac{2\pi i}{d}\left[\left(n+\ell\right)x-n\frac{aT_{1}^{2}}{2}\right]\right\} \nonumber \\
 & \quad\times\int\diff s\, B_{\ell}^{(2)}\left(\frac{s}{d}\right)\widetilde{D}\left(s+\frac{nhT_{1}}{md}\right)\exp\left[\ih\left(p-maT_{1}\right)s\right].\label{eq:w2}\end{align}
Superscript labels $(k)$ are used to distinguish the gratings G$_{k}$.
We are looking for a fringe modulation in the probability density
$w_{3}\left(x\right)=\int\diff p\, w_{3}\left(x,p\right)$ to find
a particle at position $x$ after another time $T_{2}$ behind G$_{2}$,
\begin{align}
w_{3}\left(x\right) & =\int\diff p\, w_{2}\left(x-\frac{pT_{2}}{m}+\frac{aT_{2}^{2}}{2},p-maT_{2}\right)\nonumber \\
 & =\frac{1}{\Delta x}\sum_{n,\ell}A_{n}^{(1)}B_{\ell-n}^{(2)}\left(\frac{\ell T_{2}}{T_{T}}\right)\widetilde{D}\left(\frac{\ell T_{2}+nT_{1}}{T_{T}}d\right)\nonumber \\
 & \quad\times\exp\left\{ \frac{2\pi i}{d}\left[\ell\left(x-\frac{aT_{2}^{2}}{2}-aT_{1}T_{2}\right)-n\frac{aT_{1}^{2}}{2}\right]\right\} .\end{align}
This is a $d$-periodic Fourier sum, where the sharply peaked function
$\widetilde{D}$ restricts the Fourier amplitudes to those summation
indices that fulfill $\left|\ell T_{2}+nT_{1}\right|\ll T_{T}$. There
is at most one index $n$ for each Fourier order $\ell$ that lies
within this bound, since we require times $T_{1,2}$ of the order
of the Talbot time to distinguish between quantum interference and
classical shadow fringes. A pronounced fringe modulation, which is
carried by the lowest harmonic orders $\ell=\pm1,\pm2,\ldots$, appears
only if the corresponding index $n$ of the Talbot coefficients of
the first grating is also small.

This \emph{resonance condition }selects few specific time ratios $T_{1}/T_{2}$
at which high-contrast Talbot-Lau fringes are possible; the times
before and after the second grating cannot be varied independently.
Consider for instance a deviation of $5\%$ from the obvious resonance
$T_{1}=T_{2}$, say $T_{1}/T_{2}=0.95$. Then the first possible Fourier
contribution would come from the high-order index pair $\left(\ell,n\right)=\left(\pm19,\mp20\right)$,
an undetectable fringe oscillation at $19$ times the fundamental
frequency!

We restrict our view here to the standard case $T_{1}=T_{2}\equiv T$,
as used in all molecular Talbot-Lau experiments. It yields a pronounced
fundamental fringe oscillation; the interference pattern reduces to\begin{equation}
w_{3}\left(x\right)=\frac{1}{\Delta x}\sum_{\ell}A_{-\ell}^{(1)}B_{2\ell}^{(2)}\left(\frac{\ell T}{T_{T}}\right)\exp\left[\frac{2\pi i\ell}{d}\left(x-aT^{2}\right)\right].\end{equation}
In most Talbot-Lau experiments, a movable third grating mask G$_{3}$
of the same period is used in combination with a mass-spectrometric
particle detection scheme in order to resolve the interference fringes.
The detection signal is proportional to the fraction of particles
transmitted through G$_{3}$ as a function of its lateral position
$x_{s}$ relative to the other gratings,\begin{align}
S\left(x_{s}\right) & =\sum_{\ell}A_{-\ell}^{(1)}A_{-\ell}^{(3)}B_{2\ell}^{(2)}\left(\frac{\ell T}{T_{T}}\right)\exp\left[\frac{2\pi i\ell}{d}\left(x_{s}-aT^{2}\right)\right]\nonumber \\
 & \propto\int\diff x\, w_{3}\left(x\right)\left|t^{(3)}\left(x-x_{s}\right)\right|^{2}.\label{eq:Signal}\end{align}
The fringes are recorded by varying the lateral shift $x_{s}$. Their
contrast can be quantified in terms of the visibility\begin{equation}
\V=\frac{\max_{x}\left\{ S\left(x\right)\right\} -\min_{x}\left\{ S\left(x\right)\right\} }{\max_{x}\left\{ S\left(x\right)\right\} +\min_{x}\left\{ S\left(x\right)\right\} }\in\left[0,1\right],\label{eq:Vis}\end{equation}
but in practice it suffices to measure the contrast in terms of the
sinusoidal visibility\begin{equation}
\V_{\sin}=2\left|\frac{A_{1}^{(1)}A_{1}^{(3)}B_{2}^{(2)}\left(T/T_{T}\right)}{A_{0}^{(1)}A_{0}^{(2)}A_{0}^{(3)}}\right|.\label{eq:VisSin}\end{equation}
The latter is naturally obtained as the ratio between the amplitude
and the offset of a sine curve fitted to the measurement data---a
more noise-robust quantity than $\V$ involving all data points and
not only the greatest and smallest ones.

In the following, we will apply the developed model to specific Talbot-Lau
schemes employing optical gratings. They will differ mainly in the
Talbot coefficients of the three gratings, see Section \ref{sec:gratingtrafo}.
The quantum predictions are straightforwardly compared to a hypothetical
classical model of ballistic particles: One must merely replace the
Talbot coefficients $B_{n}^{(2)}\left(\xi\right)$ of the second grating
with their classical counterparts \eqref{eq:Bn_cl}. Everything else
remains the same.

We note once again that the quantum Talbot coefficients \eqref{eq:Bn_q}
are periodic in the argument $\xi$, whereas the classical ones in
\eqref{eq:Bn_cl} are not. Specifically, we find $B_{n}^{(2)}\left(\xi\right)=\left(-\right)^{n\xi}A_{n}^{(2)}$
for integer $\xi$, which explains the recurring images of the second
grating mask at times $T=NT_{T}$.

\subsubsection{The KDTLI setup}

The KDTLI setup is a stationary configuration of two material grating
masks G$_{1,3}$ and a standing-wave phase grating G$_{2}$, see fig.~\ref{fig:KDTLI_setup}. The Talbot coefficients of the phase grating were defined
in \eqref{eq:Bn_q_KDTLI}, and the Fourier coefficients of the material
masks are $A_{n}^{(1,3)}=f_{1,3}\sinc\left(\pi nf_{1,3}\right)$,
given the slit opening fractions $f_{1,3}$. Due to the stationary
grating arrangement, the propagation time $T=L/v_{z}$ between the
gratings depends on the longitudinal velocity of the particles. The
interference signal must therefore be averaged over the velocity distribution
$\mu\left(v_{z}\right)$ of the matter-wave beam,\begin{align}
S\left(x_{s}\right) & =f_{1}f_{3}\sum_{\ell}\sinc\left(\pi\ell f_{1}\right)\sinc\left(\pi\ell f_{3}\right)\exp\left(\frac{2\pi i\ell x_{s}}{d}\right)\nonumber \\
 & \quad\times\int\diff v_{z}\,\mu\left(v_{z}\right)J_{2\ell}\left[\phi_{0}\sin\left(\pi\ell\frac{L}{L_{T}}\right)\right].\label{eq:SigKDTLI}\end{align}
The velocity dependence is contained in the phase modulation parameter
$\phi_{0}$, defined in \eqref{eq:phi0_KDTLI}, and in the Talbot
length $L_{T}=v_{z}T_{T}$. External acceleration is omitted here
as the gratings are oriented horizontally. In deflectometry experiments,
where an electric field is applied to induce a polarizability-dependent
acceleration, a velocity-dependent phase shift of the fringe pattern
must be inserted again.

Note that the interference contrast vanishes at integer multiples
of the Talbot length, $L=NL_{T}$, when the argument of the Bessel
function is zero---a consequence of the standing-wave phase grating.
Maximum fringe contrast is expected to recur between two consecutive
Talbot orders, depending on the strength of the phase modulation $\phi_{0}$.
The sinusoidal visibility reads as

\begin{equation}
\V_{\sin}=2\left|\sinc\left(\pi f_{1}\right)\sinc\left(\pi f_{3}\right)\int\diff v_{z}\,\mu\left(v_{z}\right)J_{2}\left[\phi_{0}\sin\left(\pi\frac{L}{L_{T}}\right)\right]\right|.\label{eq:VisSinKDTLI}\end{equation}
The classical shadow visibility is obtained by replacing $\sin\left(\pi L/L_{T}\right)$
with $\pi L/L_{T}$ in the argument of the Bessel function. The expressions
given here are valid for particles with vanishing absorption and scattering
cross-sections at the wavelength of the grating laser. The fringe
visibility is lower for particles that absorb or scatter a significant
number of standing-wave photons.

\subsubsection{The OTIMA setup}

The pulsed OTIMA setup is a time-domain Talbot-Lau scheme, where all
three gratings are realized by ionizing standing-wave pulses, see
fig.~\ref{fig:otimasetup}. Given the Talbot coefficients defined in \eqref{eq:Bn_q_OTIMA},
the detection signal reads as
\begin{align}
S\left(x_{s}\right) & =\exp\left(-\frac{n_{0}^{(1)}+n_{0}^{(2)}+n_{0}^{(3)}}{2}\right)\sum_{\ell}I_{\ell}\left(\frac{n_{0}^{(1)}}{2}\right)I_{\ell}\left(\frac{n_{0}^{(3)}}{2}\right)\exp\left[\frac{2\pi i\ell}{d}\left(x_{s}-aT^{2}\right)\right]\nonumber \\
 & \quad\times\left[\frac{\zeta_{{\rm coh}}\left(\ell T/T_{T}\right)-\zeta_{{\rm ion}}\left(\ell T/T_{T}\right)}{\zeta_{{\rm coh}}\left(\ell T/T_{T}\right)+\zeta_{{\rm ion}}\left(\ell T/T_{T}\right)}\right]^{\ell}J_{2\ell}\left(\sqrt{\zeta_{{\rm coh}}^{2}\left(\ell\frac{T}{T_{T}}\right)-\zeta_{{\rm ion}}^{2}\left(\ell\frac{T}{T_{T}}\right)}\right).\label{eq:SigOTIMA}
 \end{align}
In the experiment, the power of each grating pulse can be varied individually,
which admits different mean absorption numbers $n_{0}^{(1,2,3)}$,
as defined in \eqref{eq:n0_OTIMA}.

In the (current) OTIMA setup all three standing-wave
gratings are formed by the same mirror surface, and so the lateral
position $x_{s}$ of the detection grating cannot be shifted in a
straightforward manner. There are however means to implement and control
an effective shift without using an independent mirror. One practiced
method is to slightly tilt the second laser beam. It leads to a negligible
change in the standing-wave period that accumulates to a significant
phase shift $x_{s}$ many wavelengths away from the mirror surface
\cite{Haslinger2013}. Another method would be to apply an external deflection
field, which yields a controllable acceleration $a$ and grating shift
$\delta x=aT^{2}$. In the current experimental realization, the gratings
are vertically oriented and gravitational acceleration $a=g$ is actually
present. It plays no role as long as the interfered particles are
small and the interference times short. Nevertheless, one could implement
a significant grating shift for heavy particles with greater Talbot
times by varying the pulse delay time $T$.

\subsection{The influence of environmental decoherence\label{sec:dec_theory}}

The destructive influence of the environment on the Talbot-Lau interference
effect is an important factor to consider in the high-mass regime.
Decoherence by gas collisions and by the emission and absorption of
thermal radiation imposes strict vacuum and temperature conditions
in order to observe interference fringes. These conditions are described
quantitatively by incorporating the relevant decoherence processes
into our phase-space model.

As far as the center-of-mass motion of particles is concerned, all
relevant free-space decoherence effects are related to the momentum
transfer in random elastic or inelastic scattering events with environmental
degrees of freedom. Each decoherence process is characterized by an
event rate $\Gamma\left(t\right)$ (which could depend on time) and
by a probability distribution $g\left(\vq\right)$ of momentum $\vq$
transferred in a single event. The effect on the interference pattern
can be understood by simple stochastic considerations; a more formal
argument is found in \cite{Hornberger2004}.

Suppose that a scattering event occurs at the time $t\in\left[-T,T\right]$
before or after the second grating in a Talbot-Lau setup, and that
it transfers the momentum $\vq=\left(q_{x},q_{y},q_{z}\right)$ onto
the particle; only the $x$-component $q_{x}$ will influence the
fringe pattern. The Wigner function of the particle then transforms
as $w\left(x,p\right)\mapsto w\left(x,p-q_{x}\right)$ at the time
$t$ before (after) G$_{2}$. Recalling once again the free time evolution
\eqref{eq:wigner_freeprop} on rectilinear trajectories in phase space,
it is as if the first (third) grating was laterally displaced by $\delta x=q_{x}\left(T-\left|t\right|\right)/m$.
The greatest displacement occurs when the event happens immediately
at the second grating, $t=0$. In Fourier space, the displacement
is represented by phase factors $\exp\left(2\pi in\delta x/d\right)$
for each Fourier component $A_{n}^{(1,3)}$ of the first (third) grating.

Such displacements by scattering are now randomly distributed due
to the uncontrollable nature of the interaction. We do not know the
exact momentum recoil $\vq$ transferred to the particle in the scattering
event, and so we must average the phase factor in Fourier space over
the distribution $g\left(\vq\right)$. This results in modified Fourier
amplitudes, which are all (except $n=0$) reduced by a factor smaller
than unity, \begin{equation}
A_{n}^{(1,3)}\left(t\right)=A_{n}^{(1,3)}\underbrace{\int\diff^{3}q\, g\left(\vq\right)\exp\left[\frac{2\pi inq_{x}}{md}\left(T-\left|t\right|\right)\right]}_{\leq1}.\label{eq:A_decoh_single}\end{equation}
It does not matter whether it occurs before or after the second grating
in the symmetric grating arrangement considered here.

The multiplication transformation \eqref{eq:A_decoh_single} is how
a single decoherence event at time $t$ affects the fringe amplitudes
in the Talbot-Lau interference signal \eqref{eq:Signal}. Although
we cannot find out when such a random event happens either, we know
the event rate $\Gamma\left(t\right)$ and so we can model the mean
decoherence effect by a simple decay process,\begin{equation}
\frac{\diff}{\diff t}A_{n}^{(1,3)}\left(t\right)=\Gamma\left(t\right)\left\{ \int\diff^{3}q\, g\left(\vq\right)\exp\left[\frac{2\pi inq_{x}}{md}\left(T-\left|t\right|\right)\right]-1\right\} A_{n}^{(1,3)}\left(t\right).\end{equation}
The formal solution of this differential equation yields the reduction
factor \begin{equation}
R_{n}=\exp\left\{ -\int_{-T}^{T}\diff t\,\Gamma\left(t\right)\left[1-\int\diff^{3}q\, g\left(\vq\right)\exp\left(\frac{inq_{x}d}{\hbar}\,\frac{T-\left|t\right|}{T_{T}}\right)\right]\right\} \leq1.\label{eq:RedFact_decoh}\end{equation}
It represents the reduction of the $n$th-order Fourier amplitude
of the interference fringe signal \eqref{eq:Signal} by decoherence.
The sinusoidal visibility \eqref{eq:VisSin} is reduced by $R_{1}$.
If there are more than one independent types of environmental decoherence,
each contributes a factor of this form.

The most important decoherence mechanisms for hot nanoparticles are
the emission of thermal radiation and the collision with residual
gas particles. They give stringent upper bounds for the internal particle
temperature $T_{{\rm int}}$ and the background gas pressure $p_{g}$
in a Talbot-Lau experiment \cite{Hornberger2004}.

As an example, let us consider the thermal radiation of a particle that
is significantly hotter than the environment, $T_{{\rm int}}\gg T_{{\rm env}}$.
Neglecting small corrections due to the finite heat capacitance of
the particle, the spectral emission rate is given by\begin{equation}
\gamma_{{\rm emi}}\left(\omega\right)=\left(\frac{\omega}{\pi c}\right)^{2}\sigma_{{\rm abs}}\left(\omega\right)\exp\left(-\frac{\hbar\omega}{k_{B}T_{{\rm int}}}\right),\end{equation}
with $\sigma_{{\rm abs}}\left(\omega\right)$ the absorption cross
section for radiation of angular frequency $\omega$. This yields
the total emission and decoherence rate $\Gamma=\int_{0}^{\infty}\diff\omega\,\gamma_{{\rm emi}}\left(\omega\right)$,
as well as the normalized and isotropic momentum transfer distribution
$g\left(\vq\right)=\left(c/4\pi\hbar\Gamma q^{2}\right)\gamma_{{\rm emi}}\left(cq/\hbar\right)$.

The influence of thermal and collisional decoherence on Talbot-Lau interference is illustrated in fig.~\ref{fig:DruckThermo}. The theoretical prediction fits well to the measured reduction of visibility due to an increased background pressure or internal temperature.

\subsection{Spontaneous collapse models}\label{sec:collapseCSL}

High-mass interferometry offers an ideal testbed for alternative theories
on the nature of the quantum-classical transition. Commonly subsumed
under the term macrorealism \cite{Leggett2002a}, they suggest that standard
quantum mechanics ceases to be valid and must be modified on the macro-scale
in order to reconcile it with the fundamental principles of classical
physics. In particular, the quantum superposition principle is to
be eliminated from the macroscopic scales in order to resolve the
issue of definite measurement outcomes.

The best studied of such macrorealistic hypotheses is the model of
continuous spontaneous localization (CSL) \cite{Ghirardi1990,Bassi2013}. It postulates
an objective modification of the Schr\"{o}dinger equation of mechanical
systems by a nonlinear and stochastic term that induces a spontaneous
collapse of delocalized matter waves above a certain mass scale. The
model is characterized by essentially two parameters: The atomic collapse
rate $\lambda_{{\rm CSL}}$, which is amplified by the total mass
of a given system in atomic mass units, and the localization length
$r_{c}$ down to which matter waves are collapsed. The latter is conventionally
fixed at about $r_{c}=100\,$nm, the former is currently estimated
as $\lambda_{{\rm CSL}}\sim10^{-10\pm2}\,$Hz.

Fortunately, the quantitative predictions of the CSL model are easily
implemented into the theory of Talbot-Lau interferometry. In fact,
the observable consequences of the CSL modification in center-of-mass
experiments with nanoparticles smaller than $r_{c}$ mimic those of
a ficticious decoherence process, as discussed in Section \ref{sec:dec_theory}.
The decoherence rate parameter is simply substituted by $\Gamma=\left(m/1\,\text{u}\right)^{2}\lambda_{{\rm CSL}}$,
and the momentum transfer distribution $g\left(\vq\right)$ by a symmetric
Gaussian distribution of standard deviation $\sigma=\hbar/\sqrt{2}r_{c}$
\cite{Vacchini2007b}. The sinusoidal Talbot-Lau fringe contrast \eqref{eq:VisSin}
would be reduced by the factor \cite{Nimmrichter2011}
\begin{equation}
\V_{\sin}\mapsto\V_{\sin}\exp\left\{ -2\left(\frac{m}{1\,\text{u}}\right)^{2}\lambda_{{\rm CSL}}T\left[1-\frac{\sqrt{\pi}r_{c}T_{T}}{dT}\erf\left(\frac{dT}{2r_{c}T_{T}}\right)\right]\right\}
\label{eq:CSLVis}
\end{equation}
due to CSL. We notice a quadratic mass dependence in the exponential
decay rate. Moreover, the grating separation time $T$, which must
be of the order of the Talbot time $T_{T}$, must also be increased
in proportion to the particle mass. All this allows us to test the
CSL predictions by observing high-contrast interference with heavy
particles, placing an upper bound on the CSL rate parameter $\lambda_{{\rm CSL}}$
with each successful experiment.

\section{Talbot-Lau interferometry}
\label{sec:tli}
Interferometers of the Talbot-Lau geometry have various advantages over simple far-field diffraction schemes with a single grating:
First, they are more compact due to the less stringent collimation requirements.
In practice, the TLI design is about an order of magnitude shorter than its far-field counterpart for the same de Broglie wavelength.
This entails that the required coherence time can be shorter by the same factor and it renders near-field experiments less sensitive to external perturbations. Alternatively, a near-field scheme can operate with larger gratings at equal machine length, which makes it less sensitive to dispersive van der Waals in the grating slits.

Second, since the first grating in TL interferometry comprises thousands of parallel coherence preparation slits, it increases the signal throughput by four to five orders of magnitude in comparison to a far-field setup with a single pair of collimation slits. However, the alignment procedure of three-grating interferometers are clearly more demanding. The interference fringe visibility strongly depends on a precise alignment of the gratings with respect to each other and with respect to external force fields.
Given the large grating period and the even larger support structure, alignment of these grating could be conveniently done by comparing laser diffraction images.

\begin{figure}
\begin{center}
  \includegraphics[width=\columnwidth]{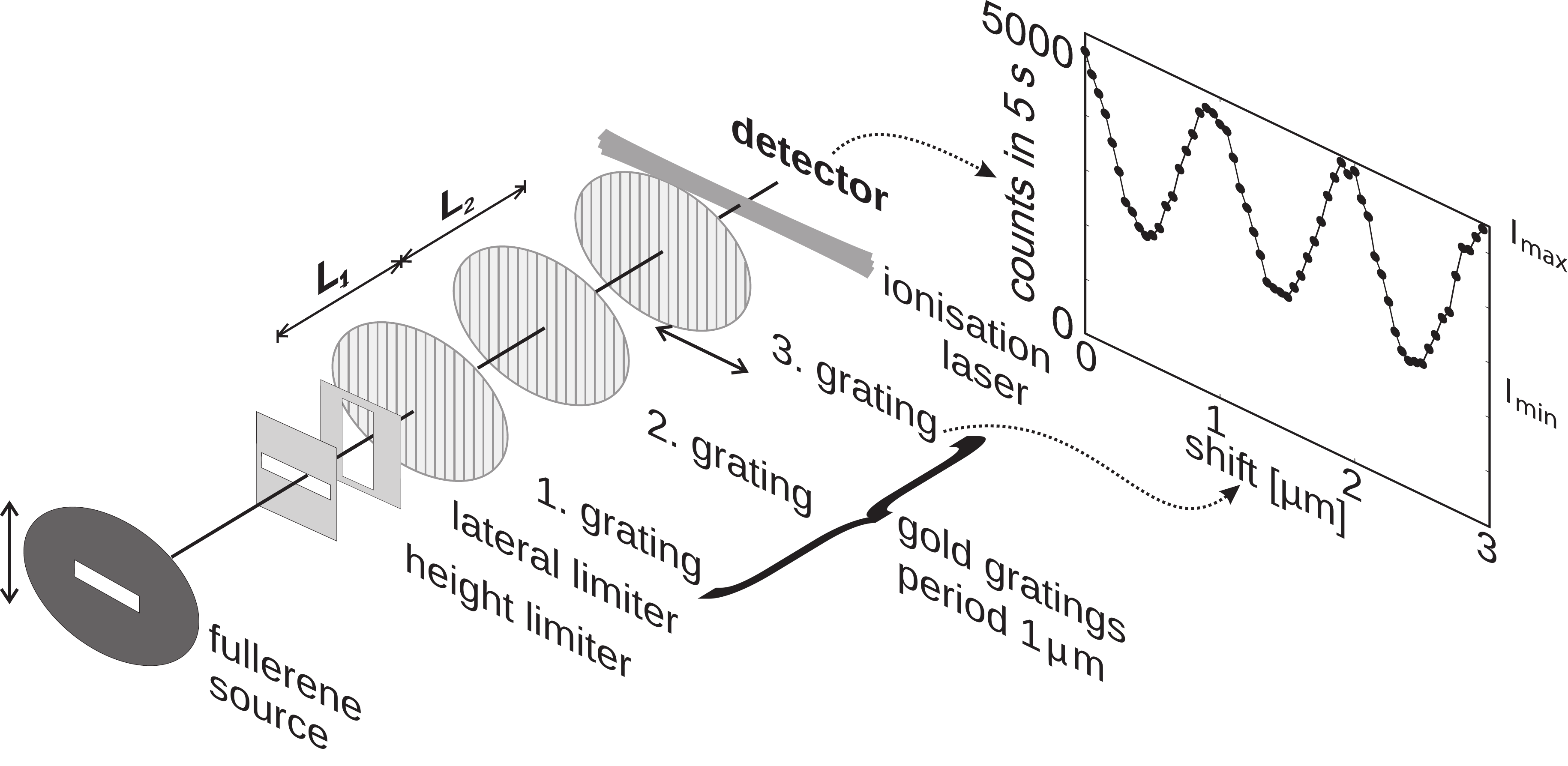}
  \caption{Setup of the first Talbot-Lau interferometer for molecules: a thermal source of fullerenes was collimated to $2\times 0.2$ mrad$^2$ to limit the alignment requirements and to allow a gravitational velocity selection by three vertical height delimiters (oven nozzle, slit, detecting laser beam waist).  The molecules were sent through a sequence of three microstructured gold gratings ($d=990$\,nm). One expects and finds a sinusoidal interferogram. The fringe visibility is then readily defined by the modulation amplitude to the signal offset. A quantitative comparison of the observed fringe visibility with the predicted dependence on the de Broglie wavelength then allows to confirm the true quantum nature of this molecular density pattern.
 This requires also to take into account that the attractive van der Waals interaction between the polarizable electron cloud of a fullerene molecule and the gold grating bar modulates the matter-wave phase and reduces the effective slit opening. ~\cite{Brezger2002}.
  }\label{fig:TLI}
\end{center}
\end{figure}

Molecular Talbot-Lau interferometry was first demonstrated with thermal beams of C$_{60}$ and C$_{70}$ \cite{Brezger2002}, and
soon extended to the biodyes tetraphenylporphyrin (TPP) and to the larger fluorofullerenes (C$_{60}$F$_{48}$) \cite{Hackermueller2003}. The setup is sketched in fig.~\ref{fig:TLI}.
Three gold gratings with a period of 990\,nm, an open fraction of $f=0.48$ and a thickness of $t=500$\,nm  were originally manufactured by Heidenhain in Traunreut/Germany for use on the X-ray satellite AXAF/Chandra. Each of them was photolithographically written into a gold membrane that spanned a free circle of 16\,mm diameter. Each membrane stretched across a thin steel ring which was then magnetically attached to the grating mount to avoid mechanical stress.
They were positioned in a mutual distance first of 23\,cm\,\cite{Brezger2002} and in later experiments at 40\,cm for explorations of longer coherence times and higher interference orders~\cite{Hackermueller2003a}.

Fullerenes were efficiently detected using delayed thermal ionization after interaction with a focused strong laser beam (here 10-30 Watt of 514\,nm or 532\,nm light), whereas other molecules, which do not photoionize, were detected by means of electron impact ionization and quadrupole mass spectrometry in the detector stage.
Since the velocity distribution of thermal sources is well described by a Maxwell-Boltzmann distribution, in some cases also with a small velocity offset~\cite{Scoles1989}, the molecular coherence length was improved by distinguishing the longitudinal velocities in the beam according to different flight times, which correspond to different free-fall distances in the gravitational field. This allowed us to establish a selectivity of $\Delta v/v \simeq 15-20\%$\,(FWHM).

\subsection{Protection from collisional and thermal decoherence}
All initial experiments were targeted at obtaining the best possible quantum coherence and the highest fringe visibility. In the given setup, the width of the slit openings and the finite velocity spread limited the interference contrast to about $30\%$, both in experiment and in theory. A variety of different effects may reduce the visibility even further: Some causes may appear 'trivial', such as vibrational disturbances of the entire interferometer or the individual gratings~\cite{Stibor2005a}, but they often decide success or failure in practice. Vibrations on the hull of the vacuum chamber as tiny as 15\,nm were observed to translate into sizeable perturbations inside the interferometer, at specific vibrational frequencies.

On the more foundational side, we may also ask how the delocalized particles may disseminate which-way information into the environment. This happens through the momentum exchange when the particles either collide with rest gas molecules in the high-vacuum chamber or when they emit thermal photons, as discussed in Section \ref{sec:dec_theory}. Decoherence theory accounts for these effects and helps in understanding the transition from quantum to classical behaviour.

\begin{figure}
\begin{center}
  \includegraphics[width=\columnwidth]{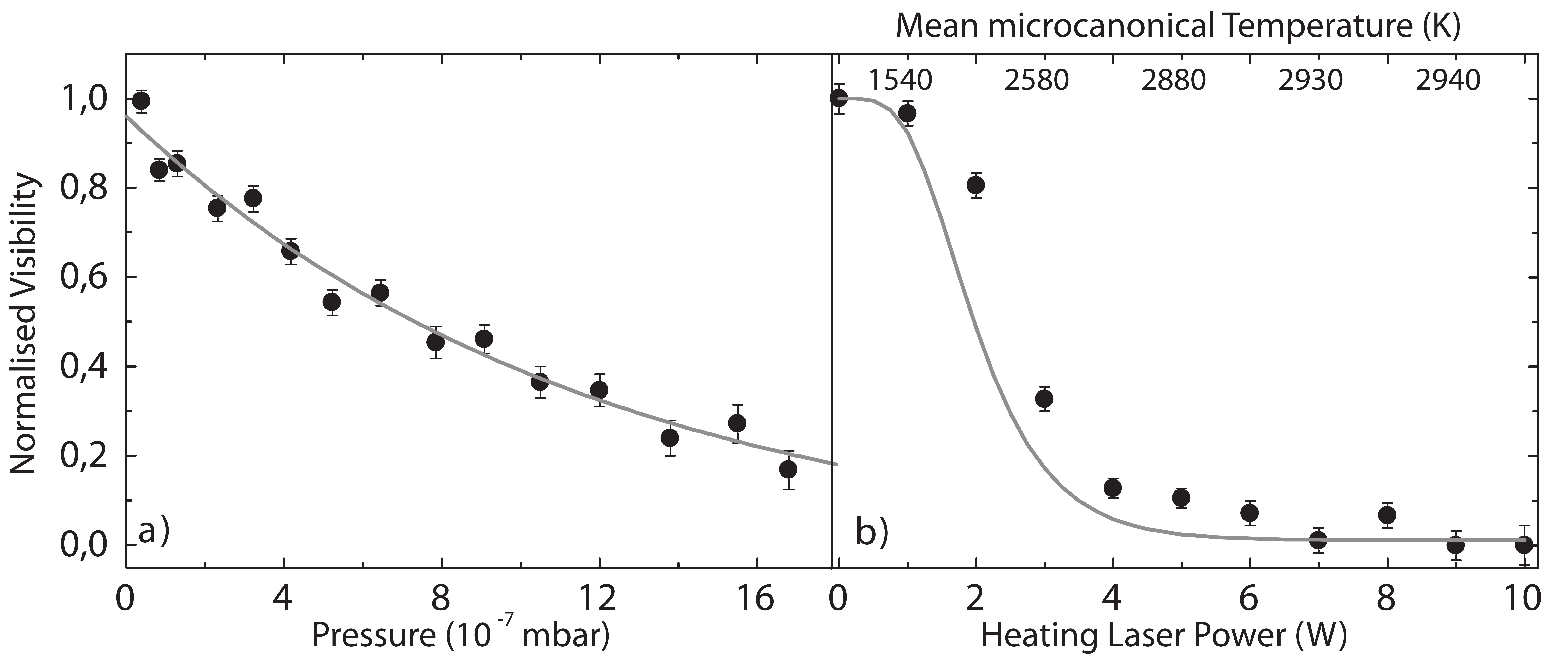}
  \caption{Experimental decoherence in Talbot-Lau interferometry: a) Interference contrast of fullerenes as a function of the residual gas pressure. The exponential decrease is in agreement with the theoretical predictions based on the expected collisional cross sections.
The small-angle van der Waals cross section, which is relevant for gas collisions, may exceed the geometrical particle cross section by two orders of magnitude~\cite{Hornberger2003}.
b) Interference fringe visibility as a function of the internal molecular temperature: Each fullerene molecule has a high number of internal degrees of freedom and acts as its own internal heat bath.
Upon emission of a visible or near-infrared photon which-path information becomes available and random momentum kicks lead to a reduction of the interference contrast~\cite{Hackermueller2004}.}\label{fig:DruckThermo}
\end{center}
\end{figure}

The quantum decoherence experiments in Vienna were focused on two questions: First, which pressure can still be tolerated in the interferometer chamber before quantum coherence is washed out? Second, what molecular temperature is still allowed before the thermal emission of radiation becomes predominant and a severe obstacle to the observation of molecular quantum interference? Exemplary results are plotted in fig.~\ref{fig:DruckThermo}.

From the exponential decrease of the fringe visibility with increasing rest-gas pressure (see fig.~\ref{fig:DruckThermo}a) one may deduce the cross section for collisional decoherence assuming that none of these collisions is sufficiently violent to remove the molecules from the detected particle beam. We found that rest gas pressures of $10^{-6}$\,mbar are clearly too high for observing interference of large molecules on the time scale of milliseconds. One can show that a technologically achievable pressure  of $10^{-10}$\,mbar would suffice to interfere particles beyond $10^6$\,u in the future \cite{Nimmrichter2011}.

The dependence of quantum coherence on the internal temperature was studied by exposing the fullerenes to intense laser light. Every single absorbed photon contributed to the internal heating of the molecule.
Although this process may lead to molecular ionization--a fact that was used in the first C$_{60}$ diffraction experiments--the emission of thermal radiation is the faster cooling  process at high temperatures.
If a visible or near-infrared photon is emitted by the interfering molecule in free flight between the gratings, its spatial coherence should be reduced by the degree of which-path information disseminated into the environment.

Again, extrapolation to complex particles shows that high-mass matter-wave interference should still be possible, provided we can cool the particles to the temperature of liquid nitrogen or ideally even liquid helium, where all electronic and vibrational
degrees of freedom are frozen into their ground states. The remaining rotational excitations will not harm the experiments, since the associated decay rates are extremely small and since the wavelength of a rotational photon is far too long to provide any information about the molecular position.

\subsection{Quantum-assisted deflectometry}
The idea of quantum-assisted deflectometry builds on established classical beam methods which were successfully applied to characterize beams of clusters \cite{Knight1985,Heer1993,Antoine1999,Heer2011} and molecules \cite{Compagnon2001a,Broyer2007}: A homogeneous force field
 $F_x=\alpha_{stat}\left(\mathbf{E}\cdot\nabla\right)E_{x}$ deflects a particle of static molecular polarizability $\alpha_{stat}$ by the amount
\begin{equation}
\Delta x=\alpha_{stat}\frac{\left(\mathbf{E}\cdot\nabla\right)E_{x}}{mv_z^2}s\left(\frac{s}{2}+l\right)\label{eq:shift}.
\end{equation}
Here $\mathbf{E}$ is the applied electric field, $m$ is the molecular mass, $v_z$ the longitudinal velocity, $s$ designates the length of the deflection electrode, and $l$ the distance of the electrode from G$_3$.

\begin{figure}
\centering
  \includegraphics[width=0.5\columnwidth]{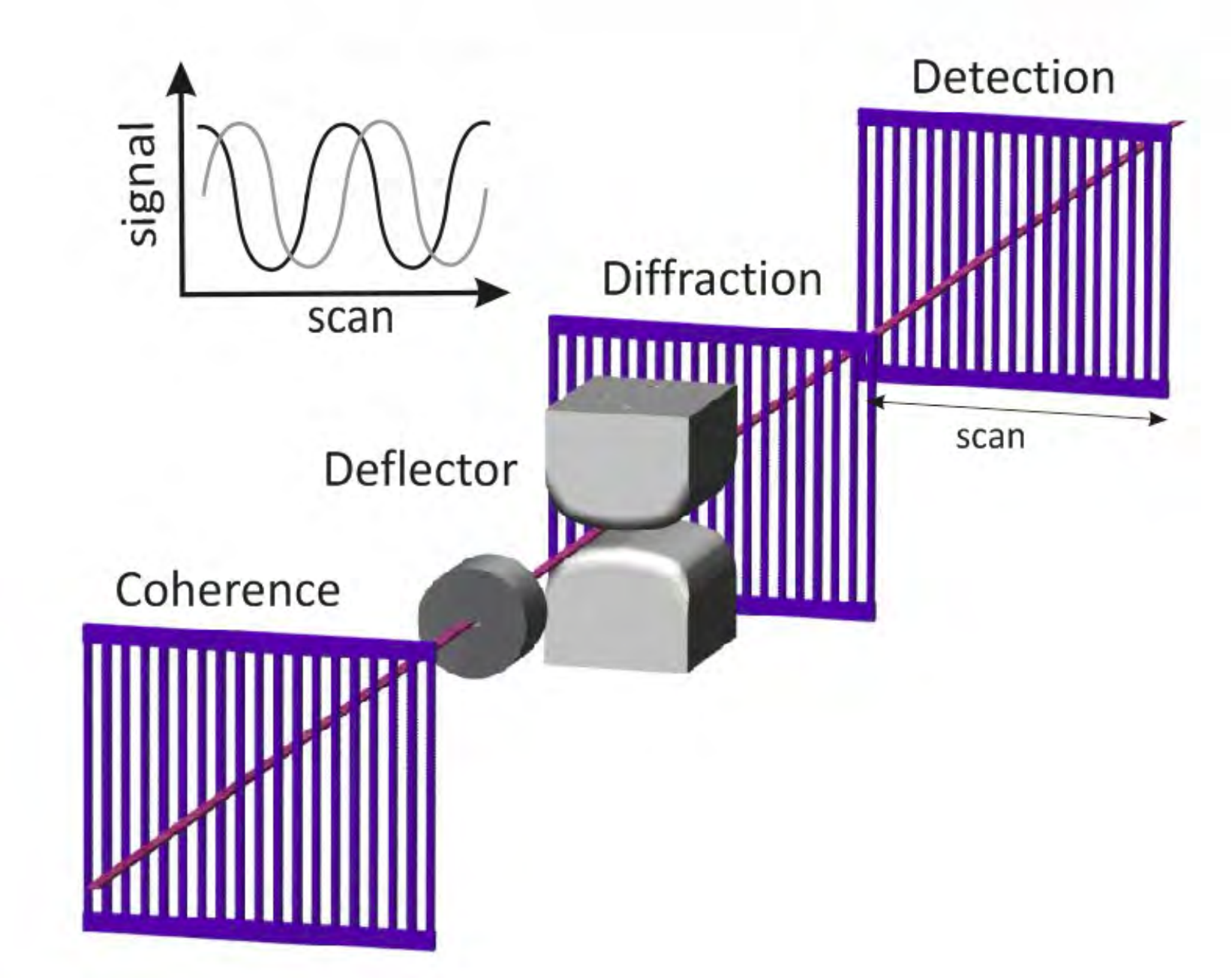}
  
  \caption{Quantum assisted deflectometry for polarizability measurements in Talbot-Lau interferometry: A pair of electrodes is designed to create a particularly homogeneous electric force field on polarizable molecules.
  In devices with a gravitational velocity selection scheme one usually operates with a rectangular molecular beam that is broader than high. This requires a suitable electrode design which ensures the required homogeneity of the deflection force field down to the level of 1\% \cite{Stefanov2008}.
  The voltage dependent deflection of the molecular interferogram then allows one to extract the polarizability. This idea has been used to measure the polarizability of the fullerenes C$_{60}$ and C$_{70}$~\cite{Berninger2007}.\label{fig:deflection}}
  
\end{figure}

This deflection can be measured with nanoscale precision in the Talbot-Lau near-field scheme since molecular interferogram serves as a tiny ruler.
Lateral shifts as small as 10\,nm can still be resolved, while most of the classical machines would have a ten thousand times lower resolution when they operate with a collimated beam.
The interferometric deflectometer was first demonstrated to retrieve the static polarizabilities of C$_{60}$ and C$_{70}$.
In fig.~\hyperref[fig:deflection]{8} the experimental setup is shown as well as a sketch of the fringe shift in the external field.

\section{Kapitza-Dirac-Talbot-Lau (KDTL) interferometry}
\label{sec:kdtli}

\begin{figure}
\begin{center}
\includegraphics[width=\columnwidth]{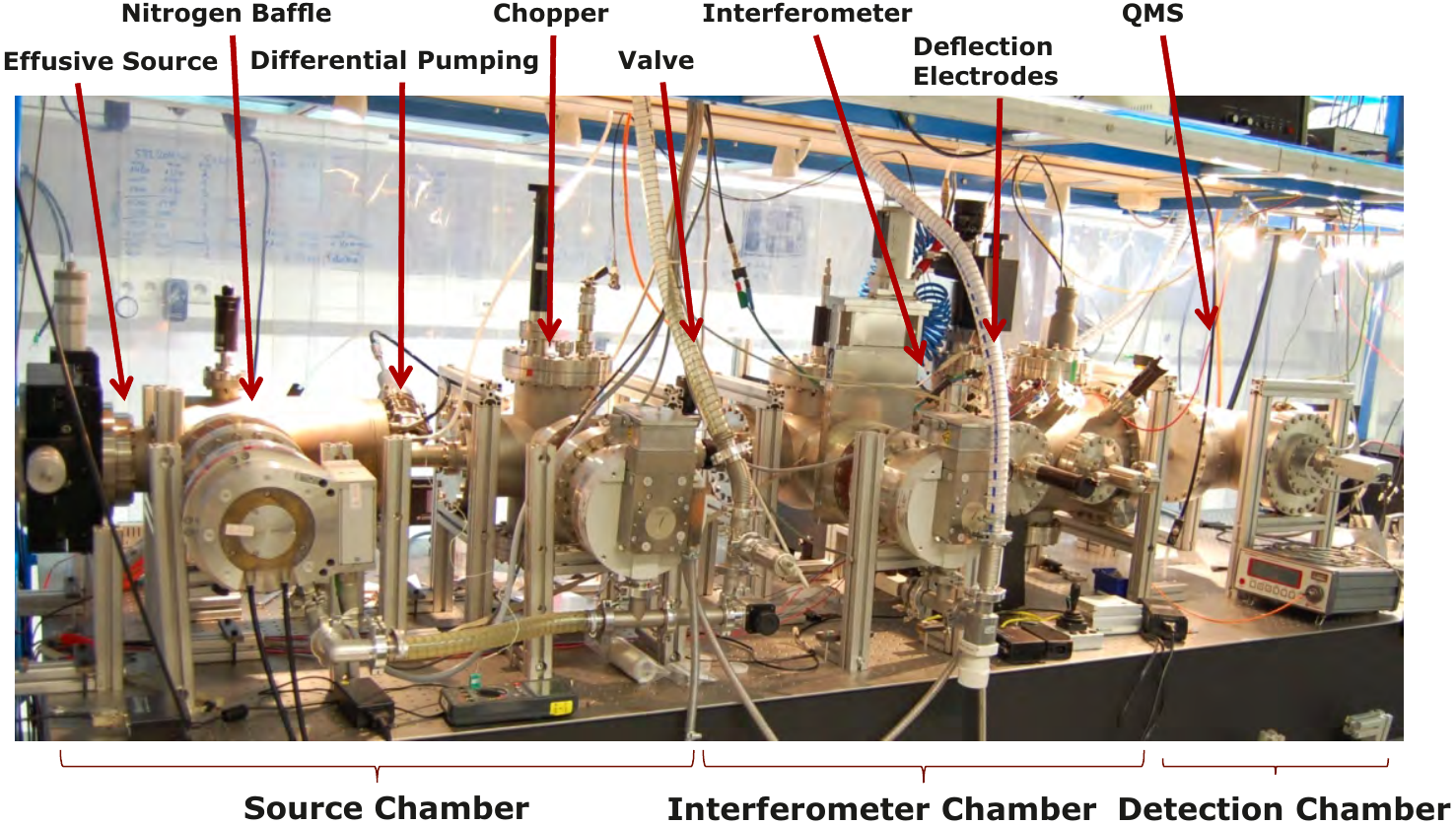}
\caption{Experimental setup of the KDTL interferometer. The vacuum chamber is mounted on an optical table for vibration isolation. It is divided into three segments for the source, the interferometer, and the detection chamber.
Collisional decoherence is avoided by pumping the entire system to about $10^{-8}$\,mbar.}
\label{fig:KDTLIphoto}
\end{center}
\end{figure}

In Kapitza-Dirac-Talbot-Lau (KDTL) interferometry the overall concept, beam sources, interference scanning and detection are done in close analogy to the TL-interferometry described above~\cite{Brezger2002}.
A key difference between the two devices is that the mechanical diffraction grating G$_2$ is substituted by a standing light wave ($\lambda_L = 532$\,nm, cw) which acts predominantly as a phase mask.
This eliminates the strongly position- and velocity-dependent phase modulation that we observed in TL-interferometry because of van der Waals forces in this central element.
In addition, in our implementation~\cite{Gerlich2007} the mechanical masks in G$_1$ and G$_3$ are substantially thinner ($t=160$\,nm instead of $500$\,nm) and finer ($d=266$\,nm) than in the previous TL interferometer. They are written into silicon nitride which is substantially stiffer than gold and also less polarizable.
\begin{figure}
\includegraphics[width=\columnwidth]{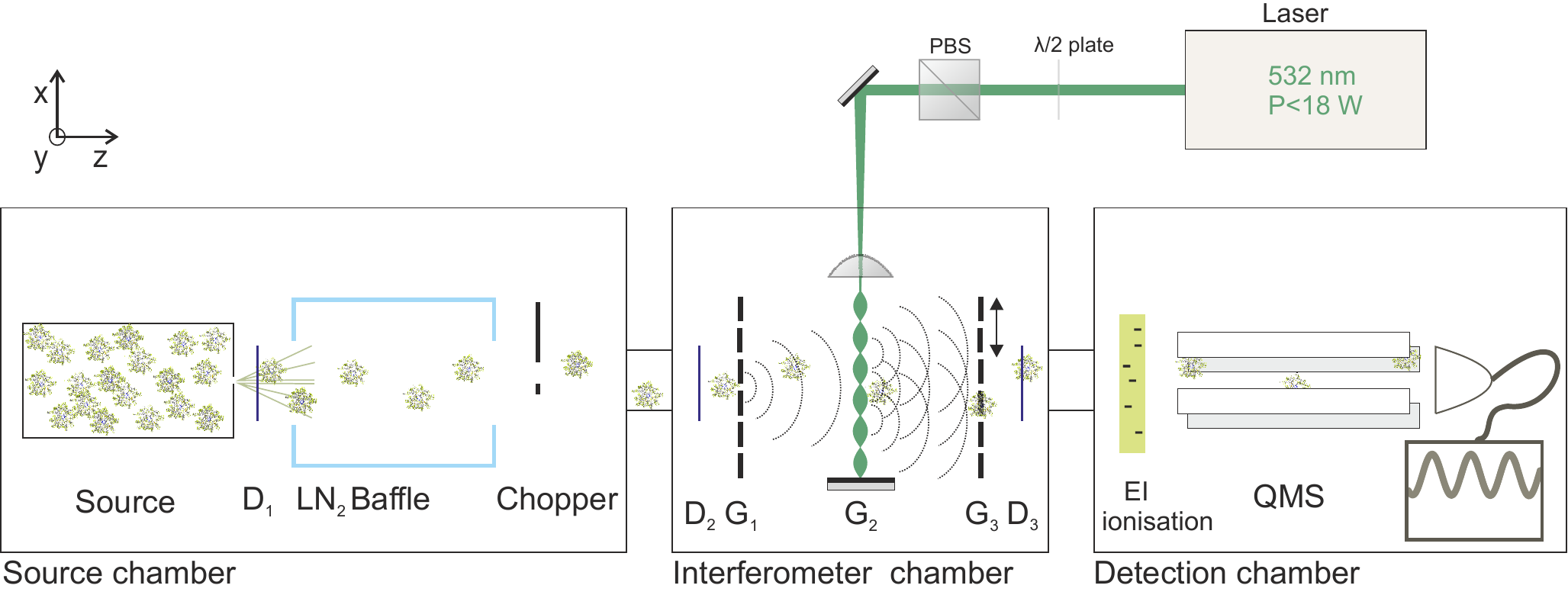}
\begin{center}
\caption{Sketch of the KDTL interferometer. The molecules are evaporated in an oven, velocity selection is performed by three height delimiters D$_1$, D$_2$, and D$_3$. The molecules traverse tree gratings - G$_1$, G$_2$, and G$_3$ - all having an equal periodicity of 266\,nm. G$_1$ and G$_3$ are SiN gratings acting as absorptive masks for coherence preparation and detection, respectively. G$_2$ is a standing light wave produced by retro-reflection of a 532\,nm laser at a plane mirror acting as the diffraction grating. G$_3$ is moved laterally across the molecular density pattern in order to scan the interference fringes. The molecules transmitted through G$_3$ are ionized by electron ionization and detected in a quadrupole mass spectrometer.}\label{fig:KDTLI_setup}
\end{center}
\end{figure}
In all KDTLI experiments performed so far the molecules were again evaporated in a stable thermal source. A sketch and photograph of the interferometer are shown in fig.~\ref{fig:KDTLI_setup} and fig.~\ref{fig:KDTLIphoto}.

\subsection{Experimental results: high-mass quantum interference}
\begin{figure}
\begin{center}
\includegraphics[width=0.85\columnwidth]{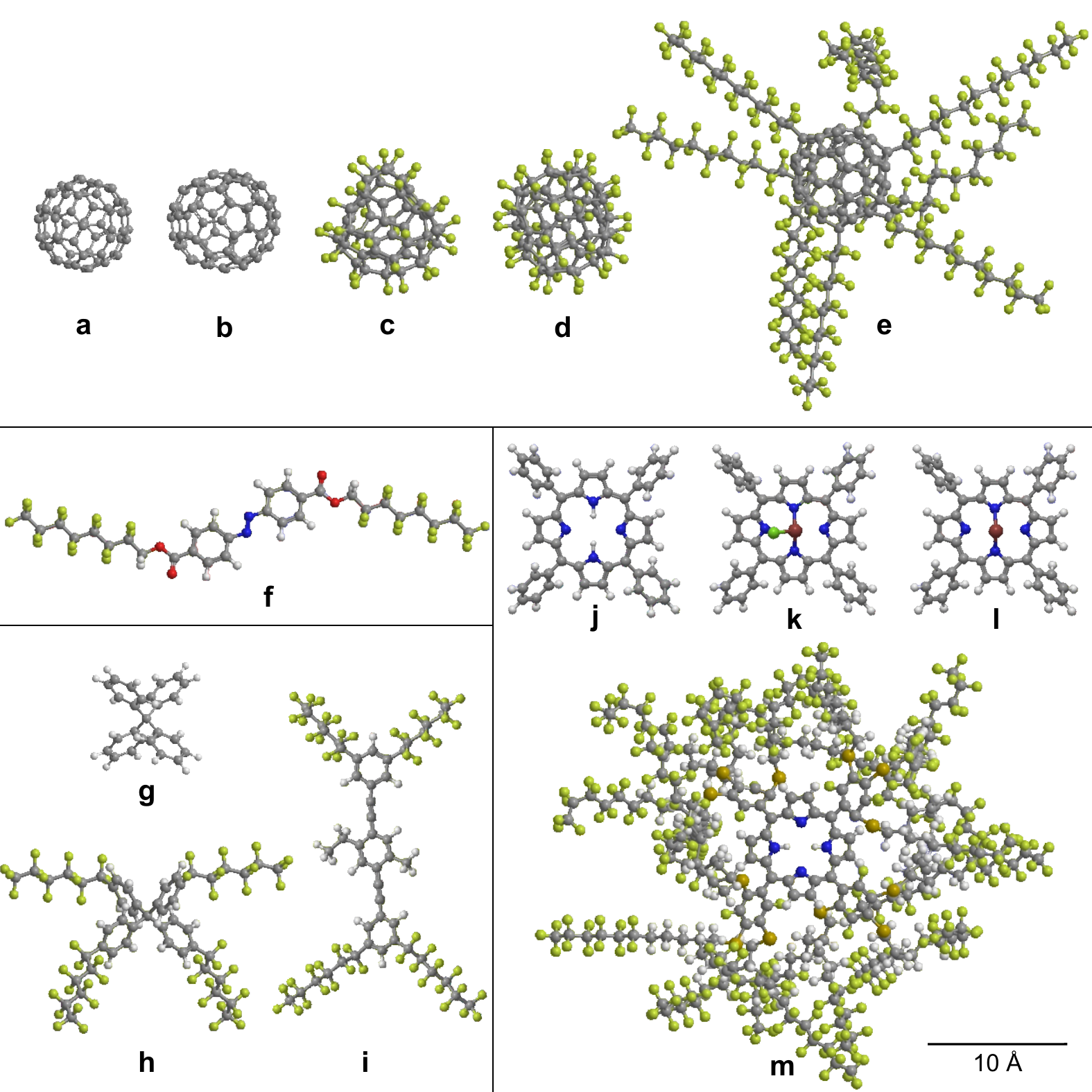}
\caption{Selection of molecules, which showed high-contrast quantum interference in our near-field matter wave interferometers:
a) and b) The fullerenes C$_\mathrm{60}$ and C$_\mathrm{70}$ were used in far-field diffraction \cite{Arndt1999}, TL interferometry \cite{Brezger2002} and
KDTL interferometry~\cite{Gerlich2007}; c) \& d) the fluorofullerene C$_{60}$F$_{48}$ showed interference in a TL setup \cite{Hackermueller2003} but its polar variant C$_{60}$F$_{36}$ only revealed good quantum interference
 in KDTLI several years later~\cite{Hornberger2009}; e) the perfluoroalkylated nanosphere C$_{60}\mathrm{[C_{12}F_{25}]_{8}}$ has a high vapor pressure and therefore guided the way to high-mass interference with thermal molecular beams~\cite{Deachapunya2008,Gerlich2011}; f) experiments with perfluoroalkyl-functionalized azobenzenes $\mathrm{C_{30}H_{12}F_{30}N_{2}O_{4}}$ demonstrated that quantum interference persists even under conformational state changes in free flight and  interference-enhanced studies could quantitatively assess the conformational dynamics~\cite{Gring2010}; g) the tailor-made molecule $\mathrm{C_{25}H_{20}}$ and its larger derivative h) $\mathrm{C_{49}H_{16}F_{52}}$ served to reveal the emergence of vibrationally induced dipole moments in floppy molecules~\cite{Eibenberger2011}; KDTL interference was capable of distinguishing the structural isomers h) \& i) in quantum-interference enhanced deflectometry; j) Tetraphenylporphyrin $\mathrm{C_{44}H_{30}N_{4}}$ which is a common dye molecule~\cite{Hackermueller2003} showed quantum coherence and in comparison with its derivatives k) $\mathrm{C_{44}H_{28}ClFeN_{4}}$  and  l)$\mathrm{C_{44}H_{28}FeN_{4}}$  it allowed us to study the influence of a permanent dipole moment on matter-wave coherence~\cite{Eibenberger2011}; m) the functionalized porphyrin $\mathrm{C_{284}H_{190}F_{320}N_{4}S_{12}}$ is currently the most massive molecule to show high-contrast quantum interference~\cite{Eibenberger2013}.}
\label{fig:Molecules}
\end{center}
\end{figure}

The KDTL interferometer has served us in many studies with about a dozen of different organic molecules. Initially demonstrated for the fullerene C$_{70}$ and a perfluoroalkyl-functionalized azobenzene~\cite{Gerlich2007}, it has proven to be well suited for quantum experiments with substantially larger compounds, up to perfluoroalkyl-functionalized nanospheres~\cite{Gerlich2011} and porphyrin derivatives~\cite{Eibenberger2013}.

A major contribution to the success of all these experiments was the ability of modern chemistry to provide tailored functional components. Already the earlier experiments with fluorofullerenes had indicated that a high fluorine content is
beneficial for the volatilization of massive compounds. It generates stable intramolecular bonds while lowering the overall molecular polarizability. This reduces the intermolecular attraction and thus the temperature required to achieve a sizeable vapor pressure. This makes it possible that even molecules in the mass range of many thousand atomic mass units can still form intense and stable molecular beams~\cite{Deachapunya2008,Tuexen2011}.

A most recent addition to the set of functionalized molecules is $\mathrm{C_{284}H_{190}F_{320}N_{4}S_{12}}$. It is built around a tetraphenylporphyrin core to which a number of perfluoroalkyl side chains were attached. It is generally demanding to synthesize particles with a well-defined number of side chains so that a 'library' of molecules with well-defined mass differences emerges quite naturally in the synthesis of these complex compounds.
The fact that it contains different numbers of side chains can actually be useful when substituting quadrupole mass spectrometry by time-of-flight detection, as used in OTIMA interferometry, where the comparison of different mass peaks is done in parallel.
We have studied quantum interference specifically for the twelve-chain compound which was post-selected from the molecular beam by electron impact ionization quadrupole mass spectrometry after interference.
Each particle has a molecular weight of 10\,123\,u and consists of 810\,atoms. An illustration of one of its possible 3D structures is shown in fig.\ref{fig:Molecules}m).

\begin{figure}
\begin{center}
  \includegraphics[width=0.7\columnwidth]{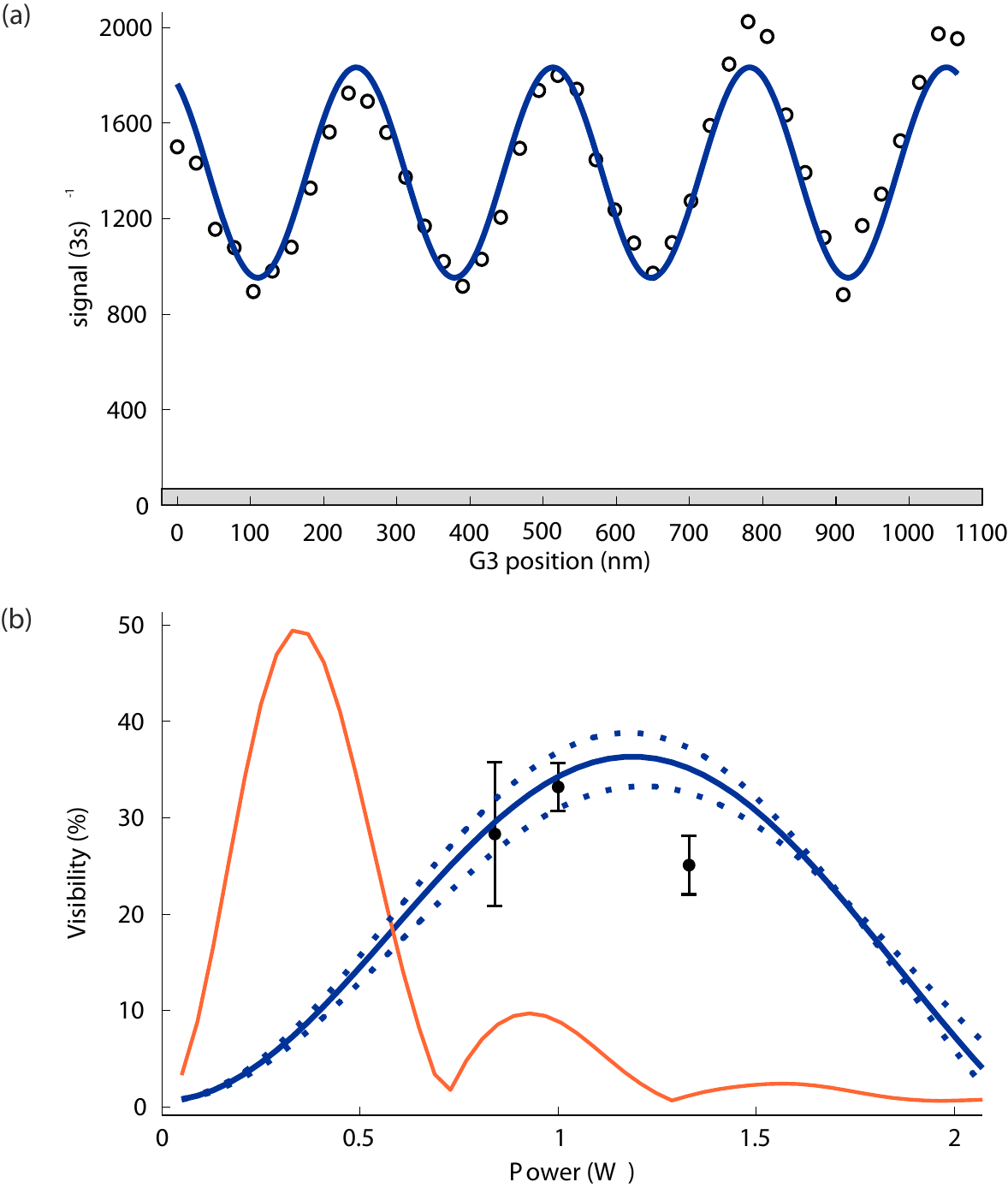}
  \caption{Quantum interference of the functionalized porphyrin $\mathrm{C_{284}H_{190}F_{320}N_{4}S_{12}}$ (\cite{Eibenberger2013} shown in figure \ref{fig:Molecules} m)),currently the most massive and complex particle to show quantum interference. It consists of 810\,atoms per particle and has a mass of 10\,123\,u. (a) Matter-wave interference pattern recorded at a laser power of $P\cong\mathrm{1\,W}$. The circles show the experimental signal as a function of the position of G$_3$ with the shaded area corresponding to the detection background signal. The blue solid line represents a sinusoidal fit to the data resulting in a fringe visibility of 33\,\%. A classical picture predicts a visibility of only 8\% for the same experimental parameters. (b) Plot of the measured interference visibility as a function of the laser power in the standing light wave. The theoretically expected contrast according to the model in \cite{Hornberger2009} is plotted as the blue (quantum) and orange (classical) lines respectively. The dashed blue lines correspond to the expected quantum contrast for an increased (reduced) mean velocity of 5\,m/s.}\label{fig:interference}
\end{center}
\end{figure}

Figure~\ref{fig:interference} shows the best interference pattern obtained for this molecule. In order to distinguish the experimental molecular density pattern from a possible classical shadow pattern (a phase-grating variant of a Moir\'{e} pattern~\cite{Oberthaler1996}) we plot also the observed fringe visibility as a function of the diffracting grating laser power which imprints a phase onto the matter wave when it passes the anti-nodes of the standing light wave.
The observed fringe visibility is in good agreement with a full quantum model, as derived in Section\,3~\cite{Hornberger2009}, and markedly different from the classical expectations~\cite{Eibenberger2011}.

Mass and atom number of our custom-made particles are already comparable to those of small proteins such as insulin (5\,700\,u) or cytochrome C (12\,000\,u).
We have opted for perfluoroalkyl-functionalized synthetic structures in all of our earlier experiments and still in many ongoing studies, since fluorination enables the volatilization of massive particles under conditions where proteins or DNA would naturally denature or fragment.

\subsection{Experimental results: quantum-interference assisted metrology}
Quantum interferometry with large molecules is highly sensitive to external forces, since the de Broglie wavelength is very small. Due to the small grating periodicity in  our the KDTL interferometer, fringe shifts smaller than 10\,nm can still be resolved.
This makes quantum-assisted metrology a powerful tool for measuring inertial forces and even more so for revealing information on internal molecular properties, including electric, magnetic or optical properties as well as internal structure and dynamics.

\subsubsection{Optical polarizability}
\begin{figure}
\begin{center}
\includegraphics[width=0.7\columnwidth]{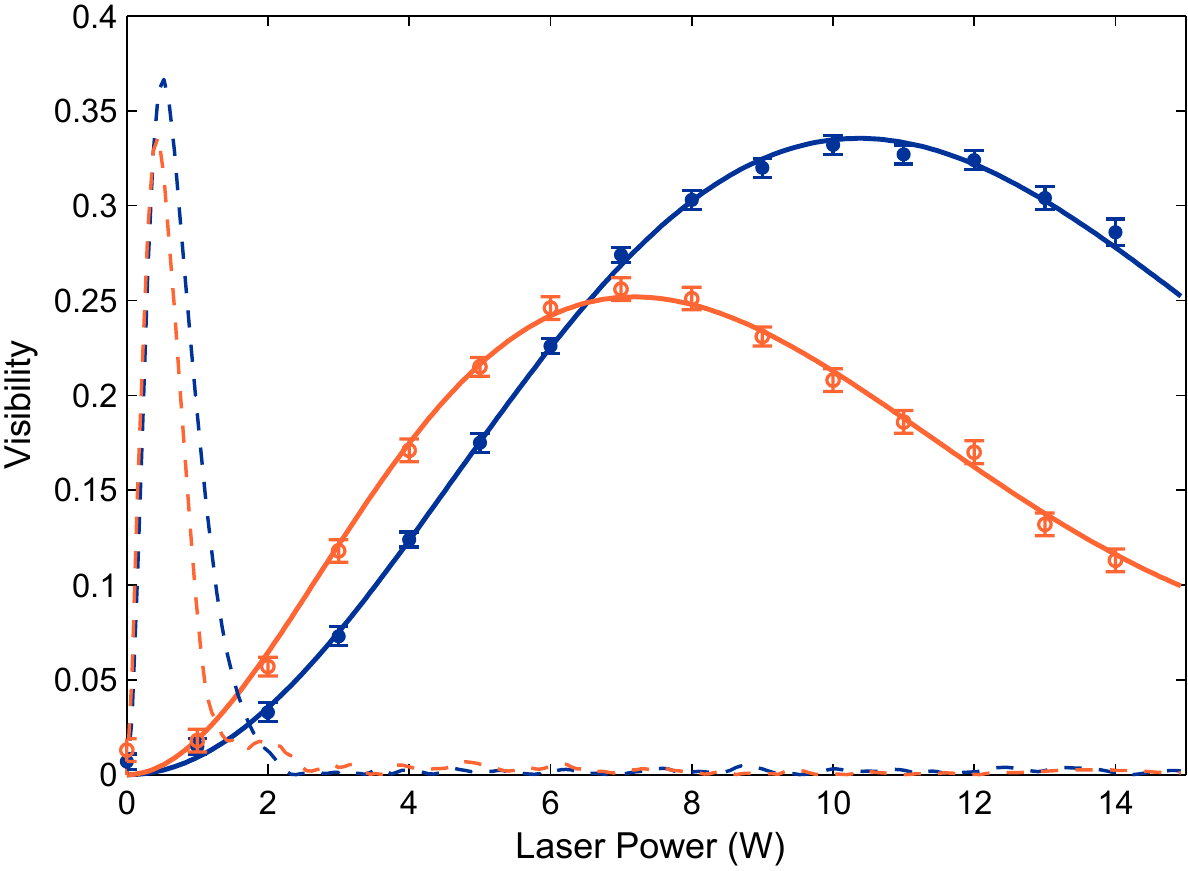}
\caption{from \cite{Hornberger2009}: Measured power dependence of the quantum interference visibility for C$_{60}$ (blue dots) and C$_{70}$ (orange dots). The solid lines represent the quantum expectation for fitted $\alpha_{\rm 532 nm}$ and $\sigma_{\rm 532 nm}$. The dashed lines represent the corresponding classical predictions.}\label{fig:C60C70}
\end{center}
\end{figure}

\begin{figure}
\begin{center}
  \includegraphics[width=0.7\columnwidth]{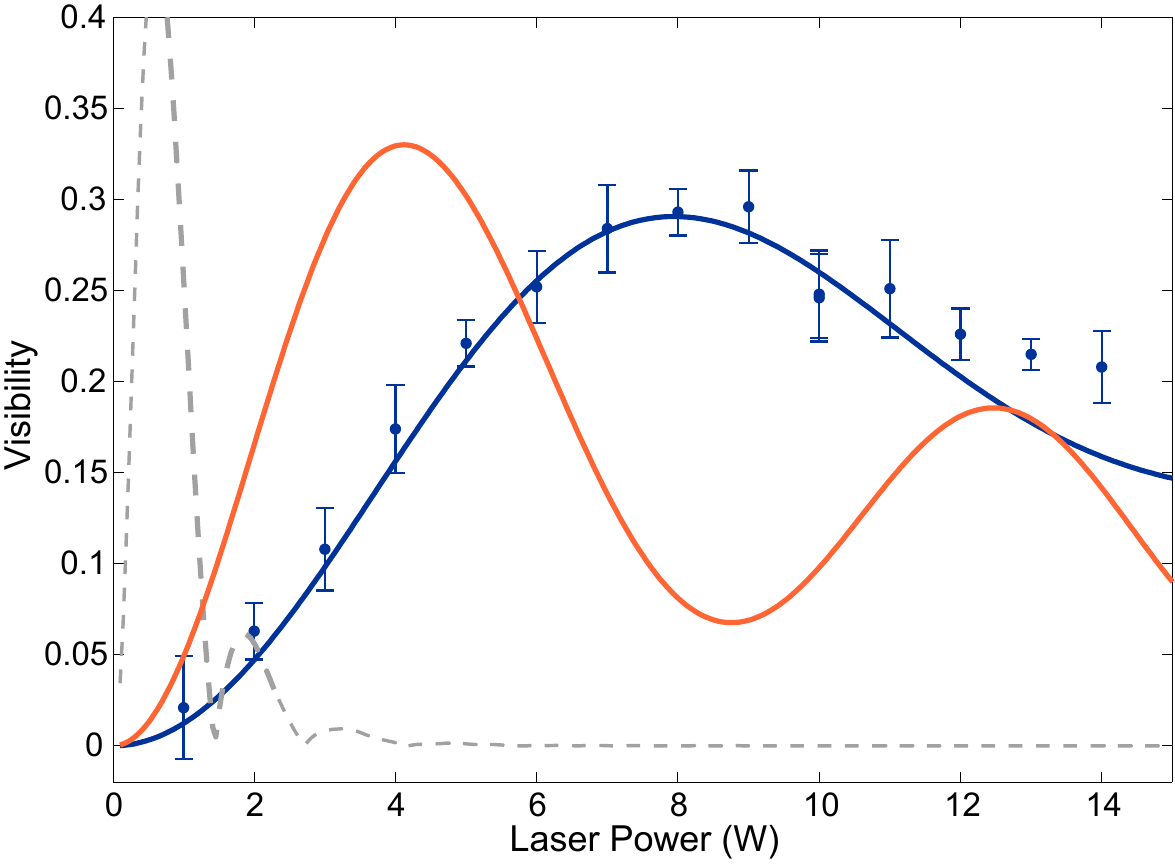}
  \caption{KDTL interference with C$_{96}$H$_{48}$Cl$_{2}$F$_{102}$P$_{2}$Pd (m=3379 amu) allows one for instance to decide whether molecular fragmentation occurs in the source (thermally) or in the detector (by electron impact).
 The interaction with the diffracting standing light wave depends crucially on the optical polarizability which is different for the parent molecule and its two equal fragments, although their polarizability-to-mass ratio is the same \cite{Gerlich2008}.}\label{fig:MassSpec}
\end{center}
\end{figure}

The fringe visibility depends on various parameters (Eq.~(\ref{eq:phi0_KDTLI})): the molecular optical polarizability $\alpha_{\rm 532 nm}$ determines the strength of the phase modulation $\phi_{0}$ in the standing light wave and hence the interference visibility (see Equation (\ref{eq:VisSinKDTLI})). In fig.\ref{fig:C60C70} the power dependence of the interference of C$_{60}$ and C$_{70}$ is displayed.
When all other experimental parameters are known, in particular the velocity distribution, the laser beam waist, the grating properties and the interferometer geometry, the value of $\alpha_{\rm 532 nm}$ can be determined from the power dependence of the fringe visibility. For a pure phase grating, where photo absorption in G$_{2}$ can be neglected, a single-parameter fit suffices to extract this value with an accuracy of a few percent.
In the presence of finite absorption, the associated cross-section $\sigma_{\rm 532 nm}$  can be fitted additionally as a second free parameter~\cite{Hornberger2009}.
The strong influence of the particle polarizability on the fringe contrast can be used to complement mass spectrometry~\cite{Gerlich2008}.
In a first demonstration experiment we asked for instance whether molecular fragmentation of a complex chemical rather occurs in the thermal source under the influence of heat or in the mass spectrometer as a result of electron impact ionization.
While standard mass spectrometry would only see the final fragment, the origin of the fragmentation process determines whether the relevant de Broglie wavelength $\lambda_{\mathrm{dB}}$ and particle polarizability $\alpha_{532}$ in the interferometer are associated with the fragment (in our example: C$_{48}$H$_{24}$F$_{51}$P, $m=1601$\,u) or the parent molecule (here: C$_{96}$H$_{48}$Cl$_{2}$F$_{102}$P$_{2}$Pd, $m=3379$\,u).
A measurement of the quantum visibility as a function of the laser power allows us to clearly decide whether the observed interference pattern is due to the parent molecule or one of its fragments (see fig.\ref{fig:MassSpec}). This is difficult to achieve in standard mass spectrometry. Even dedicated experiments in physical chemistry and classical beam deflectometry would only be sensitive to $\alpha_{\mathrm{stat}}/m$ which is the same for the two compounds~\cite{Bonin1997}.

\subsubsection{Static polarizability}
Similar to Talbot-Lau interferometry before (Section\,4), the KDTL apparatus is also equipped with a high voltage electrode for quantitative matter-wave deflectometry and it allows us to address a much wider class of molecules. The absence of the van der Waals interaction in G$_{2}$ is of particular importance for experiments with polar molecules.

\subsubsection{Vibration induced electric dipole moments}
\begin{figure}
\begin{center}
\includegraphics[width=0.6\columnwidth]{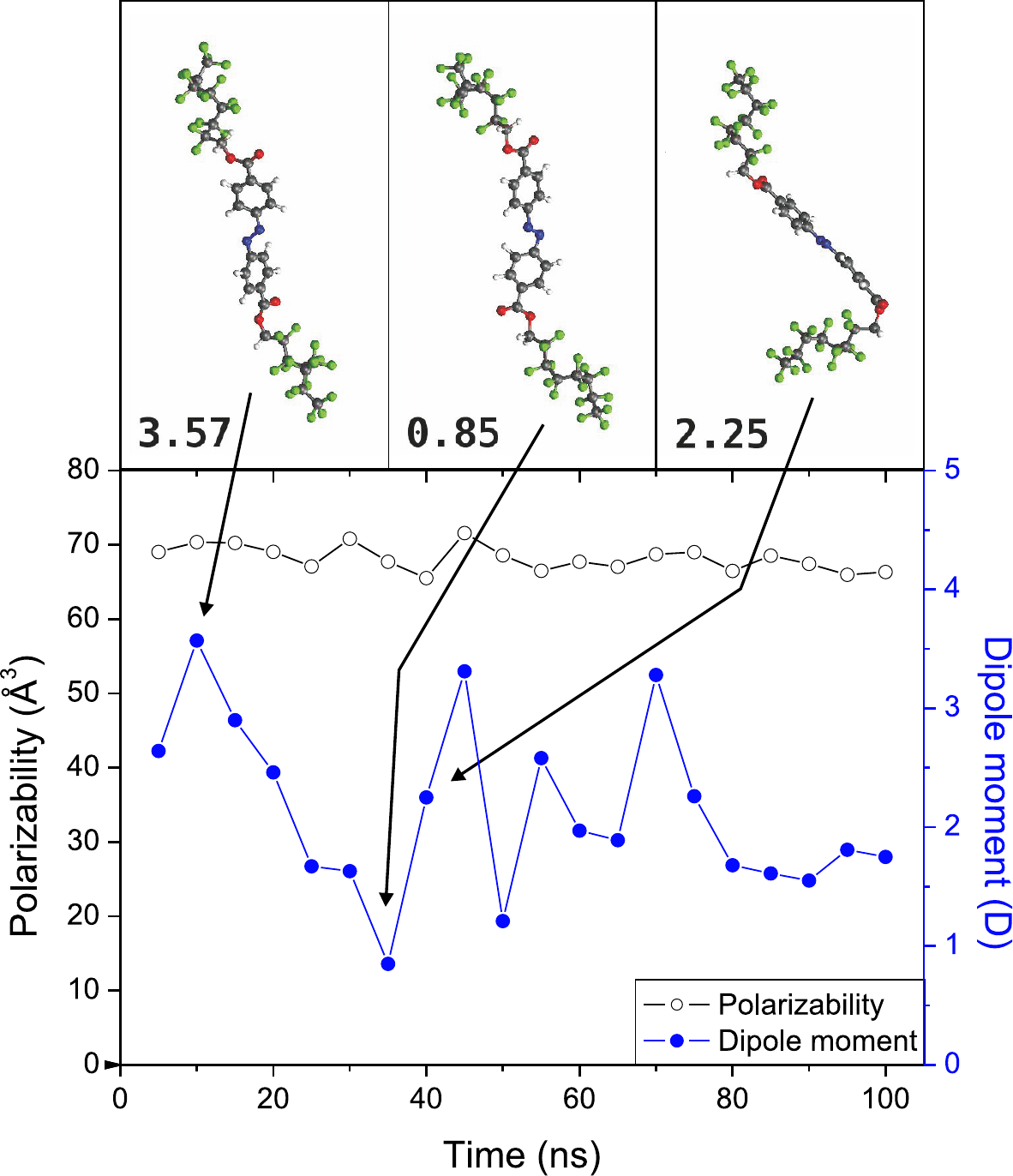}
\caption{Snapshots of a molecular dynamics simulation for the functionalized azobenzene displayed in Figure~\ref{fig:Molecules} f).
Conformation changes on the nanosecond time scale cause the electric dipole moment to vary by 300\%, even though the average dipole moment of the ground state configuration vanishes.
Changes in the nuclear positions leave the molecular polarizability largely invariant since the overall electronic bond structure is also largely unaffected~\cite{Gring2010}.}\label{fig:Azo}
\end{center}
\end{figure}
We have already seen that a molecular interference pattern may shift in response to an external electric field and that for
non-polar rigid molecules, such as the fullerenes, this response is well described by their static polarizability $\alpha_{\mathrm{stat}}$. In addition to that, floppy molecules may undergo thermally-induced conformational changes which in general also entail the development of electric dipole moments which may fluctuate on the nanosecond time scale. The interaction with the external field is then appropriately described by the electric susceptibility $\chi=\alpha_{\mathrm{stat}}+\left<d_x^2\right>/k_BT$ \cite{Vleck1965} which also includes the thermal average of all squared dipole components along the field axis $\left<d^2_x\right>$. The susceptibility replaces $\alpha_{stat}$ in Equation (\ref{eq:shift}).

This phenomenon was observed and verified with perfluoroalkyl-functionalized azobenzenes. Molecular dynamics simulations by N. Doltsinis in M\"{u}nster showed that these molecules undergo many conformational changes and develop non-zero electric dipole moments when they are heated to a temperature of 400\,K. Some snapshots of exemplary configurations are shown in fig.\ref{fig:Azo}. The resulting fringe shift in an external field agreed well with the expectations from the molecular dynamics simulation~\cite{Gring2010}.

For molecules g) and h) in fig.~\ref{fig:Molecules} we also compared the relative rigidity of substructures inside the delocalized molecules. The molecules share an identical core: Tetraphenylmethane C$_{25}$H$_{20}$ (shown as compound g) in dig.~\ref{fig:Molecules}) and the enlarged derivative C$_{49}$H$_{16}$F$_{52}$ (molecule h) in fig.~\ref{fig:Molecules}) equipped with four floppy perfluoroalkyl chains.
In consistence with our earlier results we found that for the small tetraphenylmethane molecule the electric susceptibility is dominated by the static polarizability. For the larger, floppy derivative the effective susceptibility is not sufficiently approximated by the static polarizability--vibrationally induced dipole moments play a significant role.

The experiments on floppy molecules show clearly that violent internal dynamics is fully compatible with high contrast de Broglie coherence as long as it does not provide any means to retrieve which-path information. In this case the center-of-mass motion is clearly decoupled from the internal state. At the same time, the internal properties lead to a measurable shift of the center-of-mass wave function. This is actually the principle behind quantum-assisted metrology in our case.

The important factor for the preservation of coherence is the absence of correlation between the internal and the external motion. Even in the presence of the external field there is no way to assign a specific path to the molecule. The internal state is important but it remains separable. It contributes only to the interaction potential that determines the overall molecular evolution.

\subsubsection{Permanent electric dipole moments}
\begin{figure}

\centering
  \includegraphics[width=\columnwidth]{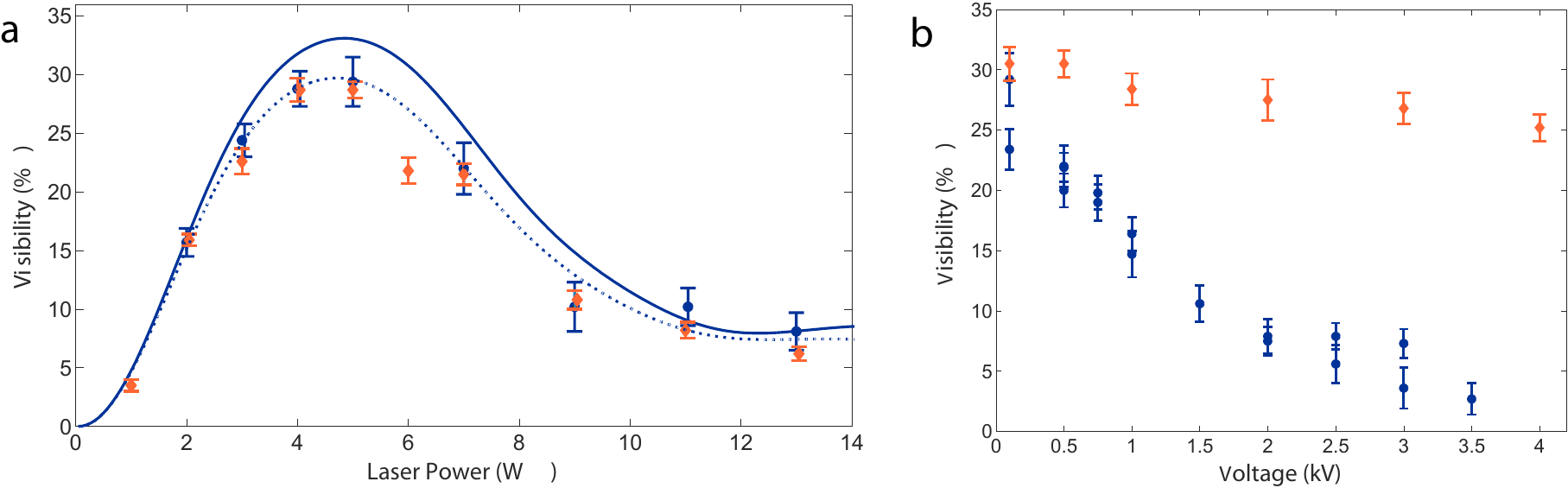}
  
  \caption{From \cite{Eibenberger2011}: a) Interference contrast as a function of the laser power in the standing light wave for FeTPP (orange squares, molecule k) in fig.~\ref{fig:Molecules}) and FeTPPCl (blue dots, molecule l) in fig.~\ref{fig:Molecules}). Due to their comparable mass, optical polarizability, mean velocity, and velocity spread they behave very similarly in the interferometer when no deflection voltage is applied. b) Quantum contrast as a function of electric deflection voltage. The (nearly) non-polar FeTPP (orange squares) is only slightly phase-averaged because of the finite velocity spread in the molecular beam. The polar FeTPPCl (blue dots) is highly sensitive to the external field as differently oriented molecules will also be deflected in different directions by different amounts.\label{fig:polar1}}

\end{figure}

In contrast to non-polar molecules which maintain their quantum contrast even in the presence of comparably strong electric fields,  we observe fringe averaging for a rigid, polar compound already at moderate fields (see fig.\ref{fig:polar1}).
This can be demonstrated using two molecular variants which are very similar in mass and polarizability but different in their static dipole moments.
The polar FeTPPCl differs from the almost non-polar FeTPP  -- see fig.~\ref{fig:Molecules} k) and l) -- by a single chlorine atom which causes an electric  dipole moment of approximately 2.7\,Debye \cite{Deachapunya2007}. Since both molecular species are very similar in mass, beam velocity and optical polarizability, their interferograms look very similar (fig.~\ref{fig:polar1} a) and they shift by the same amount in the same electric field. We observe, however, rapid fringe averaging already at moderate field values in the interference of the polar compound (fig.~\ref{fig:polar1} b). This is consistent with the view that a thermal beam source delivers molecules in random orientations and with random directions for their rotation axes. In the presence of the outer field the interferograms will be shifted depending on this orientation and the total contrast averages out.

The experiments mentioned here give only a snapshot of all what has been done and what can still be done in the future. KDTL interferometry has proven to be applicable to a wide class of atoms, organic and inorganic molecules and nanoparticles, imposing only a few constraints: The particle's optical polarizability at 532\,nm needs to be sufficiently large for the laser in G$_2$ to imprint a phase modulation on the transiting de Broglie wave of the order of $\phi=\pi$. This is for example achieved for a molecular velocity in the range of 200\,m/s, an optical polarizability of $\alpha_{532\,\mathrm{nm}}=4\pi\varepsilon_0\times 100$\AA$^3$ as well as a laser power of 10\,W focused in a beam waist of $20\,\mu$m$\times 1000\,\mu$m. Also, the absorption cross section at 532\,nm needs to be sufficiently small that the phase grating character dominates the diffraction process. For the geometry of our experiment this is warranted for an absorption cross section in the range of $\sigma_{532\,\mathrm{nm}}\simeq 10^{-17}\,$cm$^2$ and the concept works generally as long as $\alpha_{532\,\mathrm{nm}}/\sigma_{532\,\mathrm{nm}} \ge \varepsilon_0 \times 10^{-6}$\,m.

Since the KDTLI is an interferometer in real space rather than in the time-domain, coping with velocity dispersive phase shifts is a general challenge.
With progress in molecular beam source, cooling and detection techniques this challenge can, however, be largely alleviated.
The concept has been predicted to be operational for particles in the mass range of millions of atomic mass units.
Even the fact that G$_1$ and G$_3$ are material masks does not prevent the concept from working in principle.
As long as the particles are still transmitted, KDTLI is a viable universal interferometer concept also for high masses.  
\section{OTIMA interferometry}
\label{sec:otima}

\begin{figure}
\centering
\includegraphics[width=0.7\textwidth]{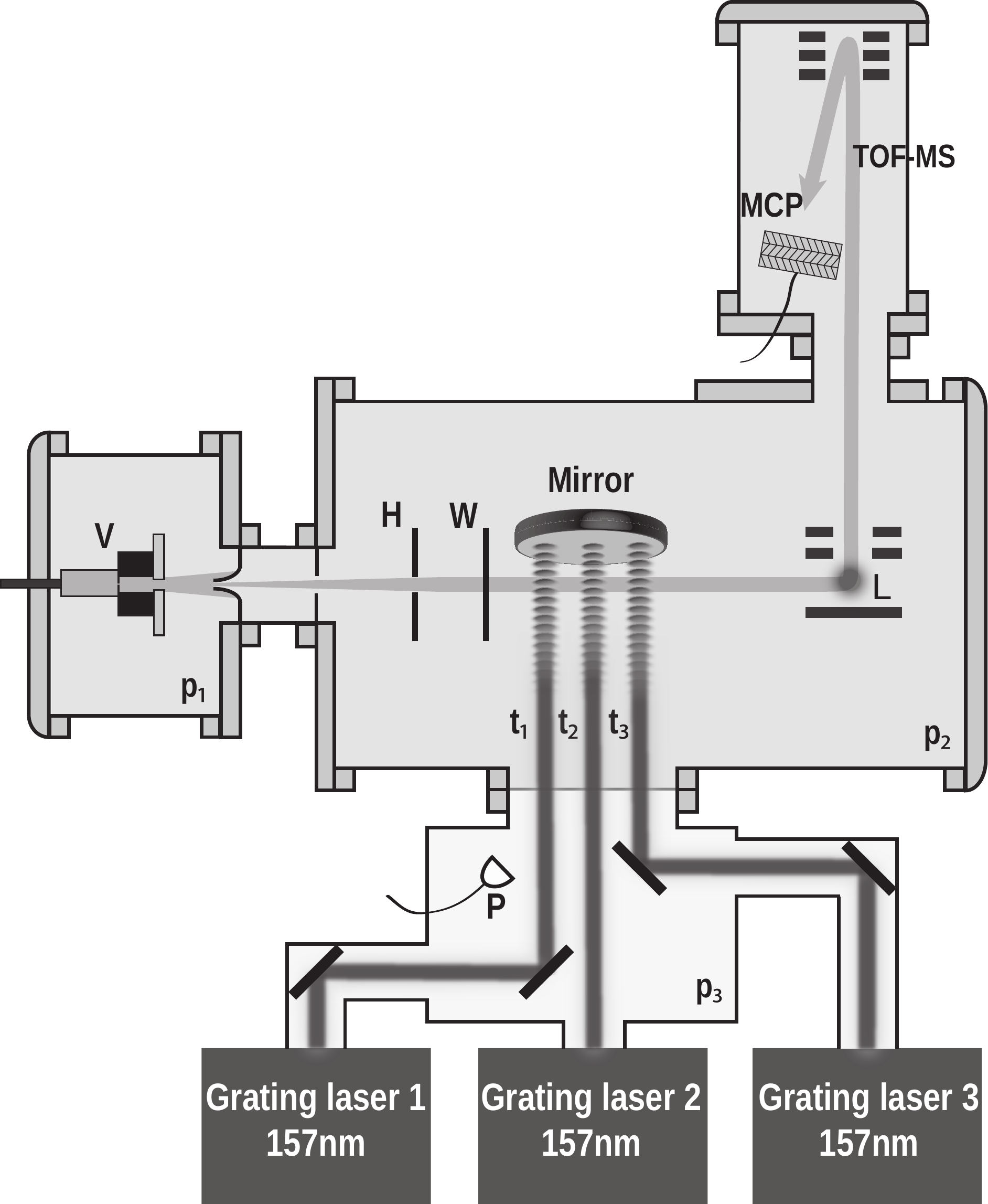}
\caption{Sketch of the OTIMA setup. A pulsed source (V) emits clouds of cold neutral particles. This particle beam is delimited in height (H) and width (W) by two adjustable slits. Three nanosecond laser pulses, which are separated in space, are back-reflected by a single ultra flat mirror and form standing light waves that ionize particles in the anti-nodes. They act on the particles with a well-defined time sequence (at $ t_{1}=0 $, $ t_{2}=T $ and $ t_{3}=2T $).
The remaining neutral particles are ionized by a pulsed laser (L) and they are detected in a time-of-flight mass-spectrometer (TOF-MS).}
\label{fig:otimasetup}
\end{figure}

\begin{figure}
\centering
\includegraphics[width=\textwidth]{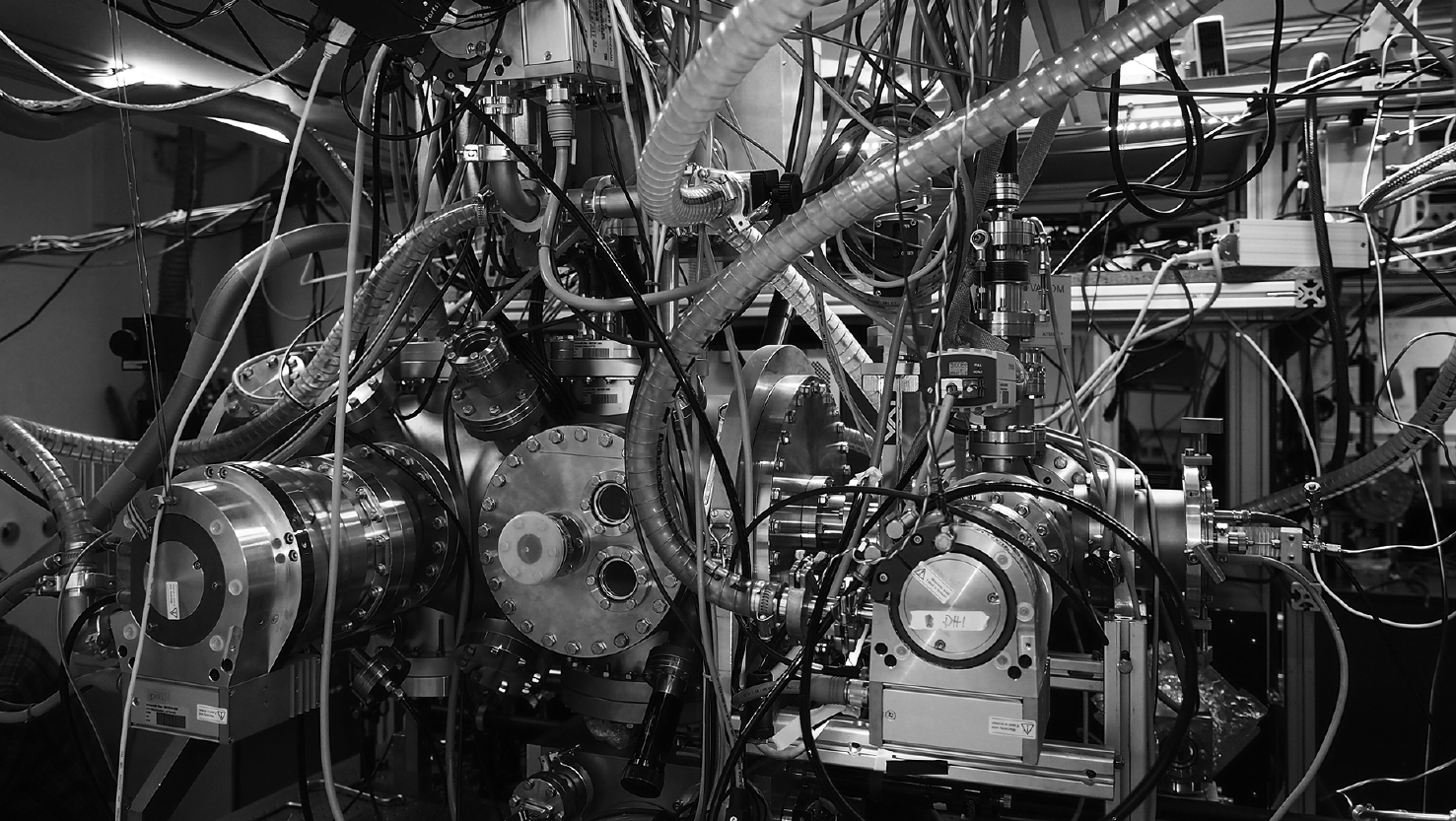}
\caption{Photo of the OTIMA setup: High vacuum is a prerequisite for all matter-wave interferometers. It is established by a set of turbomolecular pumps with vibration isolation bellows. The Even-Lavie \cite{EVEN} cluster source emits a thermal cloud of anthracene molecules, which cluster in the adiabatic co-expansion with an intense jet of noble gas. Top: the time-of-flight mass spectrometer is a central element, allowing to record interference with clusters of many masses simultaneously.}
\label{fig:otimaphoto}
\end{figure}

Talbot-Lau interferometers have proven to be suitable devices for quantum experiments with complex molecules, but it remains an outstanding goal to investigate the validity of quantum physics for truly macroscopic particles.
Future matter-wave interferometers should therefore fulfill certain criteria: they should be applicable to different types of particles. They should be able to cope with dispersive and therefore contrast-diminishing interactions between the particles and their environment. They should be robust against vibrations and ideally be fitted with a detection scheme that is sensitive on the few-particle level. Our optical time domain matter wave interferometer (OTIMA) incorporates these ideas and has demonstrated its functionality in first proof-of-principle interference experiments with clusters of organic molecules ~\cite{Haslinger2013}.
A sketch of the OTIMA setup is shown in fig.~\ref{fig:otimasetup} and a photo in fig.~\ref{fig:otimaphoto}. It is divided into three parts: a molecular beam source, the interferometer area and a detector.

The central element of the OTIMA setup is a flat $\mathrm{CaF_2}$ mirror. It retroreflects three VUV lasers pulses ($\lambda = $157\,nm, $E<5$\,mJ, $\tau =7$\,ns)
to form three standing waves with a mutual pulse delay $T$ which should be of the order of the Talbot time.
A burst of neutral particles is sent in close proximity across the mirror surface while being subjected to this pulse series. The small grating period of $d=\lambda/2=78.5$\,nm results in a Talbot time of 15\,ns/u,
and ensures that even particles of $m=10^6$\,u only require an unperturbed coherence time of 30\,ms.
The laser gratings interact with the matter waves in two different ways: On the one hand they have a phase component, as already discussed in the previous section.
 On the other hand they act as absorptive spatial filters for neutral particles, provided that single-photon ionization, or any other 'depletion' mechanism, is sufficiently probable in their anti-nodes.
A photon energy of 7.9\,eV suffices to ionize a large class of nanoparticles, such as metal and semiconductor clusters, certain molecular clusters and some biomolecules.
While G$_2$  could also be a pure phase grating, G$_1$ and G$_3$ need to be absorptive masks (as in TLI and KDTLI) to imprint the required transverse coherence onto the cluster beam and to reveal the emerging interference pattern.

Interferometry in the time domain eliminates a large range of dispersive phase shifts and one may conceive arrangements where a particle cloud even expands isotropically in all directions~\cite{Cahn1997}, with velocities of opposite sign.
This is possible since grating diffraction imparts a fixed transverse momentum onto the molecule.
In practice, a certain velocity selection is often needed to ensure that each particle sees every grating pulse.
For this reason the OTIMA interferometer is paired with a pulsed cluster source as well as a pulsed detector.
 In our current realization with anthracene molecules, adiabatic expansion from an Even-Lavie valve~\cite{EVEN} is used to produce a supersonic beam of cold neutral molecular clusters.
 The molecules are heated close to their melting point inside a noble gas environment (typically argon or neon) at high-pressure. When the valve opens, it releases a burst of molecules which cool in the presence of the co-expanding seed gas to clusters. The short (less than 30\,$\mu$s) dense cluster pulse enters the interferometer region via a skimmer as well as a horizontal and a vertical collimation slit. Behind the interferometer, the remaining neutral particles are post-ionized by another VUV laser beam and they are detected in a time-of-flight mass spectrometer (TOF-MS). This mass-resolved detection scheme allows us to compare the interferograms of all cluster masses simultaneously.

\subsection{Experimental design and conditions}
 \subsubsection{Requirements on mirror flatness and spectral purity of the grating lasers}
The dielectric surface of the interferometer mirror provides the boundary condition for all three standing light waves and thus determines their relative phases. As long as the mirror surface is smooth and flat any given separation of the laser pulses will result in a well-defined and predictable phase between the position of the molecular density pattern and G$_3$. Local corrugations on the Angstrom level and an overall flatness of 5\,nm across a mirror surface of 50\,mm would be ideally required.

A global mirror curvature caused by gravitational or thermal sag complicates the interference experiments as the local grating shift is then correlated with the particle cloud position below the mirror.
Even greater constraints are imposed by mirror corrugations that exceed about 5\,nm surface modulation over the width of a single laser beam (10\,mm). Since particles at different positions are then subjected to different grating phases the interference contrast will be diminished. This also sets stringent  requirements on the thermal stability of the mirror. CaF$_2$ exhibits minimal absorption at 157\,nm and heating by the laser pulses can be neglected. On the other hand, it is possible to illuminate the mirror locally with an intense  laser at a different wavelength to expand the material and to thermally scan the mirror phase.

At VUV wavelengths even the best available mirrors are limited to a reflectivity of  $R \simeq 96\%$. This entails that the transmitted fraction of the laser light (4\%) cannot contribute to the formation of the standing wave. The clusters are thus illuminated by a running wave background which reduces the useful signal as it ionizes all clusters with a small but mass dependent probability, independent of their position in the grating. This requires particular attention in the numerical simulations of the experiments.

As in TL and KDTL interferometry it is also important in OTIMA to match the grating periods of all three gratings over the entire width of the molecular beam.
We use three F$_2$ excimer lasers that emit predominantly on two lines, one at 157.63\,nm and one at 157.52 nm, where the latter is about ten times weaker~\cite{Sansonetti2001}.
Both lines are about 1\,pm wide and the resulting longitudinal coherence length of about 1\,cm requires the particles to pass in close proximity ($\sim 1$\,mm) to the mirror surface.
While the wavelength of many other excimer lasers, such as ArF or XeCl laser, can be varied externally, F$_2$-lasers exploit a molecular transition which varies with pressure and temperature by less than the
emission line width~\cite{Sansonetti2001}. This is the reason why three independent gas lasers are capable of generating three equally periodic light structures and thus work together to form a high-contrast matter-wave interferometer.
As in the consideration of single laser coherence, the grating nodes and antinodes of two subsequent laser beams only run out of phase when we compare more than 100\,000 grating periods, i.e. on the scale of several millimeters.

The  transverse coherence of fluorine lasers is typically limited to $40-80\,\mu$m. While at first glance this may seem to compromise the quality of a standing light wave across its 10\,mm width, it is the smoothness of the mirror surface which defines the wave fronts. Transverse coherence is only required to ensure that a standing wave can actually be formed. The formation of a standing light wave in a millimeter distance to the mirror surface is then still guaranteed as long as the divergence or the angle of incidence is limited to below 5\,mrad.

\subsubsection{Vacuum requirements}
Since collisions and interactions between interfering particles and background gas cause decoherence, the interferometer area of the OTIMA is surrounded by high vacuum of $p_{2}\textless 10^{-8}$ mbar. At this pressure even long ranging interactions between particles with a permanent or induced dipole moment are minimized to a level at which coherence times of up to 30\,ms with masses beyond $ 10^{6}$\,u are reasonable \cite{Nimmrichter2011}. In order to establish the high vacuum, the source chamber with a working pressure of $p_{1}\textless 10^{-5}$mbar is separated from the interferometer by two differential pumping stages.
All chambers are pumped by turbomolecular pumps that are separated from the chamber by dedicated vibration isolation bellows.

\subsubsection{Vibrational isolation}
 In most interferometers vibrations  are an important source of dephasing~\cite{Stibor2005a}. They result in uncontrolled and time dependent grating shifts which eliminate the interference contrast when they approach half the grating period $d/2$.
In our OTIMA setup the interferometric stability is greatly enhanced by the fact that all three grating laser pulses are reflected by the same mirror.
The total interference fringe shift in a three-grating interferometer is determined by the position shift of each individual grating. This is true for all our Talbot interferometers as well as for many of the atom interferometers described in this book:
\begin{equation}
\Delta \phi = (2\pi/d) (\Delta x_1 - 2 \Delta x_2 + \Delta x_3).
\end{equation}
Common mode perturbations, such as linear ($\Delta x_1=\Delta x_2=\Delta x_3$) or torsional pendulum oscillations (e.g. $\Delta x_2=0, \Delta x_3=-\Delta x_1$) cancel out if they  occur on a time scale longer than the transit time through the interferometer. Hence, low frequency mirror tilts and shifts can be neglected. In OTIMA interferometry, the time scale is set by the Talbot time $T_T$ and thus by the particle's mass\,$m$.
Mirror displacements larger than 10\,nm at a frequency greater than $2\pi/T_T$,
will however  cause vibrational dephasing and loss of interference contrast.
In our current experiment, the mirror vibrations were measured to be smaller than $\Delta x_i \le 5$\,nm within about $ 25\,\mu$s, i.e. on the time scale of the Talbot time for anthracene clusters.
This can be measured in situ with an optical Michelson interferometer. For experiments with higher mass particles, this optical readout can cross-correlated with the interference signal to correct for vibrations~\cite{Rodewald2011}.

\subsubsection{Alignment to gravity and the rotation of the Earth}
The reflective surface of the OTIMA interferometer mirror faces downwards and the grating axes are parallel to Earth's gravity. The vertical acceleration of the clusters does not lead to any dispersive phase shift since all particles stay in the interferometer for the same amount of time and fall by the same distance. This alleviates the alignment requirements with respect to gravity and will be important in tests of the equivalence principle with masses ranging from single atoms to clusters of $10^6$\,u, in the future~\cite{Hohensee2011b,Dimopoulos2007}. The gravitational deflection can even serve another purpose: it yields an effective fringe shift $\Delta x = gT^2$, which should the same for all particles if the equivalence principle is valid. It can therefore be used to implement a spatial scan of the interference pattern by varying the pulse delay $T$.
A limit is, however, reached when the particles are deflected to beyond the longitudinal coherence length of the grating lasers.
This limits our present interferometer to $m<10^6$\,u, since at $2T_T=30$\,ms all particles fall by 4\,mm.

In addition to that, the rotation of the Earth will induce a Coriolis phase shift $\Delta \phi=4\pi\mathbf{s}\cdot\left(\mathbf{v}\times\mathbf{\Omega}\right)T^2/d$ which cannot be eliminated even in a pulsed setup. The normal vector to the mirror surface $\mathbf{s}$ can also be aligned parallel to the axis of Earth's rotation to eliminate the Coriolis shift to first order. Higher order effects, caused by the beam splitting process and gravitational free fall remain as minor contributions~\cite{Hogan2007a,Lan2012,Dickerson2013}.

\subsubsection{Beam divergence}
\begin{figure}
 \begin{center}
 \includegraphics[width=0.51\textwidth]{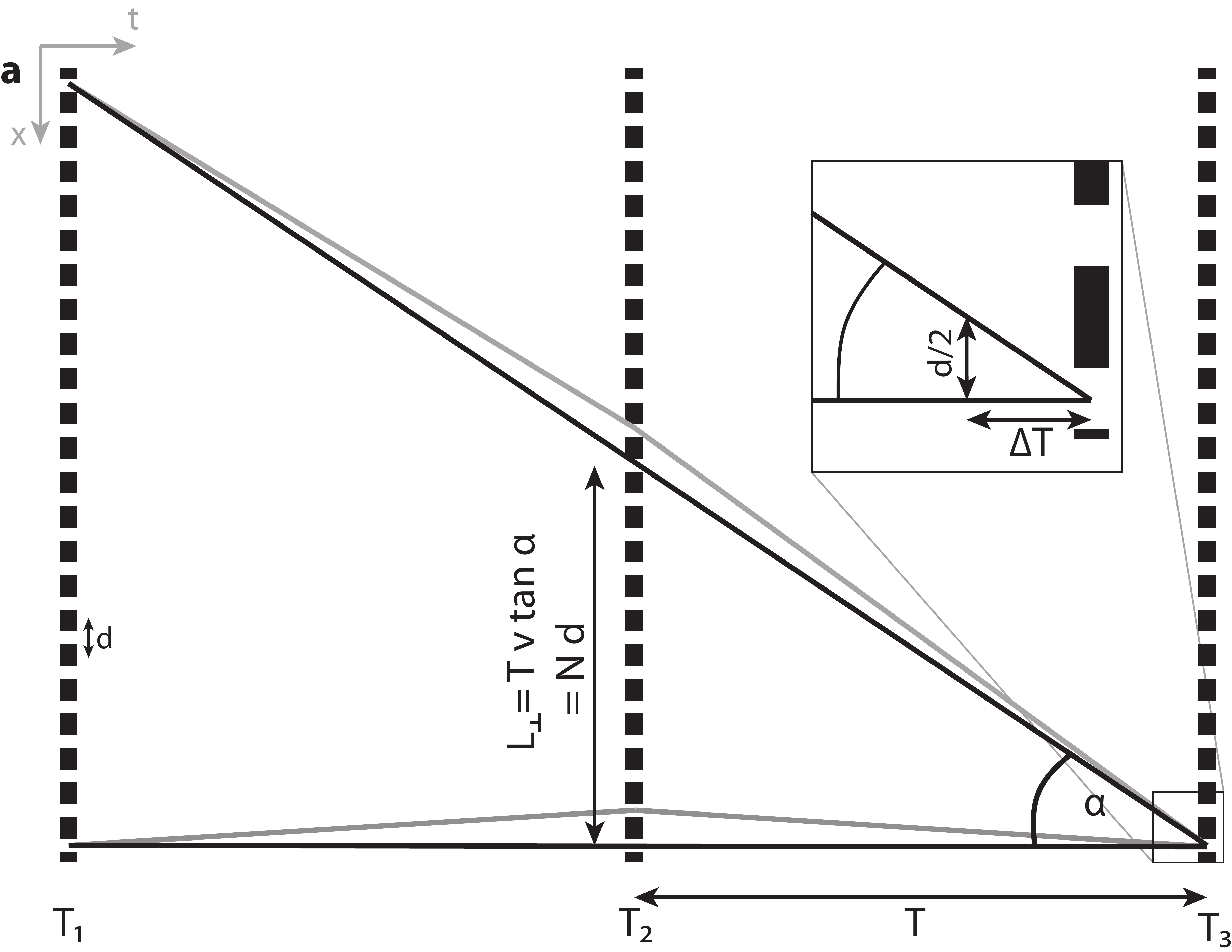} $\quad$
 \includegraphics[width=0.45\textwidth]{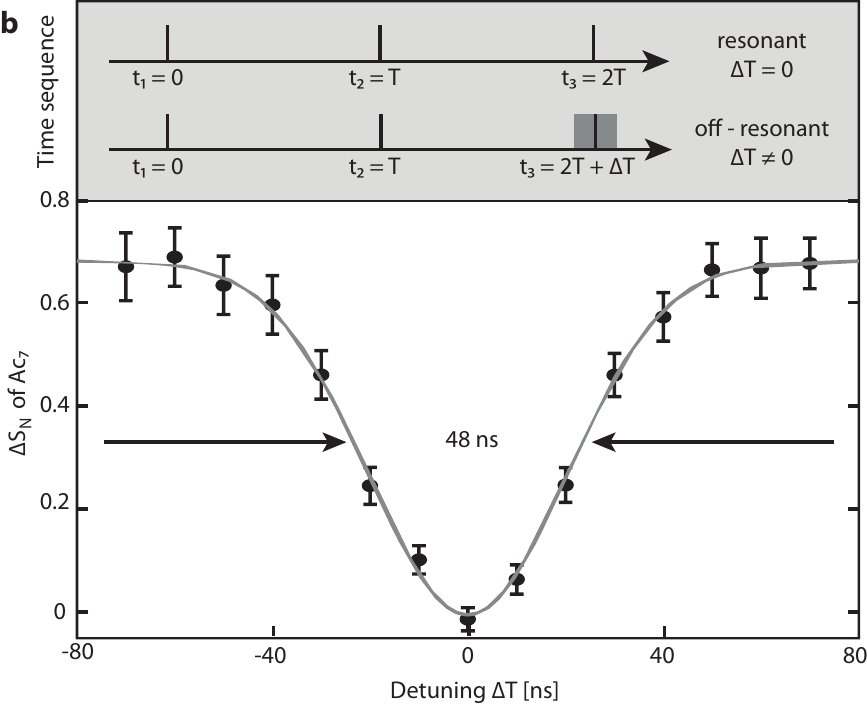}
  \caption{Influence of the beam divergence on the required timing precision.
  a) Particles that travel under different angles form the same interference patter - but only if both pulse separation times are equal.
  If the third grating pulse arrives too early or late, the particles interfering along the upper or lower trajectories will contribute to interference patterns with a relative shift.
  b) In the anthracene cluster experiment the beam divergence sets an upper limit of 40\,ns to the imbalance of pulse separation times. Since the laser jitter is fixed in the experiments but the Talbot-time grows with mass, the relative timing  requirements become less stringent for large particles.}
 \label{fig:divergence}
 \end{center}
 \end{figure}

 In the ideal case of perfect timing and negligible grating pulse duration, the molecular beam divergence has no influence on the interference contrast. In reality, however, the pulse length of the grating lasers is limited to about 8\,ns and the resolution of our timing jitter monitor ($\delta t=1$\,ns) sets an upper bound to the timing precision of the gratings. This is relevant because an asymmetry in the pulse delays causes a timing mismatch between the formation and the probing of the interference  pattern. Although each particle interferes only with itself, all particles must constructively contribute to the same interference pattern.
 For molecules arriving at G$_3$ under different angles this only works if the timing jitter of G$_3$ and the angular spread of the particle trajectories are small.
 With increasing divergence $\alpha$ the averaging over relatively shifted interference patterns leads to a reduction of the overall interference contrast.
 The beam divergence therefore imposes an upper bound on $\Delta T$. The visibility vanishes completely if the fringe  pattern is averaged over half the grating period $d$,
 \begin{equation}
  \frac{T}{L_{\perp}}<\frac{\Delta T}{d/2} \quad \rightarrow \quad
  \Delta T_{max} = \frac{d}{2v\tan\alpha} \quad \leftrightarrow \quad \frac{\Delta T}{T} < \frac{1}{2N},
  \label{eq:divergence}
 \end{equation}
 where $L_{\perp}=Nd$ is the distance that the most divergent particle with longitudinal velocity $v$ flies in the direction of the grating during time $T$ (see fig.\ref{fig:divergence}).
For a particle with a longitudinal velocity of 1000\,m/s and a divergence angle of 1\,mrad, the interference pattern will vanish once the two pulse delays $T_3-T_2$  and $T_2-T_1$ differ by more than 40\,ns.
This can be seen in fig.~\ref{fig:divergence} where the interference contrast is plotted as a function of the asymmetry in the pulse delays, as described in Sec.~\ref{sec:resonance}.

\subsection{Experimental results}
All experimental data shown in this section have been recorded with clusters of anthracene molecules. Anthracene ($C_{14}H_{10}$) is an aromatic hydrocarbon with a mass of 178\,u and its structure is shown as an inset in fig.~\ref{fig:massscan}. It has a high vapour pressure and therefore easily forms clusters in a supersonic expansion beam, in our case produced by the Even-Lavie valve. As for many clusters, the single-molecule ionization energy of 7.4\,eV decreases with the cluster size. In general, only little is known about optical properties of materials at 157\,nm. This applies, in particular to organic substances and to clusters of anthracene. The data therefore needs to be extracted from our own experiments.

\subsubsection{Quantum interference seen as a mass-dependent modulation of cluster transmission}
For sources that emit a broad mass distribution of clusters, an interference pattern can be obtained by choosing a fixed grating pulse delay and recording the interferometer transmission as a function of mass.
Two different pulse timing conditions are compared to get an unambiguous signature of quantum interference:
In the resonant mode, the delays between two subsequent laser pulses are equal to better than 1\,ns , $\Delta T_1-\Delta T_2 \equiv (t_2-t_1) - (t_3-t_2)< 1\,$ns.
For masses with $T\simeq n\cdot T_T$ (with $n \in \mathbb{N}$) constructive and destructive quantum interference should modulate the transmitted cluster signal $S_R$,  depending on the grating phases.
If the resonance condition is not met, $\Delta T_1 \neq \Delta T_2$, we do not expect any phase-dependent enhancement or reduction of the mass spectrum. We therefore compare the resonant signal to the off-resonant mode where the latter one is realized by shifting the delay of the third laser so that the relative grating delays differ by 200\,ns. Even such a small imbalance in the pulse separations suffices to wash out the interference pattern, when the cluster beam divergence exceeds 0.2\,mrad. The 'off-resonant' mass spectrum is thus a suitable reference $S_O$. In comparison to $S_R$ it has the same overall transmission but it lacks the interferometric modulation of the mass spectrum.

\begin{figure}
\includegraphics[width=\textwidth]{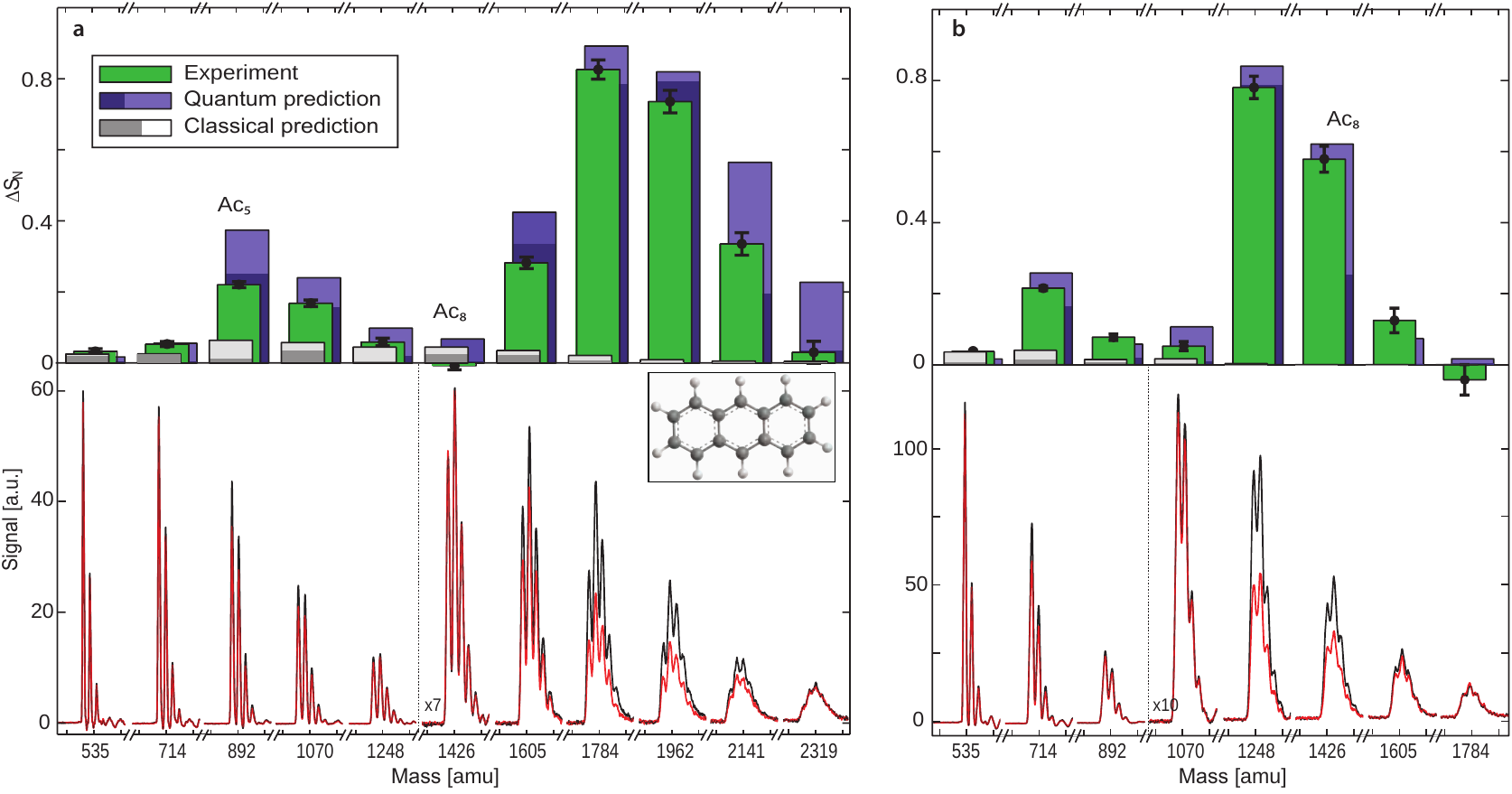}
\caption{OTIMA interference depends on the cluster mass and the Talbot time. The observed modulation on the mass spectrum peaks at the cluster mass of the anthracene decamer for $v=690$\,m/s while it shifts to the heptamer for $v=925$\,m/s. The resonances in the mass spectrum agree with the predictions of quantum mechanics. Also the observed transmission amplitudes agree with our expectations under reasonable assumptions about the cluster polarizability and absorption cross section.}
\label{fig:massscan}
\end{figure}

In fig.~\ref{fig:massscan} we plot the two signals as a function of mass and see a clear difference. To quantify the interference contrast, the normalized signal difference $\Delta S_N = (S_R-S_O)/S_O$ is introduced and compared to the quantum theoretical predictions. Figure~\ref{fig:massscan} shows two measurements which demonstrate in particular the role of the pulse separation time $T$. We can increase the most probable cluster velocity by changing the co-expanding seed gas from argon to the lighter noble gas neon, while keeping the source temperature constant. This allows us to shift the interference maximum from the 10-fold anthracene cluster Ac$_{10}$ for $T = 25.2\,\mu s$ at a velocity of  $v\simeq 690$\,m/s to the heptamer Ac$_{7}$ for $T = 18.9\,\mu$s and  $v\simeq 925$\,m/s. Theory and experiment are in good agreement. For details on the theoretical model we refer the reader to Section\,3 as well as to the literature~\cite{Nimmrichter2011a,Haslinger2013}.

\subsubsection{Interference resonance in the time domain}
\label{sec:resonance}
We can visualize the matter-wave resonance also in the time domain by systematically scanning the asymmetry $\Delta T=\Delta T_1-\Delta T_2$ in the off-resonant mode. Figure~\ref{fig:divergence} shows the interference contrast of the  anthracene heptamer Ac$_{7}$ as a function of  $\Delta T$ for a fixed pulse separation time of $T = 18.9\,\mu$s. For a cluster velocity of $v = 960$\,m/s, we can extract the beam divergence according to equation~\eqref{eq:divergence}. We find a divergence angle of  $\alpha \simeq 0.9$\,mrad which agrees well with our expectations for the given geometry of the experiment.

\subsubsection{Interference pattern in position-space}
\begin{figure}
\centering
\includegraphics[width=0.9\textwidth]{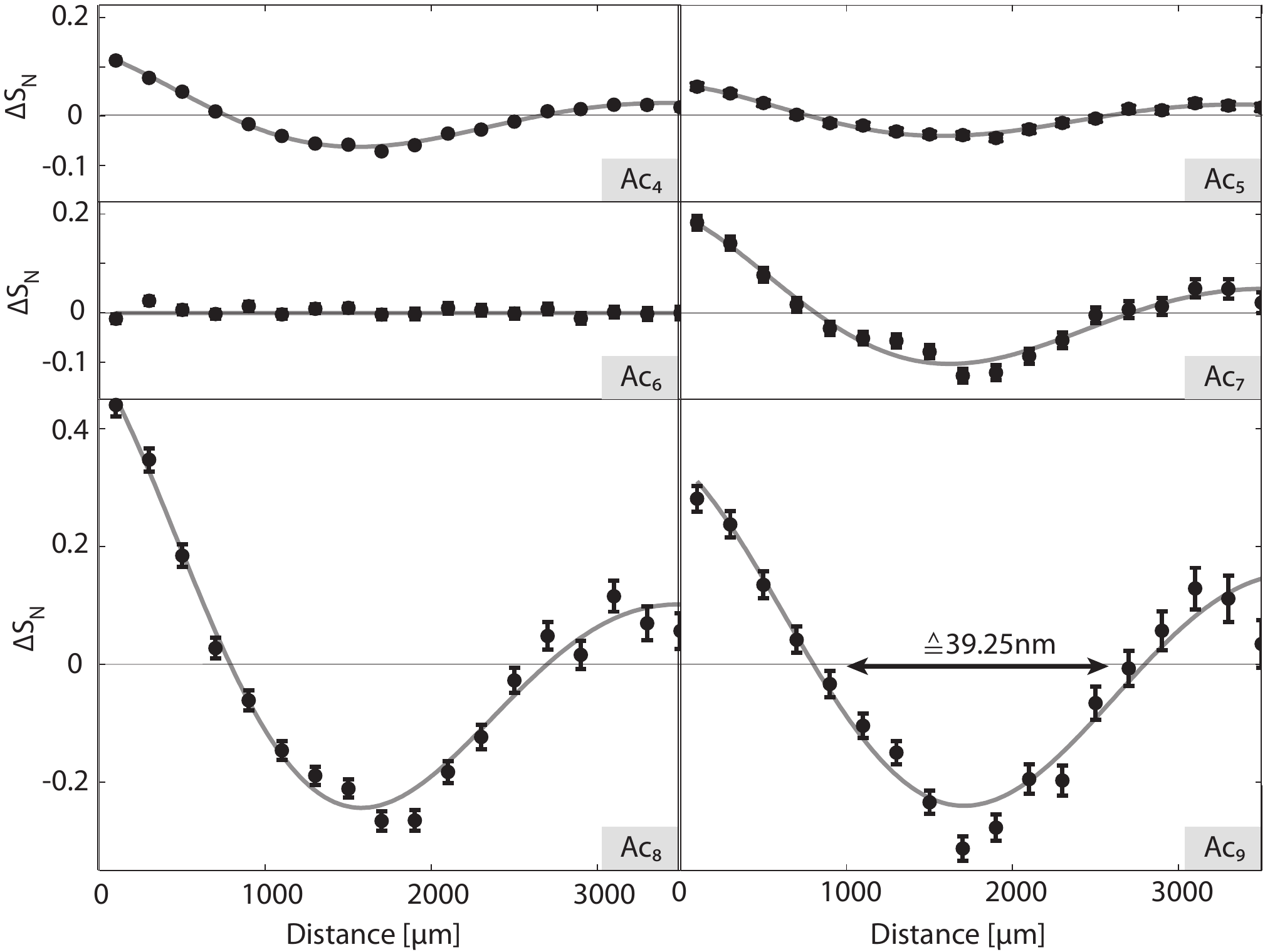}
\caption{Quantum interference in position space for the Anthracene tetramer (Ac$_4$) to the nonamer (Ac$_9$). When the laser beam in G$_2$ is tilted by 5\,mrad, a change of periodicity in the standing light wave by several picometers leads to an accumulated shift of the total grating phase which varies with the distance between the cluster beam and the mirror surface. Retracting the mirror surface by one millimeter therefore allows us to effectively sample the molecular density pattern in real space with a precision and resolution of several nanometers.}
\label{fig:scan1}
\end{figure}

In OTIMA interferometry, a simple scan of the third grating, as done in TLI and KDTLI, is impeded by the fact that all three gratings are reflected off the same rigid mirror. We may however scan the interference pattern by changing the periodicity of one of the gratings and by retracting the mirror vertically to move the cluster position in the light field.
In spite of the fact that the laser frequency is fixed, its standing wave period can be altered by tilting the grating laser beam by an angle, here specifically G$_2$ by $\theta = 5.1$\,mrad.
While the grating vector remains defined by the orientation of the mirror surface, increasing the tilt angle changes the normal wave vector $k_p = k \cos \theta$.
Many periods away from the mirror surface, in a distance of roughly 1.5\,mm, even a difference in the grating periods as small as a few picometers sums up to a relevant phase difference of the second grating with respect to the other two.
Since the position of G$_2$ influences the total fringe shift twice as strong as the values in G$_1$ and G$_3$, even a phase shift of $\pi/2$, corresponding to an effective shift of only 20\,nm is sufficient to switch the signal  from constructive to destructive interference.  By continuously increasing the separation of the mirror and the cluster beam we can observe the expected sinusoidal transmission curve, as shown in fig.~\ref{fig:scan1} for different cluster masses. The overall damping of the curves is consistent with the finite coherence length of the VUV lasers and the vertical extension of the cluster beam.

The OTIMA interferometer is consistent with our theoretical expectations in all three aspects:
We see a modulation of the transmitted mass distribution, which agrees also with the expectations of de Broglie waves assigned to each cluster mass at the known velocity.
We observe a narrow resonance in the time domain, which illustrates the potentially high sensitivity of a pulsed interferometer scheme.
And finally we see that also time domain interferometry creates a molecular density pattern in real space, which is as tiny as 80\,nm in our current implementation.
Exploiting higher-order Talbot-fringes, it is however also conceivable to create structures with a periodicity of 40 nm or smaller, with feature sizes of less than 20\,nm wide.
\section{Perspectives for quantum delocalization experiments at high masses}
\label{sec:perspectives}
Even though quantum mechanics has proven to be perfectly valid in all experimental tests so far, the number of quantum superposition experiments with increasingly macroscopic mechanical systems grows rapidly, opening a wide field of research at the interface between quantum and classical physics. Whether quantum theory must be modified in the macroworld is the essential question in this field \cite{Leggett2002a,Bassi2013}, with possible links to gravity theory \cite{Diosi1989,Penrose1996,Giulini2011}. Each new quantum experiment may then serve as another step closer to the answer by confirming the validity of quantum mechanics in a given setting of unprecedented macroscopicity. This requires, however, a unified quantitative notion of macroscopicity that allows one to compare different quantum experiments by the degree to which they pin down that answer and probe the validity of quantum mechanics.

Judging from the variety of quantum experiments with mechanical systems, different criteria could be invoked to characterize macroscopic quantum effects:
Nanomechanical oscillators in their vibrational ground state count among the most massive objects ever brought to the quantum regime \cite{Teufel2011,Chan2011}. On the other hand, if quantum superposition states were to be observed with such systems, their coherent delocalization would be limited to rather \emph{micro}scopic sizes: the width of motional ground-state wave packet is rarely larger than several dozen femtometers.

In contrast to this, neutron interferometers hold the current record with regard to the area spanned by the interference paths of a single quantum particle.
This can be as large as 100\,cm$^2$, i.e. in principle wide enough to let a single neutron be delocalized around the thickness of an arm~\cite{Zawisky2002}.
Nowadays, atom interferometers reach a high metrological sensitivity and an enclosed interferometer area in the range of square centimeters, too \cite{Lan2012,Dickerson2013}.
Modern atom interferometers employing optical beam splitters achieve superposition states with a momentum separation of up to $102\hbar k$ \cite{Chiow2011}.
State of the art macromolecule interferometers, as presented here, operate with smaller interferometer areas but at substantially higher mass.

Another macroscopic quantum phenomenon occurs in superconducting ring experiments, where superposition states of counterrotating persistent currents involving billions of electrons can be observed \cite{Friedman2000,Wal2000}. These superpositions exhibit a truly macroscopic difference in the magnetic moment associated to the superimposed loop currents. At the same time however, a many-body analysis shows that the two branches of the superposition differ by only a small number of electrons in the end \cite{Korsbakken2010}.

\begin{figure}
\includegraphics[width=10cm]{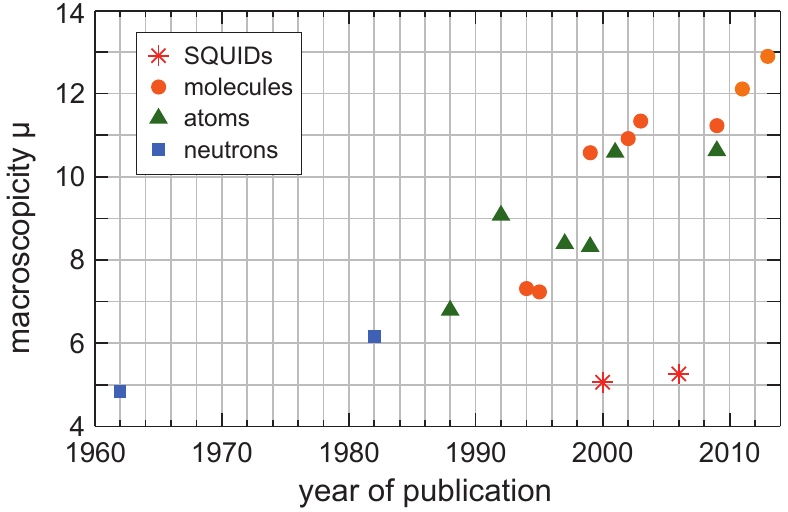}
\caption{Macroscopicities for a selection of neutron, atom, molecule and SQUID superposition experiments ordered according to their publication date. See \cite{Nimmrichter2013} and references therein for details on how the data points were obtained. The highest achieved macroscopicity values are shared by molecular Talbot-Lau interferometers and high-precision atom interferometers. SQUID experiments rank low due to the small mass of the superimposed electrons and due to the short coherence times achieved in the experiments.}
\label{fig:macro}
\end{figure}

In order to combine the various aspects of macroscopicity, and in order to compare the various experiments with mechanical systems, we have proposed a unified method to measure macroscopicity~\cite{Nimmrichter2013}. It quantifies the range of macrorealistic modifications of quantum theory that is ruled out by a given experiment. Such modifications induce classicality by adding a term to the Schr\"{o}dinger equation, which collapses delocalized wave functions at a rate that amplifies with the system size.
We have specified the mathematical form of a broad generic class of such modifications and defined a logarithmic quantity $\mu$ for the macroscopicity of mechanical quantum experiments. The microscopic reference value $\mu=0$ would be achieved by keeping an electron in a superposition state for one second. Figure~\hyperref[fig:macro]{22} shows the results for a representative selection of matter-wave experiments with neutrons, atoms, SQUIDs, and molecules. The highest macroscopicity to date, as reached in the KDTLI experiment with perfluoroalkyl-functionalized molecules~\cite{Gerlich2011,Eibenberger2013}, is the equivalent of a single electron kept in a superposition state for more than $10^{12}$ seconds.

Future experiments with nanoclusters in the OTIMA scheme may yield much higher macroscopicity values of the order of $20$, similar to what would be reached in other ambitious proposals to observe the quantum behaviour of nano- and micrometer-sized objects \cite{Marshall2003,Romero-Isart2011b}. In this regime, the most prominent example of a collapse modification, the model of continuous spontaneous localization (CSL) \cite{Ghirardi1990,Bassi2013}, should become experimentally accessible. Specifically, the CSL model predicts a mass-dependent reduction of Talbot-Lau interference contrast, as discussed in sec.~\ref{sec:collapseCSL}. We have shown that OTIMA interference with gold clusters of about a million atomic mass units would test the CSL model in its current formulation \cite{Nimmrichter2011}. That is to say, the observation of high interference visibility would place an upper bound for the rate at which the CSL effect occurs, according to eq. \hyperref[eq:CSLVis]{(37)}, which is comparable to the current estimate of the CSL rate parameter $\lambda_{{\rm CSL}}\sim10^{-10\pm2}\,$Hz.

All this confirms that de Broglie interferometry with molecules and nanoparticles remains a promising route towards testing quantum mechanics at the borderline to the classical macroworld. The quest for higher masses has led us away from elementary double-slit textbook interferometers to conceptually more advanced near-field methods, which are well understood theoretically and which facilitated the current mass and macroscopicity records in interferometry.

On the experimental side, an important key to all future experiments remains the development of suitable sources of very massive, very slow, cold and controlled nanoparticles. First success in feedback laser cooling~\cite{Gieseler2012} and cavity cooling \cite{Kiesel2013,Asenbaum2013} of particles between 70\,nm and 1000\,nm, i.e. with masses between  $10^8$\,u and $10^{12}$\,u
raise hope that high-mass quantum interference experiments will become available in a mass range where fascinating new physics lurk around the corner. Well-adapted interferometer technologies are now already available.


\newpage
\acknowledgments
We thank our collaboration partners for making nanoparticle matter-wave optics an exciting adventure.  We thank the Austrian Science Fund, FWF for financial support in the projects Z149-N16 (Wittgenstein) and DK CoQuS W1210-2 as well as the Ministry of science for support in the BM:WF project IS725001. We acknowledge funding by the European commission in the project EU NANOQUESTFIT(304886), the ERC Advanced Grant PROBIOTIQUS(320694) and the Vienna ZIT communication project (957475).

\bibliography{varenna}

\begin{thebibliography}{134}%
\makeatletter
\providecommand \@ifxundefined [1]{%
 \@ifx{#1\undefined}
}%
\providecommand \@ifnum [1]{%
 \ifnum #1\expandafter \@firstoftwo
 \else \expandafter \@secondoftwo
 \fi
}%
\providecommand \@ifx [1]{%
 \ifx #1\expandafter \@firstoftwo
 \else \expandafter \@secondoftwo
 \fi
}%
\providecommand \natexlab [1]{#1}%
\providecommand \enquote  [1]{``#1''}%
\providecommand \bibnamefont  [1]{#1}%
\providecommand \bibfnamefont [1]{#1}%
\providecommand \citenamefont [1]{#1}%
\providecommand \href@noop [0]{\@secondoftwo}%
\providecommand \href [0]{\begingroup \@sanitize@url \@href}%
\providecommand \@href[1]{\@@startlink{#1}\@@href}%
\providecommand \@@href[1]{\endgroup#1\@@endlink}%
\providecommand \@sanitize@url [0]{\catcode `\\12\catcode `\$12\catcode
  `\&12\catcode `\#12\catcode `\^12\catcode `\_12\catcode `\%12\relax}%
\providecommand \@@startlink[1]{}%
\providecommand \@@endlink[0]{}%
\providecommand \url  [0]{\begingroup\@sanitize@url \@url }%
\providecommand \@url [1]{\endgroup\@href {#1}{\urlprefix }}%
\providecommand \urlprefix  [0]{URL }%
\providecommand \Eprint [0]{\href }%
\providecommand \doibase [0]{http://dx.doi.org/}%
\providecommand \selectlanguage [0]{\@gobble}%
\providecommand \bibinfo  [0]{\@secondoftwo}%
\providecommand \bibfield  [0]{\@secondoftwo}%
\providecommand \translation [1]{[#1]}%
\providecommand \BibitemOpen [0]{}%
\providecommand \bibitemStop [0]{}%
\providecommand \bibitemNoStop [0]{.\EOS\space}%
\providecommand \EOS [0]{\spacefactor3000\relax}%
\providecommand \BibitemShut  [1]{\csname bibitem#1\endcsname}%
\let\auto@bib@innerbib\@empty
\bibitem [{\citenamefont {Hasselbach}(2010)}]{Hasselbach2010}%
  \BibitemOpen
  \bibfield  {author} {\bibinfo {author} {\bibfnamefont {F.}~\bibnamefont
  {Hasselbach}},\ }\href {\doibase doi:10.1088/0034-4885/73/1/016101}
  {\bibfield  {journal} {\bibinfo  {journal} {Rep. Prog. Phys.}\ }\textbf
  {\bibinfo {volume} {73}},\ \bibinfo {pages} {016101} (\bibinfo {year}
  {2010})}\BibitemShut {NoStop}%
\bibitem [{\citenamefont {Rauch}\ and\ \citenamefont
  {Werner}(2000)}]{Rauch2000}%
  \BibitemOpen
  \bibfield  {author} {\bibinfo {author} {\bibfnamefont {H.}~\bibnamefont
  {Rauch}}\ and\ \bibinfo {author} {\bibfnamefont {A.}~\bibnamefont {Werner}},\
  }\href@noop {} {\emph {\bibinfo {title} {Neutron Interferometry: Lessons in
  Experimental Quantum Mechanics}}}\ (\bibinfo  {publisher} {Oxford Univ.
  Press},\ \bibinfo {year} {2000})\BibitemShut {NoStop}%
\bibitem [{\citenamefont {Cronin}\ \emph {et~al.}(2009)\citenamefont {Cronin},
  \citenamefont {Schmiedmayer},\ and\ \citenamefont {Pritchard}}]{Cronin2009}%
  \BibitemOpen
  \bibfield  {author} {\bibinfo {author} {\bibfnamefont {A.~D.}\ \bibnamefont
  {Cronin}}, \bibinfo {author} {\bibfnamefont {J.}~\bibnamefont
  {Schmiedmayer}}, \ and\ \bibinfo {author} {\bibfnamefont {D.~E.}\
  \bibnamefont {Pritchard}},\ }\href {\doibase 10.1103/RevModPhys.81.1051}
  {\bibfield  {journal} {\bibinfo  {journal} {Rev. Mod. Phys.}\ }\textbf
  {\bibinfo {volume} {81}},\ \bibinfo {pages} {1051} (\bibinfo {year}
  {2009})}\BibitemShut {NoStop}%
\bibitem [{\citenamefont {Estermann}\ and\ \citenamefont
  {Stern}(1930)}]{Estermann1930}%
  \BibitemOpen
  \bibfield  {author} {\bibinfo {author} {\bibfnamefont {I.}~\bibnamefont
  {Estermann}}\ and\ \bibinfo {author} {\bibfnamefont {O.}~\bibnamefont
  {Stern}},\ }\href {\doibase 10.1007/BF01340293} {\bibfield  {journal}
  {\bibinfo  {journal} {Z. Phys.}\ }\textbf {\bibinfo {volume} {61}},\ \bibinfo
  {pages} {95} (\bibinfo {year} {1930})}\BibitemShut {NoStop}%
\bibitem [{\citenamefont {Bordé}\ \emph {et~al.}(1981)\citenamefont {Bordé},
  \citenamefont {Avrillier}, \citenamefont {Van~Lerberghe}, \citenamefont
  {Salomon}, \citenamefont {Bassi},\ and\ \citenamefont {Scoles}}]{Borde1981}%
  \BibitemOpen
  \bibfield  {author} {\bibinfo {author} {\bibfnamefont {C.~J.}\ \bibnamefont
  {Bordé}}, \bibinfo {author} {\bibfnamefont {S.}~\bibnamefont {Avrillier}},
  \bibinfo {author} {\bibfnamefont {A.}~\bibnamefont {Van~Lerberghe}}, \bibinfo
  {author} {\bibfnamefont {C.}~\bibnamefont {Salomon}}, \bibinfo {author}
  {\bibfnamefont {D.}~\bibnamefont {Bassi}}, \ and\ \bibinfo {author}
  {\bibfnamefont {G.}~\bibnamefont {Scoles}},\ }\href {\doibase
  10.1051/jphyscol:1981802} {\bibfield  {journal} {\bibinfo  {journal} {J.
  Phys. Colloq.}\ }\textbf {\bibinfo {volume} {42}},\ \bibinfo {pages} {C8}
  (\bibinfo {year} {1981})}\BibitemShut {NoStop}%
\bibitem [{\citenamefont {Bordé}\ \emph {et~al.}(1994)\citenamefont {Bordé},
  \citenamefont {Courtier}, \citenamefont {Burck}, \citenamefont {Goncharov},\
  and\ \citenamefont {Gorlicki}}]{Borde1994}%
  \BibitemOpen
  \bibfield  {author} {\bibinfo {author} {\bibfnamefont {C.}~\bibnamefont
  {Bordé}}, \bibinfo {author} {\bibfnamefont {N.}~\bibnamefont {Courtier}},
  \bibinfo {author} {\bibfnamefont {F.~D.}\ \bibnamefont {Burck}}, \bibinfo
  {author} {\bibfnamefont {A.}~\bibnamefont {Goncharov}}, \ and\ \bibinfo
  {author} {\bibfnamefont {M.}~\bibnamefont {Gorlicki}},\ }\href {\doibase
  10.1016/0375-9601(94)90437-5} {\bibfield  {journal} {\bibinfo  {journal}
  {Phys. Lett. A}\ }\textbf {\bibinfo {volume} {188}},\ \bibinfo {pages} {187}
  (\bibinfo {year} {1994})}\BibitemShut {NoStop}%
\bibitem [{\citenamefont {Bordé}(1989)}]{Borde1989}%
  \BibitemOpen
  \bibfield  {author} {\bibinfo {author} {\bibfnamefont {C.}~\bibnamefont
  {Bordé}},\ }\href {\doibase 10.1016/0375-9601(89)90537-9} {\bibfield
  {journal} {\bibinfo  {journal} {Phys. Lett. A}\ }\textbf {\bibinfo {volume}
  {140}},\ \bibinfo {pages} {10} (\bibinfo {year} {1989})}\BibitemShut
  {NoStop}%
\bibitem [{\citenamefont {Schöllkopf}\ and\ \citenamefont
  {Toennies}(1994)}]{Schoellkopf1994}%
  \BibitemOpen
  \bibfield  {author} {\bibinfo {author} {\bibfnamefont {W.}~\bibnamefont
  {Schöllkopf}}\ and\ \bibinfo {author} {\bibfnamefont {J.~P.}\ \bibnamefont
  {Toennies}},\ }\href {\doibase 10.1126/science.266.5189.1345} {\bibfield
  {journal} {\bibinfo  {journal} {Science}\ }\textbf {\bibinfo {volume}
  {266}},\ \bibinfo {pages} {1345} (\bibinfo {year} {1994})}\BibitemShut
  {NoStop}%
\bibitem [{\citenamefont {Chapman}\ \emph {et~al.}(1995)\citenamefont
  {Chapman}, \citenamefont {Hammond}, \citenamefont {Lenef}, \citenamefont
  {Schmiedmayer}, \citenamefont {Rubenstein}, \citenamefont {Smith},\ and\
  \citenamefont {Pritchard}}]{Chapman1995}%
  \BibitemOpen
  \bibfield  {author} {\bibinfo {author} {\bibfnamefont {M.~S.}\ \bibnamefont
  {Chapman}}, \bibinfo {author} {\bibfnamefont {T.~D.}\ \bibnamefont
  {Hammond}}, \bibinfo {author} {\bibfnamefont {A.}~\bibnamefont {Lenef}},
  \bibinfo {author} {\bibfnamefont {J.}~\bibnamefont {Schmiedmayer}}, \bibinfo
  {author} {\bibfnamefont {R.~A.}\ \bibnamefont {Rubenstein}}, \bibinfo
  {author} {\bibfnamefont {E.}~\bibnamefont {Smith}}, \ and\ \bibinfo {author}
  {\bibfnamefont {D.~E.}\ \bibnamefont {Pritchard}},\ }\href {\doibase
  10.1103/PhysRevLett.75.3783} {\bibfield  {journal} {\bibinfo  {journal}
  {Phys. Rev. Lett.}\ }\textbf {\bibinfo {volume} {75}},\ \bibinfo {pages}
  {3783 } (\bibinfo {year} {1995})}\BibitemShut {NoStop}%
\bibitem [{\citenamefont {Arndt}\ \emph {et~al.}(1999)\citenamefont {Arndt},
  \citenamefont {Nairz}, \citenamefont {Voss-Andreae}, \citenamefont {Keller},
  \citenamefont {van~der Zouw},\ and\ \citenamefont {Zeilinger}}]{Arndt1999}%
  \BibitemOpen
  \bibfield  {author} {\bibinfo {author} {\bibfnamefont {M.}~\bibnamefont
  {Arndt}}, \bibinfo {author} {\bibfnamefont {O.}~\bibnamefont {Nairz}},
  \bibinfo {author} {\bibfnamefont {J.}~\bibnamefont {Voss-Andreae}}, \bibinfo
  {author} {\bibfnamefont {C.}~\bibnamefont {Keller}}, \bibinfo {author}
  {\bibfnamefont {G.}~\bibnamefont {van~der Zouw}}, \ and\ \bibinfo {author}
  {\bibfnamefont {A.}~\bibnamefont {Zeilinger}},\ }\href {\doibase
  10.1038/44348} {\bibfield  {journal} {\bibinfo  {journal} {Nature}\ }\textbf
  {\bibinfo {volume} {401}},\ \bibinfo {pages} {680} (\bibinfo {year}
  {1999})}\BibitemShut {NoStop}%
\bibitem [{\citenamefont {Kolodney}\ \emph {et~al.}(1995)\citenamefont
  {Kolodney}, \citenamefont {Tsipinyuk},\ and\ \citenamefont
  {Budrevich}}]{Kolodney1995}%
  \BibitemOpen
  \bibfield  {author} {\bibinfo {author} {\bibfnamefont {E.}~\bibnamefont
  {Kolodney}}, \bibinfo {author} {\bibfnamefont {B.}~\bibnamefont {Tsipinyuk}},
  \ and\ \bibinfo {author} {\bibfnamefont {A.}~\bibnamefont {Budrevich}},\
  }\href {\doibase 10.1063/1.469235} {\bibfield  {journal} {\bibinfo  {journal}
  {J. Chem. Phys.}\ }\textbf {\bibinfo {volume} {102}},\ \bibinfo {pages} {9263
  } (\bibinfo {year} {1995})}\BibitemShut {NoStop}%
\bibitem [{\citenamefont {Hackermüller}\ \emph {et~al.}(2004)\citenamefont
  {Hackermüller}, \citenamefont {Hornberger}, \citenamefont {Brezger},
  \citenamefont {Zeilinger},\ and\ \citenamefont {Arndt}}]{Hackermueller2004}%
  \BibitemOpen
  \bibfield  {author} {\bibinfo {author} {\bibfnamefont {L.}~\bibnamefont
  {Hackermüller}}, \bibinfo {author} {\bibfnamefont {K.}~\bibnamefont
  {Hornberger}}, \bibinfo {author} {\bibfnamefont {B.}~\bibnamefont {Brezger}},
  \bibinfo {author} {\bibfnamefont {A.}~\bibnamefont {Zeilinger}}, \ and\
  \bibinfo {author} {\bibfnamefont {M.}~\bibnamefont {Arndt}},\ }\href
  {\doibase 10.1038/Nature02276} {\bibfield  {journal} {\bibinfo  {journal}
  {Nature}\ }\textbf {\bibinfo {volume} {427}},\ \bibinfo {pages} {711}
  (\bibinfo {year} {2004})}\BibitemShut {NoStop}%
\bibitem [{\citenamefont {Hornberger}\ \emph {et~al.}(2003)\citenamefont
  {Hornberger}, \citenamefont {Uttenthaler}, \citenamefont {Brezger},
  \citenamefont {Hackermüller}, \citenamefont {Arndt},\ and\ \citenamefont
  {Zeilinger}}]{Hornberger2003}%
  \BibitemOpen
  \bibfield  {author} {\bibinfo {author} {\bibfnamefont {K.}~\bibnamefont
  {Hornberger}}, \bibinfo {author} {\bibfnamefont {S.}~\bibnamefont
  {Uttenthaler}}, \bibinfo {author} {\bibfnamefont {B.}~\bibnamefont
  {Brezger}}, \bibinfo {author} {\bibfnamefont {L.}~\bibnamefont
  {Hackermüller}}, \bibinfo {author} {\bibfnamefont {M.}~\bibnamefont {Arndt}},
  \ and\ \bibinfo {author} {\bibfnamefont {A.}~\bibnamefont {Zeilinger}},\
  }\href {\doibase 10.1103/PhysRevLett.90.160401} {\bibfield  {journal}
  {\bibinfo  {journal} {Phys. Rev. Lett.}\ }\textbf {\bibinfo {volume} {90}},\
  \bibinfo {pages} {160401} (\bibinfo {year} {2003})}\BibitemShut {NoStop}%
\bibitem [{\citenamefont {de~Broglie}(1923)}]{Broglie1923a}%
  \BibitemOpen
  \bibfield  {author} {\bibinfo {author} {\bibfnamefont {L.}~\bibnamefont
  {de~Broglie}},\ }\href {\doibase 10.1038/112540a0} {\bibfield  {journal}
  {\bibinfo  {journal} {Nature}\ }\textbf {\bibinfo {volume} {112}},\ \bibinfo
  {pages} {540} (\bibinfo {year} {1923})}\BibitemShut {NoStop}%
\bibitem [{\citenamefont {Davisson}\ and\ \citenamefont
  {Germer}(1927)}]{Davisson1927a}%
  \BibitemOpen
  \bibfield  {author} {\bibinfo {author} {\bibfnamefont {C.}~\bibnamefont
  {Davisson}}\ and\ \bibinfo {author} {\bibfnamefont {L.~H.}\ \bibnamefont
  {Germer}},\ }\href {\doibase 10.1038/119558a0} {\bibfield  {journal}
  {\bibinfo  {journal} {Nature}\ }\textbf {\bibinfo {volume} {119}},\ \bibinfo
  {pages} {558} (\bibinfo {year} {1927})}\BibitemShut {NoStop}%
\bibitem [{\citenamefont {von Halban}\ and\ \citenamefont
  {Preiswerk}(1936)}]{Halban1936}%
  \BibitemOpen
  \bibfield  {author} {\bibinfo {author} {\bibfnamefont {H.}~\bibnamefont {von
  Halban}}\ and\ \bibinfo {author} {\bibfnamefont {P.}~\bibnamefont
  {Preiswerk}},\ }\href@noop {} {\bibfield  {journal} {\bibinfo  {journal} {C.\
  R.\ Hebd.\ S\'{e}ances Acad.}\ }\textbf {\bibinfo {volume} {203}},\ \bibinfo
  {pages} {73} (\bibinfo {year} {1936})}\BibitemShut {NoStop}%
\bibitem [{\citenamefont {Keith}\ \emph {et~al.}(1988)\citenamefont {Keith},
  \citenamefont {Schattenburg}, \citenamefont {Smith},\ and\ \citenamefont
  {Pritchard}}]{Keith1988}%
  \BibitemOpen
  \bibfield  {author} {\bibinfo {author} {\bibfnamefont {D.~W.}\ \bibnamefont
  {Keith}}, \bibinfo {author} {\bibfnamefont {M.~L.}\ \bibnamefont
  {Schattenburg}}, \bibinfo {author} {\bibfnamefont {H.~I.}\ \bibnamefont
  {Smith}}, \ and\ \bibinfo {author} {\bibfnamefont {D.~E.}\ \bibnamefont
  {Pritchard}},\ }\href {\doibase 10.1103/PhysRevLett.61.1580} {\bibfield
  {journal} {\bibinfo  {journal} {Phys. Rev. Lett.}\ }\textbf {\bibinfo
  {volume} {61}},\ \bibinfo {pages} {1580} (\bibinfo {year}
  {1988})}\BibitemShut {NoStop}%
\bibitem [{\citenamefont {Keith}\ \emph {et~al.}(1991)\citenamefont {Keith},
  \citenamefont {Ekstrom}, \citenamefont {Turchette},\ and\ \citenamefont
  {Pritchard}}]{Keith1991}%
  \BibitemOpen
  \bibfield  {author} {\bibinfo {author} {\bibfnamefont {D.~W.}\ \bibnamefont
  {Keith}}, \bibinfo {author} {\bibfnamefont {C.~R.}\ \bibnamefont {Ekstrom}},
  \bibinfo {author} {\bibfnamefont {Q.~A.}\ \bibnamefont {Turchette}}, \ and\
  \bibinfo {author} {\bibfnamefont {D.~E.}\ \bibnamefont {Pritchard}},\ }\href
  {\doibase 10.1103/PhysRevLett.66.2693} {\bibfield  {journal} {\bibinfo
  {journal} {Phys. Rev. Lett.}\ }\textbf {\bibinfo {volume} {66}},\ \bibinfo
  {pages} {2693} (\bibinfo {year} {1991})}\BibitemShut {NoStop}%
\bibitem [{\citenamefont {Kasevich}\ \emph {et~al.}(1991)\citenamefont
  {Kasevich}, \citenamefont {Weiss}, \citenamefont {Riis}, \citenamefont
  {Moler}, \citenamefont {Kasapi},\ and\ \citenamefont {Chu}}]{Kasevich1991}%
  \BibitemOpen
  \bibfield  {author} {\bibinfo {author} {\bibfnamefont {M.}~\bibnamefont
  {Kasevich}}, \bibinfo {author} {\bibfnamefont {D.~S.}\ \bibnamefont {Weiss}},
  \bibinfo {author} {\bibfnamefont {E.}~\bibnamefont {Riis}}, \bibinfo {author}
  {\bibfnamefont {K.}~\bibnamefont {Moler}}, \bibinfo {author} {\bibfnamefont
  {S.}~\bibnamefont {Kasapi}}, \ and\ \bibinfo {author} {\bibfnamefont
  {S.}~\bibnamefont {Chu}},\ }\href {\doibase 10.1103/PhysRevLett.66.2297}
  {\bibfield  {journal} {\bibinfo  {journal} {Phys. Rev. Lett.}\ }\textbf
  {\bibinfo {volume} {66}},\ \bibinfo {pages} {2297} (\bibinfo {year}
  {1991})}\BibitemShut {NoStop}%
\bibitem [{\citenamefont {Anderson}\ \emph {et~al.}(1995)\citenamefont
  {Anderson}, \citenamefont {Ensher}, \citenamefont {Matthews}, \citenamefont
  {Wieman},\ and\ \citenamefont {Cornell}}]{Anderson1995}%
  \BibitemOpen
  \bibfield  {author} {\bibinfo {author} {\bibfnamefont {M.~H.}\ \bibnamefont
  {Anderson}}, \bibinfo {author} {\bibfnamefont {J.~R.}\ \bibnamefont
  {Ensher}}, \bibinfo {author} {\bibfnamefont {M.~R.}\ \bibnamefont
  {Matthews}}, \bibinfo {author} {\bibfnamefont {C.~E.}\ \bibnamefont
  {Wieman}}, \ and\ \bibinfo {author} {\bibfnamefont {E.~A.}\ \bibnamefont
  {Cornell}},\ }\href {\doibase 10.1126/science.269.5221.198} {\bibfield
  {journal} {\bibinfo  {journal} {Science}\ }\textbf {\bibinfo {volume}
  {269}},\ \bibinfo {pages} {198} (\bibinfo {year} {1995})}\BibitemShut
  {NoStop}%
\bibitem [{\citenamefont {Davis}\ \emph {et~al.}(1995)\citenamefont {Davis},
  \citenamefont {Calongne},\ and\ \citenamefont {Rodriguez}}]{Davis1995}%
  \BibitemOpen
  \bibfield  {author} {\bibinfo {author} {\bibfnamefont {W.}~\bibnamefont
  {Davis}}, \bibinfo {author} {\bibfnamefont {S.}~\bibnamefont {Calongne}}, \
  and\ \bibinfo {author} {\bibfnamefont {J.}~\bibnamefont {Rodriguez}},\ }\href
  {\doibase 10.1063/1.469184} {\bibfield  {journal} {\bibinfo  {journal} {J.
  Chem. Phys.}\ }\textbf {\bibinfo {volume} {102}},\ \bibinfo {pages} {716 }
  (\bibinfo {year} {1995})}\BibitemShut {NoStop}%
\bibitem [{\citenamefont {Eibenberger}\ \emph {et~al.}(2013)\citenamefont
  {Eibenberger}, \citenamefont {Gerlich}, \citenamefont {Arndt}, \citenamefont
  {Mayor},\ and\ \citenamefont {Tüxen}}]{Eibenberger2013}%
  \BibitemOpen
  \bibfield  {author} {\bibinfo {author} {\bibfnamefont {S.}~\bibnamefont
  {Eibenberger}}, \bibinfo {author} {\bibfnamefont {S.}~\bibnamefont
  {Gerlich}}, \bibinfo {author} {\bibfnamefont {M.}~\bibnamefont {Arndt}},
  \bibinfo {author} {\bibfnamefont {M.}~\bibnamefont {Mayor}}, \ and\ \bibinfo
  {author} {\bibfnamefont {J.}~\bibnamefont {Tüxen}},\ }\href {\doibase
  10.1039/C3CP51500A} {\bibfield  {journal} {\bibinfo  {journal} {Phys. Chem.
  Chem. Phys.}\ }\textbf {\bibinfo {volume} {15}},\ \bibinfo {pages} {14696}
  (\bibinfo {year} {2013})}\BibitemShut {NoStop}%
\bibitem [{\citenamefont {Arndt}\ \emph {et~al.}(2005)\citenamefont {Arndt},
  \citenamefont {Hornberger},\ and\ \citenamefont {Zeilinger}}]{Arndt2005}%
  \BibitemOpen
  \bibfield  {author} {\bibinfo {author} {\bibfnamefont {M.}~\bibnamefont
  {Arndt}}, \bibinfo {author} {\bibfnamefont {K.}~\bibnamefont {Hornberger}}, \
  and\ \bibinfo {author} {\bibfnamefont {A.}~\bibnamefont {Zeilinger}},\
  }\href@noop {} {\bibfield  {journal} {\bibinfo  {journal} {Phys. World}\
  }\textbf {\bibinfo {volume} {18}},\ \bibinfo {pages} {35} (\bibinfo {year}
  {2005})}\BibitemShut {NoStop}%
\bibitem [{\citenamefont {Zeh}(1970)}]{Zeh1970}%
  \BibitemOpen
  \bibfield  {author} {\bibinfo {author} {\bibfnamefont {H.~D.}\ \bibnamefont
  {Zeh}},\ }\href {\doibase 10.1007/BF00708656} {\bibfield  {journal} {\bibinfo
   {journal} {Found. Phys.}\ }\textbf {\bibinfo {volume} {1}},\ \bibinfo
  {pages} {69} (\bibinfo {year} {1970})}\BibitemShut {NoStop}%
\bibitem [{\citenamefont {Caldeira}\ and\ \citenamefont
  {Leggett}(1985)}]{Caldeira1985}%
  \BibitemOpen
  \bibfield  {author} {\bibinfo {author} {\bibfnamefont {A.~O.}\ \bibnamefont
  {Caldeira}}\ and\ \bibinfo {author} {\bibfnamefont {A.~J.}\ \bibnamefont
  {Leggett}},\ }\href {\doibase 10.1103/PhysRevA.31.1059} {\bibfield  {journal}
  {\bibinfo  {journal} {Phys. Rev. A}\ }\textbf {\bibinfo {volume} {31}},\
  \bibinfo {pages} {1059} (\bibinfo {year} {1985})}\BibitemShut {NoStop}%
\bibitem [{\citenamefont {Joos}\ and\ \citenamefont {Zeh}(1985)}]{Joos1985}%
  \BibitemOpen
  \bibfield  {author} {\bibinfo {author} {\bibfnamefont {E.}~\bibnamefont
  {Joos}}\ and\ \bibinfo {author} {\bibfnamefont {H.~D.}\ \bibnamefont {Zeh}},\
  }\href {\doibase 10.1007/BF01725541} {\bibfield  {journal} {\bibinfo
  {journal} {Z. Phys. B.}\ }\textbf {\bibinfo {volume} {59}},\ \bibinfo {pages}
  {223} (\bibinfo {year} {1985})}\BibitemShut {NoStop}%
\bibitem [{\citenamefont {Zurek}(1991)}]{Zurek1991}%
  \BibitemOpen
  \bibfield  {author} {\bibinfo {author} {\bibfnamefont {W.~H.}\ \bibnamefont
  {Zurek}},\ }\href {\doibase 10.1063/1.881293} {\bibfield  {journal} {\bibinfo
   {journal} {Phys. Today}\ }\textbf {\bibinfo {volume} {44}},\ \bibinfo
  {pages} {36} (\bibinfo {year} {1991})}\BibitemShut {NoStop}%
\bibitem [{\citenamefont {Haroche}(2013)}]{Haroche2013}%
  \BibitemOpen
  \bibfield  {author} {\bibinfo {author} {\bibfnamefont {S.}~\bibnamefont
  {Haroche}},\ }\href {\doibase 10.1103/RevModPhys.85.1083} {\bibfield
  {journal} {\bibinfo  {journal} {Rev. Mod. Phys.}\ }\textbf {\bibinfo {volume}
  {85}},\ \bibinfo {pages} {1083} (\bibinfo {year} {2013})}\BibitemShut
  {NoStop}%
\bibitem [{\citenamefont {Romero-Isart}\ \emph {et~al.}(2010)\citenamefont
  {Romero-Isart}, \citenamefont {Juan}, \citenamefont {Quidant},\ and\
  \citenamefont {Cirac}}]{Romero-Isart2010}%
  \BibitemOpen
  \bibfield  {author} {\bibinfo {author} {\bibfnamefont {O.}~\bibnamefont
  {Romero-Isart}}, \bibinfo {author} {\bibfnamefont {M.}~\bibnamefont {Juan}},
  \bibinfo {author} {\bibfnamefont {R.}~\bibnamefont {Quidant}}, \ and\
  \bibinfo {author} {\bibfnamefont {J.}~\bibnamefont {Cirac}},\ }\href
  {\doibase 10.1088/1367-2630/12/3/033015} {\bibfield  {journal} {\bibinfo
  {journal} {New J. Phys.}\ }\textbf {\bibinfo {volume} {12}},\ \bibinfo
  {pages} {033015} (\bibinfo {year} {2010})}\BibitemShut {NoStop}%
\bibitem [{\citenamefont {Chang}\ \emph {et~al.}(2010)\citenamefont {Chang},
  \citenamefont {Regal}, \citenamefont {Papp}, \citenamefont {Wilson},
  \citenamefont {Ye}, \citenamefont {Painter}, \citenamefont {Kimble},\ and\
  \citenamefont {Zoller}}]{Chang2010}%
  \BibitemOpen
  \bibfield  {author} {\bibinfo {author} {\bibfnamefont {D.~E.}\ \bibnamefont
  {Chang}}, \bibinfo {author} {\bibfnamefont {C.~A.}\ \bibnamefont {Regal}},
  \bibinfo {author} {\bibfnamefont {S.~B.}\ \bibnamefont {Papp}}, \bibinfo
  {author} {\bibfnamefont {D.~J.}\ \bibnamefont {Wilson}}, \bibinfo {author}
  {\bibfnamefont {J.}~\bibnamefont {Ye}}, \bibinfo {author} {\bibfnamefont
  {O.}~\bibnamefont {Painter}}, \bibinfo {author} {\bibfnamefont {H.~J.}\
  \bibnamefont {Kimble}}, \ and\ \bibinfo {author} {\bibfnamefont
  {P.}~\bibnamefont {Zoller}},\ }\href {\doibase 10.1073/pnas.0912969107}
  {\bibfield  {journal} {\bibinfo  {journal} {Proc. Nat. Acad. Sc.}\ }\textbf
  {\bibinfo {volume} {107}},\ \bibinfo {pages} {1005} (\bibinfo {year}
  {2010})}\BibitemShut {NoStop}%
\bibitem [{\citenamefont {Nimmrichter}\ \emph
  {et~al.}(2011{\natexlab{a}})\citenamefont {Nimmrichter}, \citenamefont
  {Hornberger}, \citenamefont {Haslinger},\ and\ \citenamefont
  {Arndt}}]{Nimmrichter2011}%
  \BibitemOpen
  \bibfield  {author} {\bibinfo {author} {\bibfnamefont {S.}~\bibnamefont
  {Nimmrichter}}, \bibinfo {author} {\bibfnamefont {K.}~\bibnamefont
  {Hornberger}}, \bibinfo {author} {\bibfnamefont {P.}~\bibnamefont
  {Haslinger}}, \ and\ \bibinfo {author} {\bibfnamefont {M.}~\bibnamefont
  {Arndt}},\ }\href {\doibase 10.1103/PhysRevA.83.043621} {\bibfield  {journal}
  {\bibinfo  {journal} {Phys. Rev. A}\ }\textbf {\bibinfo {volume} {83}},\
  \bibinfo {pages} {043621} (\bibinfo {year} {2011}{\natexlab{a}})}\BibitemShut
  {NoStop}%
\bibitem [{\citenamefont {Leggett}(2002)}]{Leggett2002a}%
  \BibitemOpen
  \bibfield  {author} {\bibinfo {author} {\bibfnamefont {A.~J.}\ \bibnamefont
  {Leggett}},\ }\href {\doibase 10.1088/0953-8984/14/15/201} {\bibfield
  {journal} {\bibinfo  {journal} {J. Phys.: Condens. Matter}\ }\textbf
  {\bibinfo {volume} {14}},\ \bibinfo {pages} {R415} (\bibinfo {year}
  {2002})}\BibitemShut {NoStop}%
\bibitem [{\citenamefont {Ghirardi}\ \emph {et~al.}(1990)\citenamefont
  {Ghirardi}, \citenamefont {Pearle},\ and\ \citenamefont
  {Rimini}}]{Ghirardi1990}%
  \BibitemOpen
  \bibfield  {author} {\bibinfo {author} {\bibfnamefont {G.~C.}\ \bibnamefont
  {Ghirardi}}, \bibinfo {author} {\bibfnamefont {P.}~\bibnamefont {Pearle}}, \
  and\ \bibinfo {author} {\bibfnamefont {A.}~\bibnamefont {Rimini}},\ }\href
  {\doibase 10.1103/PhysRevA.42.78} {\bibfield  {journal} {\bibinfo  {journal}
  {Phys. Rev. A}\ }\textbf {\bibinfo {volume} {42}},\ \bibinfo {pages} {78 }
  (\bibinfo {year} {1990})}\BibitemShut {NoStop}%
\bibitem [{\citenamefont {Bassi}\ \emph {et~al.}(2013)\citenamefont {Bassi},
  \citenamefont {Lochan}, \citenamefont {Satin}, \citenamefont {Singh},\ and\
  \citenamefont {Ulbricht}}]{Bassi2013}%
  \BibitemOpen
  \bibfield  {author} {\bibinfo {author} {\bibfnamefont {A.}~\bibnamefont
  {Bassi}}, \bibinfo {author} {\bibfnamefont {K.}~\bibnamefont {Lochan}},
  \bibinfo {author} {\bibfnamefont {S.}~\bibnamefont {Satin}}, \bibinfo
  {author} {\bibfnamefont {T.~P.}\ \bibnamefont {Singh}}, \ and\ \bibinfo
  {author} {\bibfnamefont {H.}~\bibnamefont {Ulbricht}},\ }\href {\doibase
  10.1103/RevModPhys.85.471} {\bibfield  {journal} {\bibinfo  {journal} {Rev.
  Mod. Phys.}\ }\textbf {\bibinfo {volume} {85}},\ \bibinfo {pages} {471}
  (\bibinfo {year} {2013})}\BibitemShut {NoStop}%
\bibitem [{\citenamefont {Diosi}(1989)}]{Diosi1989}%
  \BibitemOpen
  \bibfield  {author} {\bibinfo {author} {\bibfnamefont {L.}~\bibnamefont
  {Diosi}},\ }\href {\doibase 10.1103/PhysRevA.40.1165} {\bibfield  {journal}
  {\bibinfo  {journal} {Phys. Rev. A}\ }\textbf {\bibinfo {volume} {40}},\
  \bibinfo {pages} {1165 } (\bibinfo {year} {1989})}\BibitemShut {NoStop}%
\bibitem [{\citenamefont {Penrose}(1996)}]{Penrose1996}%
  \BibitemOpen
  \bibfield  {author} {\bibinfo {author} {\bibfnamefont {R.}~\bibnamefont
  {Penrose}},\ }\href {\doibase 10.1007/BF02105068} {\bibfield  {journal}
  {\bibinfo  {journal} {Gen. Rel. Grav.}\ }\textbf {\bibinfo {volume} {28}},\
  \bibinfo {pages} {581} (\bibinfo {year} {1996})}\BibitemShut {NoStop}%
\bibitem [{\citenamefont {Giulini}\ and\ \citenamefont
  {Gro{\ss}ardt}(2011)}]{Giulini2011}%
  \BibitemOpen
  \bibfield  {author} {\bibinfo {author} {\bibfnamefont {D.}~\bibnamefont
  {Giulini}}\ and\ \bibinfo {author} {\bibfnamefont {A.}~\bibnamefont
  {Gro{\ss}ardt}},\ }\href {\doibase 10.1088/0264-9381/28/19/195026} {\bibfield
   {journal} {\bibinfo  {journal} {Class. Quant. Grav.}\ }\textbf {\bibinfo
  {volume} {28}},\ \bibinfo {pages} {195026} (\bibinfo {year}
  {2011})}\BibitemShut {NoStop}%
\bibitem [{\citenamefont {Nimmrichter}\ \emph {et~al.}(2008)\citenamefont
  {Nimmrichter}, \citenamefont {Hornberger}, \citenamefont {Ulbricht},\ and\
  \citenamefont {Arndt}}]{Nimmrichter2008}%
  \BibitemOpen
  \bibfield  {author} {\bibinfo {author} {\bibfnamefont {S.}~\bibnamefont
  {Nimmrichter}}, \bibinfo {author} {\bibfnamefont {K.}~\bibnamefont
  {Hornberger}}, \bibinfo {author} {\bibfnamefont {H.}~\bibnamefont
  {Ulbricht}}, \ and\ \bibinfo {author} {\bibfnamefont {M.}~\bibnamefont
  {Arndt}},\ }\href {\doibase 10.1103/PhysRevA.78.063607} {\bibfield  {journal}
  {\bibinfo  {journal} {Phys. Rev. A}\ }\textbf {\bibinfo {volume} {78}},\
  \bibinfo {pages} {063607} (\bibinfo {year} {2008})}\BibitemShut {NoStop}%
\bibitem [{\citenamefont {Gerlich}\ \emph {et~al.}(2008)\citenamefont
  {Gerlich}, \citenamefont {Gring}, \citenamefont {Ulbricht}, \citenamefont
  {Hornberger}, \citenamefont {Tüxen}, \citenamefont {Mayor},\ and\
  \citenamefont {Arndt}}]{Gerlich2008}%
  \BibitemOpen
  \bibfield  {author} {\bibinfo {author} {\bibfnamefont {S.}~\bibnamefont
  {Gerlich}}, \bibinfo {author} {\bibfnamefont {M.}~\bibnamefont {Gring}},
  \bibinfo {author} {\bibfnamefont {H.}~\bibnamefont {Ulbricht}}, \bibinfo
  {author} {\bibfnamefont {K.}~\bibnamefont {Hornberger}}, \bibinfo {author}
  {\bibfnamefont {J.}~\bibnamefont {Tüxen}}, \bibinfo {author} {\bibfnamefont
  {M.}~\bibnamefont {Mayor}}, \ and\ \bibinfo {author} {\bibfnamefont
  {M.}~\bibnamefont {Arndt}},\ }\href {\doibase 10.1002/anie.200801942}
  {\bibfield  {journal} {\bibinfo  {journal} {Angew. Chem. Int. Ed.}\ }\textbf
  {\bibinfo {volume} {47}},\ \bibinfo {pages} {6195 } (\bibinfo {year}
  {2008})}\BibitemShut {NoStop}%
\bibitem [{\citenamefont {Gring}\ \emph {et~al.}(2010)\citenamefont {Gring},
  \citenamefont {Gerlich}, \citenamefont {Eibenberger}, \citenamefont
  {Nimmrichter}, \citenamefont {Berrada}, \citenamefont {Arndt}, \citenamefont
  {Ulbricht}, \citenamefont {Hornberger}, \citenamefont {Müri}, \citenamefont
  {Mayor}, \citenamefont {Böckmann},\ and\ \citenamefont
  {Doltsinis}}]{Gring2010}%
  \BibitemOpen
  \bibfield  {author} {\bibinfo {author} {\bibfnamefont {M.}~\bibnamefont
  {Gring}}, \bibinfo {author} {\bibfnamefont {S.}~\bibnamefont {Gerlich}},
  \bibinfo {author} {\bibfnamefont {S.}~\bibnamefont {Eibenberger}}, \bibinfo
  {author} {\bibfnamefont {S.}~\bibnamefont {Nimmrichter}}, \bibinfo {author}
  {\bibfnamefont {T.}~\bibnamefont {Berrada}}, \bibinfo {author} {\bibfnamefont
  {M.}~\bibnamefont {Arndt}}, \bibinfo {author} {\bibfnamefont
  {H.}~\bibnamefont {Ulbricht}}, \bibinfo {author} {\bibfnamefont
  {K.}~\bibnamefont {Hornberger}}, \bibinfo {author} {\bibfnamefont
  {M.}~\bibnamefont {Müri}}, \bibinfo {author} {\bibfnamefont {M.}~\bibnamefont
  {Mayor}}, \bibinfo {author} {\bibfnamefont {M.}~\bibnamefont {Böckmann}}, \
  and\ \bibinfo {author} {\bibfnamefont {N.}~\bibnamefont {Doltsinis}},\ }\href
  {\doibase 10.1103/PhysRevA.81.031604} {\bibfield  {journal} {\bibinfo
  {journal} {Phys. Rev. A}\ }\textbf {\bibinfo {volume} {81}},\ \bibinfo
  {pages} {031604} (\bibinfo {year} {2010})}\BibitemShut {NoStop}%
\bibitem [{\citenamefont {Juffmann}\ \emph {et~al.}(2013)\citenamefont
  {Juffmann}, \citenamefont {Ulbricht},\ and\ \citenamefont
  {Arndt}}]{Juffmann2013}%
  \BibitemOpen
  \bibfield  {author} {\bibinfo {author} {\bibfnamefont {T.}~\bibnamefont
  {Juffmann}}, \bibinfo {author} {\bibfnamefont {H.}~\bibnamefont {Ulbricht}},
  \ and\ \bibinfo {author} {\bibfnamefont {M.}~\bibnamefont {Arndt}},\ }\href
  {\doibase 10.1088/0034-4885/76/8/086402} {\bibfield  {journal} {\bibinfo
  {journal} {Rep. Prog. Phys.}\ }\textbf {\bibinfo {volume} {76}},\ \bibinfo
  {pages} {086402} (\bibinfo {year} {2013})}\BibitemShut {NoStop}%
\bibitem [{\citenamefont {Juffmann}\ \emph {et~al.}(2009)\citenamefont
  {Juffmann}, \citenamefont {Truppe}, \citenamefont {Geyer}, \citenamefont
  {Mayor}, \citenamefont {Deachapunya}, \citenamefont {Ulbricht},\ and\
  \citenamefont {Arndt}}]{Juffmann2009}%
  \BibitemOpen
  \bibfield  {author} {\bibinfo {author} {\bibfnamefont {T.}~\bibnamefont
  {Juffmann}}, \bibinfo {author} {\bibfnamefont {S.}~\bibnamefont {Truppe}},
  \bibinfo {author} {\bibfnamefont {P.}~\bibnamefont {Geyer}}, \bibinfo
  {author} {\bibfnamefont {A.}~\bibnamefont {Mayor}}, \bibinfo {author}
  {\bibfnamefont {S.}~\bibnamefont {Deachapunya}}, \bibinfo {author}
  {\bibfnamefont {H.}~\bibnamefont {Ulbricht}}, \ and\ \bibinfo {author}
  {\bibfnamefont {M.}~\bibnamefont {Arndt}},\ }\href {\doibase
  10.1103/PhysRevLett.103.263601} {\bibfield  {journal} {\bibinfo  {journal}
  {Phys. Rev. Lett.}\ }\textbf {\bibinfo {volume} {103}},\ \bibinfo {pages}
  {263601} (\bibinfo {year} {2009})}\BibitemShut {NoStop}%
\bibitem [{\citenamefont {Eiermann}\ \emph {et~al.}(2003)\citenamefont
  {Eiermann}, \citenamefont {Treutlein}, \citenamefont {Anker}, \citenamefont
  {Albiez}, \citenamefont {M.Taglieber}, \citenamefont {Marzlin},\ and\
  \citenamefont {Oberthaler}}]{Eiermann2003}%
  \BibitemOpen
  \bibfield  {author} {\bibinfo {author} {\bibfnamefont {B.}~\bibnamefont
  {Eiermann}}, \bibinfo {author} {\bibfnamefont {P.}~\bibnamefont {Treutlein}},
  \bibinfo {author} {\bibfnamefont {T.}~\bibnamefont {Anker}}, \bibinfo
  {author} {\bibfnamefont {M.}~\bibnamefont {Albiez}}, \bibinfo {author}
  {\bibnamefont {M.Taglieber}}, \bibinfo {author} {\bibfnamefont {K.-P.}\
  \bibnamefont {Marzlin}}, \ and\ \bibinfo {author} {\bibfnamefont {M.~K.}\
  \bibnamefont {Oberthaler}},\ }\href {\doibase 10.1103/PhysRevLett.91.060402}
  {\bibfield  {journal} {\bibinfo  {journal} {Phys. Rev. Lett.}\ }\textbf
  {\bibinfo {volume} {91}},\ \bibinfo {pages} {060402} (\bibinfo {year}
  {2003})}\BibitemShut {NoStop}%
\bibitem [{\citenamefont {Ockeloen}\ \emph {et~al.}(2013)\citenamefont
  {Ockeloen}, \citenamefont {Schmied}, \citenamefont {Riedel},\ and\
  \citenamefont {Treutlein}}]{Ockeloen2013}%
  \BibitemOpen
  \bibfield  {author} {\bibinfo {author} {\bibfnamefont {C.~F.}\ \bibnamefont
  {Ockeloen}}, \bibinfo {author} {\bibfnamefont {R.}~\bibnamefont {Schmied}},
  \bibinfo {author} {\bibfnamefont {M.~F.}\ \bibnamefont {Riedel}}, \ and\
  \bibinfo {author} {\bibfnamefont {P.}~\bibnamefont {Treutlein}},\ }\href
  {\doibase 10.1103/PhysRevLett.111.143001} {\bibfield  {journal} {\bibinfo
  {journal} {Phys. Rev. Lett.}\ }\textbf {\bibinfo {volume} {111}},\ \bibinfo
  {pages} {143001} (\bibinfo {year} {2013})}\BibitemShut {NoStop}%
\bibitem [{\citenamefont {Born}\ and\ \citenamefont {Wolf}(1993)}]{Born1993}%
  \BibitemOpen
  \bibfield  {author} {\bibinfo {author} {\bibfnamefont {M.}~\bibnamefont
  {Born}}\ and\ \bibinfo {author} {\bibfnamefont {E.}~\bibnamefont {Wolf}},\
  }\href@noop {} {\emph {\bibinfo {title} {Principles of Optics}}}\ (\bibinfo
  {publisher} {Pergamon Press},\ \bibinfo {year} {1993})\BibitemShut {NoStop}%
\bibitem [{\citenamefont {Scoles}\ \emph {et~al.}(1989)\citenamefont {Scoles},
  \citenamefont {Bassi}, \citenamefont {Buck}, \citenamefont {Laine},\ and\
  \citenamefont {Braun}}]{Scoles1989}%
  \BibitemOpen
  \bibfield  {author} {\bibinfo {author} {\bibfnamefont {G.}~\bibnamefont
  {Scoles}}, \bibinfo {author} {\bibfnamefont {D.}~\bibnamefont {Bassi}},
  \bibinfo {author} {\bibfnamefont {U.}~\bibnamefont {Buck}}, \bibinfo {author}
  {\bibfnamefont {D.}~\bibnamefont {Laine}}, \ and\ \bibinfo {author}
  {\bibfnamefont {C.}~\bibnamefont {Braun}},\ }\href@noop {} {\bibfield
  {journal} {\bibinfo  {journal} {Applied Optics}\ }\textbf {\bibinfo {volume}
  {28}},\ \bibinfo {pages} {3258} (\bibinfo {year} {1989})}\BibitemShut
  {NoStop}%
\bibitem [{\citenamefont {Szewc}\ \emph {et~al.}(2010)\citenamefont {Szewc},
  \citenamefont {Collier},\ and\ \citenamefont {Ulbricht}}]{Szewc2010}%
  \BibitemOpen
  \bibfield  {author} {\bibinfo {author} {\bibfnamefont {C.}~\bibnamefont
  {Szewc}}, \bibinfo {author} {\bibfnamefont {J.~D.}\ \bibnamefont {Collier}},
  \ and\ \bibinfo {author} {\bibfnamefont {H.}~\bibnamefont {Ulbricht}},\
  }\href {\doibase 10.1063/1.3499254} {\bibfield  {journal} {\bibinfo
  {journal} {Rev. Sci. Instrum.}\ }\textbf {\bibinfo {volume} {81}},\ \bibinfo
  {pages} {106107} (\bibinfo {year} {2010})}\BibitemShut {NoStop}%
\bibitem [{\citenamefont {Nairz}\ \emph {et~al.}(2000)\citenamefont {Nairz},
  \citenamefont {Arndt},\ and\ \citenamefont {Zeilinger}}]{Nairz2000}%
  \BibitemOpen
  \bibfield  {author} {\bibinfo {author} {\bibfnamefont {O.}~\bibnamefont
  {Nairz}}, \bibinfo {author} {\bibfnamefont {M.}~\bibnamefont {Arndt}}, \ and\
  \bibinfo {author} {\bibfnamefont {A.}~\bibnamefont {Zeilinger}},\ }\href
  {\doibase 10.1080/09500340008232198} {\bibfield  {journal} {\bibinfo
  {journal} {J. Mod. Opt.}\ }\textbf {\bibinfo {volume} {47}},\ \bibinfo
  {pages} {2811} (\bibinfo {year} {2000})}\BibitemShut {NoStop}%
\bibitem [{\citenamefont {Brezger}\ \emph {et~al.}(2002)\citenamefont
  {Brezger}, \citenamefont {Hackermüller}, \citenamefont {Uttenthaler},
  \citenamefont {Petschinka}, \citenamefont {Arndt},\ and\ \citenamefont
  {Zeilinger}}]{Brezger2002}%
  \BibitemOpen
  \bibfield  {author} {\bibinfo {author} {\bibfnamefont {B.}~\bibnamefont
  {Brezger}}, \bibinfo {author} {\bibfnamefont {L.}~\bibnamefont
  {Hackermüller}}, \bibinfo {author} {\bibfnamefont {S.}~\bibnamefont
  {Uttenthaler}}, \bibinfo {author} {\bibfnamefont {J.}~\bibnamefont
  {Petschinka}}, \bibinfo {author} {\bibfnamefont {M.}~\bibnamefont {Arndt}}, \
  and\ \bibinfo {author} {\bibfnamefont {A.}~\bibnamefont {Zeilinger}},\ }\href
  {\doibase 10.1103/PhysRevLett.88.100404} {\bibfield  {journal} {\bibinfo
  {journal} {Phys. Rev. Lett.}\ }\textbf {\bibinfo {volume} {88}},\ \bibinfo
  {pages} {100404} (\bibinfo {year} {2002})}\BibitemShut {NoStop}%
\bibitem [{\citenamefont {Juffmann}\ \emph
  {et~al.}(2012{\natexlab{a}})\citenamefont {Juffmann}, \citenamefont
  {Nimmrichter}, \citenamefont {Arndt}, \citenamefont {Gleiter},\ and\
  \citenamefont {Hornberger}}]{Juffmann2012}%
  \BibitemOpen
  \bibfield  {author} {\bibinfo {author} {\bibfnamefont {T.}~\bibnamefont
  {Juffmann}}, \bibinfo {author} {\bibfnamefont {S.}~\bibnamefont
  {Nimmrichter}}, \bibinfo {author} {\bibfnamefont {M.}~\bibnamefont {Arndt}},
  \bibinfo {author} {\bibfnamefont {H.}~\bibnamefont {Gleiter}}, \ and\
  \bibinfo {author} {\bibfnamefont {K.}~\bibnamefont {Hornberger}},\ }\href
  {\doibase 10.1007/s10701-010-9520-5} {\bibfield  {journal} {\bibinfo
  {journal} {Found. Phys.}\ }\textbf {\bibinfo {volume} {42}},\ \bibinfo
  {pages} {98} (\bibinfo {year} {2012}{\natexlab{a}})}\BibitemShut {NoStop}%
\bibitem [{\citenamefont {Gieseler}\ \emph {et~al.}(2012)\citenamefont
  {Gieseler}, \citenamefont {Deutsch}, \citenamefont {Quidant},\ and\
  \citenamefont {Novotny}}]{Gieseler2012}%
  \BibitemOpen
  \bibfield  {author} {\bibinfo {author} {\bibfnamefont {J.}~\bibnamefont
  {Gieseler}}, \bibinfo {author} {\bibfnamefont {B.}~\bibnamefont {Deutsch}},
  \bibinfo {author} {\bibfnamefont {R.}~\bibnamefont {Quidant}}, \ and\
  \bibinfo {author} {\bibfnamefont {L.}~\bibnamefont {Novotny}},\ }\href
  {\doibase 10.1103/PhysRevLett.109.103603} {\bibfield  {journal} {\bibinfo
  {journal} {Phys. Rev. Lett.}\ }\textbf {\bibinfo {volume} {109}},\ \bibinfo
  {pages} {103603} (\bibinfo {year} {2012})}\BibitemShut {NoStop}%
\bibitem [{\citenamefont {Kiesel}\ \emph {et~al.}(2013)\citenamefont {Kiesel},
  \citenamefont {Blaser}, \citenamefont {Delic}, \citenamefont {Grass},
  \citenamefont {Kaltenbaek},\ and\ \citenamefont {Aspelmeyer}}]{Kiesel2013}%
  \BibitemOpen
  \bibfield  {author} {\bibinfo {author} {\bibfnamefont {N.}~\bibnamefont
  {Kiesel}}, \bibinfo {author} {\bibfnamefont {F.}~\bibnamefont {Blaser}},
  \bibinfo {author} {\bibfnamefont {U.}~\bibnamefont {Delic}}, \bibinfo
  {author} {\bibfnamefont {D.}~\bibnamefont {Grass}}, \bibinfo {author}
  {\bibfnamefont {R.}~\bibnamefont {Kaltenbaek}}, \ and\ \bibinfo {author}
  {\bibfnamefont {M.}~\bibnamefont {Aspelmeyer}},\ }\href {\doibase
  10.1073/pnas.1309167110} {\bibfield  {journal} {\bibinfo  {journal} {PNAS}\
  }\textbf {\bibinfo {volume} {110}},\ \bibinfo {pages} {14180} (\bibinfo
  {year} {2013})}\BibitemShut {NoStop}%
\bibitem [{\citenamefont {Asenbaum}\ \emph {et~al.}(2013)\citenamefont
  {Asenbaum}, \citenamefont {Kuhn}, \citenamefont {Nimmrichter}, \citenamefont
  {Sezer},\ and\ \citenamefont {Arndt}}]{Asenbaum2013}%
  \BibitemOpen
  \bibfield  {author} {\bibinfo {author} {\bibfnamefont {P.}~\bibnamefont
  {Asenbaum}}, \bibinfo {author} {\bibfnamefont {S.}~\bibnamefont {Kuhn}},
  \bibinfo {author} {\bibfnamefont {S.}~\bibnamefont {Nimmrichter}}, \bibinfo
  {author} {\bibfnamefont {U.}~\bibnamefont {Sezer}}, \ and\ \bibinfo {author}
  {\bibfnamefont {M.}~\bibnamefont {Arndt}},\ }\href {\doibase
  10.1038/ncomms3743} {\bibfield  {journal} {\bibinfo  {journal} {Nature
  Communications}\ }\textbf {\bibinfo {volume} {4}},\ \bibinfo {pages} {2743}
  (\bibinfo {year} {2013})}\BibitemShut {NoStop}%
\bibitem [{\citenamefont {Savas}\ \emph {et~al.}(1995)\citenamefont {Savas},
  \citenamefont {Shah}, \citenamefont {Schattenburg}, \citenamefont {Carter},\
  and\ \citenamefont {Smith}}]{Savas1995}%
  \BibitemOpen
  \bibfield  {author} {\bibinfo {author} {\bibfnamefont {T.~A.}\ \bibnamefont
  {Savas}}, \bibinfo {author} {\bibfnamefont {S.~N.}\ \bibnamefont {Shah}},
  \bibinfo {author} {\bibfnamefont {M.~L.}\ \bibnamefont {Schattenburg}},
  \bibinfo {author} {\bibfnamefont {J.~M.}\ \bibnamefont {Carter}}, \ and\
  \bibinfo {author} {\bibfnamefont {H.~I.}\ \bibnamefont {Smith}},\ }\href
  {\doibase 10.1116/1.588255} {\bibfield  {journal} {\bibinfo  {journal} {J.
  Vac. Sci. Technol.}\ }\textbf {\bibinfo {volume} {B 13}},\ \bibinfo {pages}
  {2732} (\bibinfo {year} {1995})}\BibitemShut {NoStop}%
\bibitem [{\citenamefont {Carnal}\ and\ \citenamefont
  {Mlynek}(1991)}]{Carnal1991}%
  \BibitemOpen
  \bibfield  {author} {\bibinfo {author} {\bibfnamefont {O.}~\bibnamefont
  {Carnal}}\ and\ \bibinfo {author} {\bibfnamefont {J.}~\bibnamefont
  {Mlynek}},\ }\href {\doibase 10.1103/PhysRevLett.66.2689} {\bibfield
  {journal} {\bibinfo  {journal} {Phys. Rev. Lett.}\ }\textbf {\bibinfo
  {volume} {66}},\ \bibinfo {pages} {2689} (\bibinfo {year}
  {1991})}\BibitemShut {NoStop}%
\bibitem [{\citenamefont {Sclafani}\ \emph {et~al.}(2013)\citenamefont
  {Sclafani}, \citenamefont {Juffmann}, \citenamefont {Knobloch},\ and\
  \citenamefont {Arndt}}]{Sclafani2013}%
  \BibitemOpen
  \bibfield  {author} {\bibinfo {author} {\bibfnamefont {M.}~\bibnamefont
  {Sclafani}}, \bibinfo {author} {\bibfnamefont {T.}~\bibnamefont {Juffmann}},
  \bibinfo {author} {\bibfnamefont {C.}~\bibnamefont {Knobloch}}, \ and\
  \bibinfo {author} {\bibfnamefont {M.}~\bibnamefont {Arndt}},\ }\href
  {\doibase 10.1088/1367-2630/15/8/083004} {\bibfield  {journal} {\bibinfo
  {journal} {New J. Phys.}\ }\textbf {\bibinfo {volume} {15}},\ \bibinfo
  {pages} {083004} (\bibinfo {year} {2013})}\BibitemShut {NoStop}%
\bibitem [{\citenamefont {Grisenti}\ \emph {et~al.}(1999)\citenamefont
  {Grisenti}, \citenamefont {Schöllkopf}, \citenamefont {Toennies},
  \citenamefont {Hegerfeldt},\ and\ \citenamefont {Köhler}}]{Grisenti1999}%
  \BibitemOpen
  \bibfield  {author} {\bibinfo {author} {\bibfnamefont {R.~E.}\ \bibnamefont
  {Grisenti}}, \bibinfo {author} {\bibfnamefont {W.}~\bibnamefont
  {Schöllkopf}}, \bibinfo {author} {\bibfnamefont {J.~P.}\ \bibnamefont
  {Toennies}}, \bibinfo {author} {\bibfnamefont {G.~C.}\ \bibnamefont
  {Hegerfeldt}}, \ and\ \bibinfo {author} {\bibfnamefont {T.}~\bibnamefont
  {Köhler}},\ }\href {\doibase 10.1103/PhysRevLett.83.1755} {\bibfield
  {journal} {\bibinfo  {journal} {Phys. Rev. Lett.}\ }\textbf {\bibinfo
  {volume} {83}},\ \bibinfo {pages} {1755} (\bibinfo {year}
  {1999})}\BibitemShut {NoStop}%
\bibitem [{\citenamefont {Nairz}\ \emph {et~al.}(2003)\citenamefont {Nairz},
  \citenamefont {Arndt},\ and\ \citenamefont {Zeilinger}}]{Nairz2003}%
  \BibitemOpen
  \bibfield  {author} {\bibinfo {author} {\bibfnamefont {O.}~\bibnamefont
  {Nairz}}, \bibinfo {author} {\bibfnamefont {M.}~\bibnamefont {Arndt}}, \ and\
  \bibinfo {author} {\bibfnamefont {A.}~\bibnamefont {Zeilinger}},\ }\href
  {\doibase 10.1119/1.1531580} {\bibfield  {journal} {\bibinfo  {journal} {Am.
  J. Phys.}\ }\textbf {\bibinfo {volume} {71}},\ \bibinfo {pages} {319}
  (\bibinfo {year} {2003})}\BibitemShut {NoStop}%
\bibitem [{\citenamefont {Juffmann}\ \emph
  {et~al.}(2012{\natexlab{b}})\citenamefont {Juffmann}, \citenamefont {Milic},
  \citenamefont {Müllneritsch}, \citenamefont {Asenbaum}, \citenamefont
  {Tsukernik}, \citenamefont {Tüxen}, \citenamefont {Mayor}, \citenamefont
  {Cheshnovsky},\ and\ \citenamefont {Arndt}}]{Juffmann2012a}%
  \BibitemOpen
  \bibfield  {author} {\bibinfo {author} {\bibfnamefont {T.}~\bibnamefont
  {Juffmann}}, \bibinfo {author} {\bibfnamefont {A.}~\bibnamefont {Milic}},
  \bibinfo {author} {\bibfnamefont {M.}~\bibnamefont {Müllneritsch}}, \bibinfo
  {author} {\bibfnamefont {P.}~\bibnamefont {Asenbaum}}, \bibinfo {author}
  {\bibfnamefont {A.}~\bibnamefont {Tsukernik}}, \bibinfo {author}
  {\bibfnamefont {J.}~\bibnamefont {Tüxen}}, \bibinfo {author} {\bibfnamefont
  {M.}~\bibnamefont {Mayor}}, \bibinfo {author} {\bibfnamefont
  {O.}~\bibnamefont {Cheshnovsky}}, \ and\ \bibinfo {author} {\bibfnamefont
  {M.}~\bibnamefont {Arndt}},\ }\href {\doibase 10.1038/nnano.2012.34}
  {\bibfield  {journal} {\bibinfo  {journal} {Nature Nanotechn.}\ }\textbf
  {\bibinfo {volume} {7}},\ \bibinfo {pages} {297 } (\bibinfo {year}
  {2012}{\natexlab{b}})}\BibitemShut {NoStop}%
\bibitem [{\citenamefont {Storey}\ \emph {et~al.}(1992)\citenamefont {Storey},
  \citenamefont {Collett},\ and\ \citenamefont {Walls}}]{Storey1992}%
  \BibitemOpen
  \bibfield  {author} {\bibinfo {author} {\bibfnamefont {P.}~\bibnamefont
  {Storey}}, \bibinfo {author} {\bibfnamefont {M.}~\bibnamefont {Collett}}, \
  and\ \bibinfo {author} {\bibfnamefont {D.}~\bibnamefont {Walls}},\ }\href
  {\doibase 10.1103/PhysRevLett.68.472} {\bibfield  {journal} {\bibinfo
  {journal} {Phys. Rev. Lett.}\ }\textbf {\bibinfo {volume} {68}},\ \bibinfo
  {pages} {472} (\bibinfo {year} {1992})}\BibitemShut {NoStop}%
\bibitem [{\citenamefont {Abfalterer}\ \emph {et~al.}(1997)\citenamefont
  {Abfalterer}, \citenamefont {Keller}, \citenamefont {Bernet}, \citenamefont
  {Oberthaler}, \citenamefont {Schmiedmayer},\ and\ \citenamefont
  {Zeilinger}}]{Abfalterer1997}%
  \BibitemOpen
  \bibfield  {author} {\bibinfo {author} {\bibfnamefont {R.}~\bibnamefont
  {Abfalterer}}, \bibinfo {author} {\bibfnamefont {C.}~\bibnamefont {Keller}},
  \bibinfo {author} {\bibfnamefont {S.}~\bibnamefont {Bernet}}, \bibinfo
  {author} {\bibfnamefont {M.~K.}\ \bibnamefont {Oberthaler}}, \bibinfo
  {author} {\bibfnamefont {J.}~\bibnamefont {Schmiedmayer}}, \ and\ \bibinfo
  {author} {\bibfnamefont {A.}~\bibnamefont {Zeilinger}},\ }\href {\doibase
  10.1103/PhysRevA.56.R4365} {\bibfield  {journal} {\bibinfo  {journal} {Phys.
  Rev. A}\ }\textbf {\bibinfo {volume} {56}},\ \bibinfo {pages} {R4365}
  (\bibinfo {year} {1997})}\BibitemShut {NoStop}%
\bibitem [{\citenamefont {Fray}\ \emph {et~al.}(2004)\citenamefont {Fray},
  \citenamefont {Diez}, \citenamefont {Hänsch},\ and\ \citenamefont
  {Weitz}}]{Fray2004}%
  \BibitemOpen
  \bibfield  {author} {\bibinfo {author} {\bibfnamefont {S.}~\bibnamefont
  {Fray}}, \bibinfo {author} {\bibfnamefont {C.}~\bibnamefont {Diez}}, \bibinfo
  {author} {\bibfnamefont {T.}~\bibnamefont {Hänsch}}, \ and\ \bibinfo {author}
  {\bibfnamefont {M.}~\bibnamefont {Weitz}},\ }\href {\doibase
  10.1103/PhysRevLett.93.240404} {\bibfield  {journal} {\bibinfo  {journal}
  {Phys. Rev. Lett.}\ }\textbf {\bibinfo {volume} {93}},\ \bibinfo {pages}
  {240404} (\bibinfo {year} {2004})}\BibitemShut {NoStop}%
\bibitem [{\citenamefont {Reiger}\ \emph {et~al.}(2006)\citenamefont {Reiger},
  \citenamefont {Hackermüller}, \citenamefont {Berninger},\ and\ \citenamefont
  {Arndt}}]{Reiger2006}%
  \BibitemOpen
  \bibfield  {author} {\bibinfo {author} {\bibfnamefont {E.}~\bibnamefont
  {Reiger}}, \bibinfo {author} {\bibfnamefont {L.}~\bibnamefont
  {Hackermüller}}, \bibinfo {author} {\bibfnamefont {M.}~\bibnamefont
  {Berninger}}, \ and\ \bibinfo {author} {\bibfnamefont {M.}~\bibnamefont
  {Arndt}},\ }\href {\doibase 10.1016/j.optcom.2006.02.060} {\bibfield
  {journal} {\bibinfo  {journal} {Opt. Comm.}\ }\textbf {\bibinfo {volume}
  {264}},\ \bibinfo {pages} {326} (\bibinfo {year} {2006})}\BibitemShut
  {NoStop}%
\bibitem [{\citenamefont {Haslinger}\ \emph {et~al.}(2013)\citenamefont
  {Haslinger}, \citenamefont {Dörre}, \citenamefont {Geyer}, \citenamefont
  {Rodewald}, \citenamefont {Nimmrichter},\ and\ \citenamefont
  {Arndt}}]{Haslinger2013}%
  \BibitemOpen
  \bibfield  {author} {\bibinfo {author} {\bibfnamefont {P.}~\bibnamefont
  {Haslinger}}, \bibinfo {author} {\bibfnamefont {N.}~\bibnamefont {Dörre}},
  \bibinfo {author} {\bibfnamefont {P.}~\bibnamefont {Geyer}}, \bibinfo
  {author} {\bibfnamefont {J.}~\bibnamefont {Rodewald}}, \bibinfo {author}
  {\bibfnamefont {S.}~\bibnamefont {Nimmrichter}}, \ and\ \bibinfo {author}
  {\bibfnamefont {M.}~\bibnamefont {Arndt}},\ }\href {\doibase
  10.1038/nphys2542} {\bibfield  {journal} {\bibinfo  {journal} {Nature
  Physics}\ }\textbf {\bibinfo {volume} {9}},\ \bibinfo {pages} {144} (\bibinfo
  {year} {2013})}\BibitemShut {NoStop}%
\bibitem [{\citenamefont {Kapitza}\ and\ \citenamefont
  {Dirac}(1933)}]{Kapitza1933}%
  \BibitemOpen
  \bibfield  {author} {\bibinfo {author} {\bibfnamefont {P.~L.}\ \bibnamefont
  {Kapitza}}\ and\ \bibinfo {author} {\bibfnamefont {P.~A.~M.}\ \bibnamefont
  {Dirac}},\ }\href {\doibase 10.1017/S0305004100011105} {\bibfield  {journal}
  {\bibinfo  {journal} {Proc. Camb. Philos. Soc.}\ }\textbf {\bibinfo {volume}
  {29}},\ \bibinfo {pages} {297 } (\bibinfo {year} {1933})}\BibitemShut
  {NoStop}%
\bibitem [{\citenamefont {Moskowitz}\ \emph {et~al.}(1983)\citenamefont
  {Moskowitz}, \citenamefont {Gould}, \citenamefont {Atlas},\ and\
  \citenamefont {Pritchard}}]{Moskowitz1983}%
  \BibitemOpen
  \bibfield  {author} {\bibinfo {author} {\bibfnamefont {P.~E.}\ \bibnamefont
  {Moskowitz}}, \bibinfo {author} {\bibfnamefont {P.~L.}\ \bibnamefont
  {Gould}}, \bibinfo {author} {\bibfnamefont {S.~R.}\ \bibnamefont {Atlas}}, \
  and\ \bibinfo {author} {\bibfnamefont {D.~E.}\ \bibnamefont {Pritchard}},\
  }\href {\doibase 10.1103/PhysRevLett.51.370} {\bibfield  {journal} {\bibinfo
  {journal} {Phys. Rev. Lett.}\ }\textbf {\bibinfo {volume} {51}},\ \bibinfo
  {pages} {370 } (\bibinfo {year} {1983})}\BibitemShut {NoStop}%
\bibitem [{\citenamefont {Gould}\ \emph {et~al.}(1986)\citenamefont {Gould},
  \citenamefont {Ruff},\ and\ \citenamefont {Pritchard}}]{Gould1986}%
  \BibitemOpen
  \bibfield  {author} {\bibinfo {author} {\bibfnamefont {P.~L.}\ \bibnamefont
  {Gould}}, \bibinfo {author} {\bibfnamefont {G.~A.}\ \bibnamefont {Ruff}}, \
  and\ \bibinfo {author} {\bibfnamefont {D.~E.}\ \bibnamefont {Pritchard}},\
  }\href {\doibase 10.1103/PhysRevLett.56.827} {\bibfield  {journal} {\bibinfo
  {journal} {Phys. Rev. Lett.}\ }\textbf {\bibinfo {volume} {56}},\ \bibinfo
  {pages} {827 } (\bibinfo {year} {1986})}\BibitemShut {NoStop}%
\bibitem [{\citenamefont {Martin}\ \emph {et~al.}(1988)\citenamefont {Martin},
  \citenamefont {Oldaker}, \citenamefont {Miklich},\ and\ \citenamefont
  {Pritchard}}]{Martin1988}%
  \BibitemOpen
  \bibfield  {author} {\bibinfo {author} {\bibfnamefont {P.~J.}\ \bibnamefont
  {Martin}}, \bibinfo {author} {\bibfnamefont {B.~G.}\ \bibnamefont {Oldaker}},
  \bibinfo {author} {\bibfnamefont {A.~H.}\ \bibnamefont {Miklich}}, \ and\
  \bibinfo {author} {\bibfnamefont {D.~E.}\ \bibnamefont {Pritchard}},\ }\href
  {\doibase 10.1103/PhysRevLett.60.515} {\bibfield  {journal} {\bibinfo
  {journal} {Phys. Rev. Lett.}\ }\textbf {\bibinfo {volume} {60}},\ \bibinfo
  {pages} {515} (\bibinfo {year} {1988})}\BibitemShut {NoStop}%
\bibitem [{\citenamefont {Nairz}\ \emph {et~al.}(2001)\citenamefont {Nairz},
  \citenamefont {Brezger}, \citenamefont {Arndt},\ and\ \citenamefont
  {Zeilinger}}]{Nairz2001}%
  \BibitemOpen
  \bibfield  {author} {\bibinfo {author} {\bibfnamefont {O.}~\bibnamefont
  {Nairz}}, \bibinfo {author} {\bibfnamefont {B.}~\bibnamefont {Brezger}},
  \bibinfo {author} {\bibfnamefont {M.}~\bibnamefont {Arndt}}, \ and\ \bibinfo
  {author} {\bibfnamefont {A.}~\bibnamefont {Zeilinger}},\ }\href {\doibase
  10.1103/PhysRevLett.87.160401} {\bibfield  {journal} {\bibinfo  {journal}
  {Phys. Rev. Lett.}\ }\textbf {\bibinfo {volume} {87}},\ \bibinfo {pages}
  {160401} (\bibinfo {year} {2001})}\BibitemShut {NoStop}%
\bibitem [{\citenamefont {Dresselhaus}\ \emph {et~al.}(1998)\citenamefont
  {Dresselhaus}, \citenamefont {Dresselhaus},\ and\ \citenamefont
  {Eklund}}]{Dresselhaus1998}%
  \BibitemOpen
  \bibfield  {author} {\bibinfo {author} {\bibfnamefont {M.~S.}\ \bibnamefont
  {Dresselhaus}}, \bibinfo {author} {\bibfnamefont {G.}~\bibnamefont
  {Dresselhaus}}, \ and\ \bibinfo {author} {\bibfnamefont {P.~C.}\ \bibnamefont
  {Eklund}},\ }\href@noop {} {\emph {\bibinfo {title} {Science of Fullerenes
  and Carbon Nanotubes}}},\ \bibinfo {edition} {2nd}\ ed.\ (\bibinfo
  {publisher} {Acad. Press},\ \bibinfo {address} {San Diego},\ \bibinfo {year}
  {1998})\BibitemShut {NoStop}%
\bibitem [{\citenamefont {Moshinsky}(1952)}]{Moshinski1952}%
  \BibitemOpen
  \bibfield  {author} {\bibinfo {author} {\bibfnamefont {M.}~\bibnamefont
  {Moshinsky}},\ }\href {\doibase 10.1103/PhysRev.88.625} {\bibfield  {journal}
  {\bibinfo  {journal} {Phys. Rev.}\ }\textbf {\bibinfo {volume} {88}},\
  \bibinfo {pages} {625} (\bibinfo {year} {1952})}\BibitemShut {NoStop}%
\bibitem [{\citenamefont {Hils}\ \emph {et~al.}(1998)\citenamefont {Hils},
  \citenamefont {Felber}, \citenamefont {Gahler}, \citenamefont {Glaser},
  \citenamefont {Golub}, \citenamefont {Habicht},\ and\ \citenamefont
  {Wille}}]{Hils1998}%
  \BibitemOpen
  \bibfield  {author} {\bibinfo {author} {\bibfnamefont {T.}~\bibnamefont
  {Hils}}, \bibinfo {author} {\bibfnamefont {J.}~\bibnamefont {Felber}},
  \bibinfo {author} {\bibfnamefont {R.}~\bibnamefont {Gahler}}, \bibinfo
  {author} {\bibfnamefont {W.}~\bibnamefont {Glaser}}, \bibinfo {author}
  {\bibfnamefont {R.}~\bibnamefont {Golub}}, \bibinfo {author} {\bibfnamefont
  {K.}~\bibnamefont {Habicht}}, \ and\ \bibinfo {author} {\bibfnamefont
  {P.}~\bibnamefont {Wille}},\ }\href {\doibase 10.1103/PhysRevA.58.4784}
  {\bibfield  {journal} {\bibinfo  {journal} {Phys. Rev. A}\ }\textbf {\bibinfo
  {volume} {58}},\ \bibinfo {pages} {4784} (\bibinfo {year}
  {1998})}\BibitemShut {NoStop}%
\bibitem [{\citenamefont {Steane}\ \emph {et~al.}(1995)\citenamefont {Steane},
  \citenamefont {Szriftgiser}, \citenamefont {Desbiolles},\ and\ \citenamefont
  {Dalibard}}]{Steane1995}%
  \BibitemOpen
  \bibfield  {author} {\bibinfo {author} {\bibfnamefont {A.}~\bibnamefont
  {Steane}}, \bibinfo {author} {\bibfnamefont {P.}~\bibnamefont {Szriftgiser}},
  \bibinfo {author} {\bibfnamefont {P.}~\bibnamefont {Desbiolles}}, \ and\
  \bibinfo {author} {\bibfnamefont {J.}~\bibnamefont {Dalibard}},\ }\href
  {\doibase 10.1103/PhysRevLett.74.4972} {\bibfield  {journal} {\bibinfo
  {journal} {Phys. Rev. Lett.}\ }\textbf {\bibinfo {volume} {74}},\ \bibinfo
  {pages} {4972 } (\bibinfo {year} {1995})}\BibitemShut {NoStop}%
\bibitem [{\citenamefont {Szriftgiser}\ \emph {et~al.}(1996)\citenamefont
  {Szriftgiser}, \citenamefont {Guéry-Odelin}, \citenamefont {Arndt},\ and\
  \citenamefont {Dalibard}}]{Szriftgiser1996}%
  \BibitemOpen
  \bibfield  {author} {\bibinfo {author} {\bibfnamefont {P.}~\bibnamefont
  {Szriftgiser}}, \bibinfo {author} {\bibfnamefont {D.}~\bibnamefont
  {Guéry-Odelin}}, \bibinfo {author} {\bibfnamefont {M.}~\bibnamefont {Arndt}},
  \ and\ \bibinfo {author} {\bibfnamefont {J.}~\bibnamefont {Dalibard}},\
  }\href {\doibase 10.1103/PhysRevLett.77.4} {\bibfield  {journal} {\bibinfo
  {journal} {Phys. Rev. Lett.}\ }\textbf {\bibinfo {volume} {77}},\ \bibinfo
  {pages} {4} (\bibinfo {year} {1996})}\BibitemShut {NoStop}%
\bibitem [{\citenamefont {Kasevich}\ and\ \citenamefont
  {Chu}(1991)}]{Kasevich1991a}%
  \BibitemOpen
  \bibfield  {author} {\bibinfo {author} {\bibfnamefont {M.}~\bibnamefont
  {Kasevich}}\ and\ \bibinfo {author} {\bibfnamefont {S.}~\bibnamefont {Chu}},\
  }\href {\doibase 10.1103/PhysRevLett.67.181} {\bibfield  {journal} {\bibinfo
  {journal} {Phys. Rev. Lett.}\ }\textbf {\bibinfo {volume} {67}},\ \bibinfo
  {pages} {181} (\bibinfo {year} {1991})}\BibitemShut {NoStop}%
\bibitem [{\citenamefont {Cahn}\ \emph {et~al.}(1997)\citenamefont {Cahn},
  \citenamefont {Kumarakrishnan}, \citenamefont {Shim}, \citenamefont
  {Sleator}, \citenamefont {Berman},\ and\ \citenamefont
  {Dubetsky}}]{Cahn1997}%
  \BibitemOpen
  \bibfield  {author} {\bibinfo {author} {\bibfnamefont {S.~B.}\ \bibnamefont
  {Cahn}}, \bibinfo {author} {\bibfnamefont {A.}~\bibnamefont
  {Kumarakrishnan}}, \bibinfo {author} {\bibfnamefont {U.}~\bibnamefont
  {Shim}}, \bibinfo {author} {\bibfnamefont {T.}~\bibnamefont {Sleator}},
  \bibinfo {author} {\bibfnamefont {P.~R.}\ \bibnamefont {Berman}}, \ and\
  \bibinfo {author} {\bibfnamefont {B.}~\bibnamefont {Dubetsky}},\ }\href
  {\doibase 10.1103/PhysRevLett.79.784} {\bibfield  {journal} {\bibinfo
  {journal} {Phys. Rev. Lett.}\ }\textbf {\bibinfo {volume} {79}},\ \bibinfo
  {pages} {784} (\bibinfo {year} {1997})}\BibitemShut {NoStop}%
\bibitem [{\citenamefont {Turlapov}\ \emph {et~al.}(2005)\citenamefont
  {Turlapov}, \citenamefont {Tonyushkin},\ and\ \citenamefont
  {Sleator}}]{Turlapov2005}%
  \BibitemOpen
  \bibfield  {author} {\bibinfo {author} {\bibfnamefont {A.}~\bibnamefont
  {Turlapov}}, \bibinfo {author} {\bibfnamefont {A.}~\bibnamefont
  {Tonyushkin}}, \ and\ \bibinfo {author} {\bibfnamefont {T.}~\bibnamefont
  {Sleator}},\ }\href {\doibase 10.1103/PhysRevA.71.043612} {\bibfield
  {journal} {\bibinfo  {journal} {Phys. Rev. A}\ }\textbf {\bibinfo {volume}
  {71}},\ \bibinfo {pages} {43612} (\bibinfo {year} {2005})}\BibitemShut
  {NoStop}%
\bibitem [{\citenamefont {Deng}\ \emph {et~al.}(1999)\citenamefont {Deng},
  \citenamefont {Hagley}, \citenamefont {Denschlag}, \citenamefont {Simsarian},
  \citenamefont {Edwards}, \citenamefont {Clark}, \citenamefont {Helmerson},
  \citenamefont {Rolston},\ and\ \citenamefont {Phillips}}]{Deng1999}%
  \BibitemOpen
  \bibfield  {author} {\bibinfo {author} {\bibfnamefont {L.}~\bibnamefont
  {Deng}}, \bibinfo {author} {\bibfnamefont {E.~W.}\ \bibnamefont {Hagley}},
  \bibinfo {author} {\bibfnamefont {J.}~\bibnamefont {Denschlag}}, \bibinfo
  {author} {\bibfnamefont {J.}~\bibnamefont {Simsarian}}, \bibinfo {author}
  {\bibfnamefont {M.}~\bibnamefont {Edwards}}, \bibinfo {author} {\bibfnamefont
  {C.}~\bibnamefont {Clark}}, \bibinfo {author} {\bibfnamefont
  {K.}~\bibnamefont {Helmerson}}, \bibinfo {author} {\bibfnamefont
  {S.}~\bibnamefont {Rolston}}, \ and\ \bibinfo {author} {\bibfnamefont
  {W.}~\bibnamefont {Phillips}},\ }\href {\doibase 10.1103/PhysRevLett.83.5407}
  {\bibfield  {journal} {\bibinfo  {journal} {Phys. Rev. Lett.}\ }\textbf
  {\bibinfo {volume} {83}},\ \bibinfo {pages} {5407} (\bibinfo {year}
  {1999})}\BibitemShut {NoStop}%
\bibitem [{\citenamefont {Patorski}(1989)}]{Patorski1989}%
  \BibitemOpen
  \bibfield  {author} {\bibinfo {author} {\bibfnamefont {K.}~\bibnamefont
  {Patorski}},\ }\enquote {\bibinfo {title} {Self-imaging and its
  applications},}\ in\ \href@noop {} {\emph {\bibinfo {booktitle} {Progress in
  Optics {XXVII}}}},\ \bibinfo {editor} {edited by\ \bibinfo {editor}
  {\bibfnamefont {E.}~\bibnamefont {Wolf}}}\ (\bibinfo  {publisher}
  {Elsevier},\ \bibinfo {address} {Amsterdam},\ \bibinfo {year} {1989})\ pp.\
  \bibinfo {pages} {2--108}\BibitemShut {NoStop}%
\bibitem [{\citenamefont {Pfeiffer}\ \emph {et~al.}(2006)\citenamefont
  {Pfeiffer}, \citenamefont {Weitkamp}, \citenamefont {Bunk},\ and\
  \citenamefont {David}}]{Pfeiffer2006}%
  \BibitemOpen
  \bibfield  {author} {\bibinfo {author} {\bibfnamefont {F.}~\bibnamefont
  {Pfeiffer}}, \bibinfo {author} {\bibfnamefont {T.}~\bibnamefont {Weitkamp}},
  \bibinfo {author} {\bibfnamefont {O.}~\bibnamefont {Bunk}}, \ and\ \bibinfo
  {author} {\bibfnamefont {C.}~\bibnamefont {David}},\ }\href {\doibase
  10.1038/nphys265} {\bibfield  {journal} {\bibinfo  {journal} {Nature
  Physics}\ }\textbf {\bibinfo {volume} {2}},\ \bibinfo {pages} {258} (\bibinfo
  {year} {2006})}\BibitemShut {NoStop}%
\bibitem [{\citenamefont {Clauser}\ and\ \citenamefont
  {Li}(1994)}]{Clauser1994}%
  \BibitemOpen
  \bibfield  {author} {\bibinfo {author} {\bibfnamefont {J.~F.}\ \bibnamefont
  {Clauser}}\ and\ \bibinfo {author} {\bibfnamefont {S.}~\bibnamefont {Li}},\
  }\href {\doibase 10.1103/PhysRevA.50.2430} {\bibfield  {journal} {\bibinfo
  {journal} {Phys. Rev. A}\ }\textbf {\bibinfo {volume} {50}},\ \bibinfo
  {pages} {2430} (\bibinfo {year} {1994})}\BibitemShut {NoStop}%
\bibitem [{\citenamefont {Gerlich}\ \emph {et~al.}(2007)\citenamefont
  {Gerlich}, \citenamefont {Hackermüller}, \citenamefont {Hornberger},
  \citenamefont {Stibor}, \citenamefont {Ulbricht}, \citenamefont {Gring},
  \citenamefont {Goldfarb}, \citenamefont {Savas}, \citenamefont {Müri},
  \citenamefont {Mayor},\ and\ \citenamefont {Arndt}}]{Gerlich2007}%
  \BibitemOpen
  \bibfield  {author} {\bibinfo {author} {\bibfnamefont {S.}~\bibnamefont
  {Gerlich}}, \bibinfo {author} {\bibfnamefont {L.}~\bibnamefont
  {Hackermüller}}, \bibinfo {author} {\bibfnamefont {K.}~\bibnamefont
  {Hornberger}}, \bibinfo {author} {\bibfnamefont {A.}~\bibnamefont {Stibor}},
  \bibinfo {author} {\bibfnamefont {H.}~\bibnamefont {Ulbricht}}, \bibinfo
  {author} {\bibfnamefont {M.}~\bibnamefont {Gring}}, \bibinfo {author}
  {\bibfnamefont {F.}~\bibnamefont {Goldfarb}}, \bibinfo {author}
  {\bibfnamefont {T.}~\bibnamefont {Savas}}, \bibinfo {author} {\bibfnamefont
  {M.}~\bibnamefont {Müri}}, \bibinfo {author} {\bibfnamefont {M.}~\bibnamefont
  {Mayor}}, \ and\ \bibinfo {author} {\bibfnamefont {M.}~\bibnamefont
  {Arndt}},\ }\href {\doibase 10.1038/Nphys701} {\bibfield  {journal} {\bibinfo
   {journal} {Nature Physics}\ }\textbf {\bibinfo {volume} {3}},\ \bibinfo
  {pages} {711} (\bibinfo {year} {2007})}\BibitemShut {NoStop}%
\bibitem [{\citenamefont {Talbot}(1836)}]{Talbot1836}%
  \BibitemOpen
  \bibfield  {author} {\bibinfo {author} {\bibfnamefont {H.~F.}\ \bibnamefont
  {Talbot}},\ }\href {\doibase 10.1080/14786443608649032} {\bibfield  {journal}
  {\bibinfo  {journal} {Philos. Mag.}\ }\textbf {\bibinfo {volume} {9}},\
  \bibinfo {pages} {401} (\bibinfo {year} {1836})}\BibitemShut {NoStop}%
\bibitem [{\citenamefont {Nowak}\ \emph {et~al.}(1997)\citenamefont {Nowak},
  \citenamefont {Kurtsiefer}, \citenamefont {Pfau},\ and\ \citenamefont
  {David}}]{Nowak1997}%
  \BibitemOpen
  \bibfield  {author} {\bibinfo {author} {\bibfnamefont {S.}~\bibnamefont
  {Nowak}}, \bibinfo {author} {\bibfnamefont {C.}~\bibnamefont {Kurtsiefer}},
  \bibinfo {author} {\bibfnamefont {T.}~\bibnamefont {Pfau}}, \ and\ \bibinfo
  {author} {\bibfnamefont {C.}~\bibnamefont {David}},\ }\href {\doibase
  10.1364/OL.22.001430} {\bibfield  {journal} {\bibinfo  {journal} {Opt.
  Lett.}\ }\textbf {\bibinfo {volume} {22}},\ \bibinfo {pages} {1430} (\bibinfo
  {year} {1997})}\BibitemShut {NoStop}%
\bibitem [{\citenamefont {Clauser}(1997)}]{Clauser1997}%
  \BibitemOpen
  \bibfield  {author} {\bibinfo {author} {\bibfnamefont {J.}~\bibnamefont
  {Clauser}},\ }\enquote {\bibinfo {title} {Experimental metaphysics},}\ \
  (\bibinfo  {publisher} {Kluwer Academic},\ \bibinfo {year} {1997})\ \bibinfo
  {type} {Book section}\ \bibinfo {chapter} {De Broglie-wave interference of
  small rocks and live viruses}, pp.\ \bibinfo {pages} {1--11}\BibitemShut
  {NoStop}%
\bibitem [{\citenamefont {Hornberger}\ \emph {et~al.}(2012)\citenamefont
  {Hornberger}, \citenamefont {Gerlich}, \citenamefont {Haslinger},
  \citenamefont {Nimmrichter},\ and\ \citenamefont {Arndt}}]{Hornberger2012}%
  \BibitemOpen
  \bibfield  {author} {\bibinfo {author} {\bibfnamefont {K.}~\bibnamefont
  {Hornberger}}, \bibinfo {author} {\bibfnamefont {S.}~\bibnamefont {Gerlich}},
  \bibinfo {author} {\bibfnamefont {P.}~\bibnamefont {Haslinger}}, \bibinfo
  {author} {\bibfnamefont {S.}~\bibnamefont {Nimmrichter}}, \ and\ \bibinfo
  {author} {\bibfnamefont {M.}~\bibnamefont {Arndt}},\ }\href {\doibase
  10.1103/RevModPhys.84.157} {\bibfield  {journal} {\bibinfo  {journal} {Rev.
  Mod. Phys.}\ }\textbf {\bibinfo {volume} {84}},\ \bibinfo {pages} {157}
  (\bibinfo {year} {2012})}\BibitemShut {NoStop}%
\bibitem [{\citenamefont {Hackermüller}\ \emph
  {et~al.}(2003{\natexlab{a}})\citenamefont {Hackermüller}, \citenamefont
  {Uttenthaler}, \citenamefont {Hornberger}, \citenamefont {Reiger},
  \citenamefont {Brezger}, \citenamefont {Zeilinger},\ and\ \citenamefont
  {Arndt}}]{Hackermueller2003}%
  \BibitemOpen
  \bibfield  {author} {\bibinfo {author} {\bibfnamefont {L.}~\bibnamefont
  {Hackermüller}}, \bibinfo {author} {\bibfnamefont {S.}~\bibnamefont
  {Uttenthaler}}, \bibinfo {author} {\bibfnamefont {K.}~\bibnamefont
  {Hornberger}}, \bibinfo {author} {\bibfnamefont {E.}~\bibnamefont {Reiger}},
  \bibinfo {author} {\bibfnamefont {B.}~\bibnamefont {Brezger}}, \bibinfo
  {author} {\bibfnamefont {A.}~\bibnamefont {Zeilinger}}, \ and\ \bibinfo
  {author} {\bibfnamefont {M.}~\bibnamefont {Arndt}},\ }\href {\doibase
  10.1103/PhysRevLett.91.090408} {\bibfield  {journal} {\bibinfo  {journal}
  {Phys. Rev. Lett.}\ }\textbf {\bibinfo {volume} {91}},\ \bibinfo {pages}
  {90408} (\bibinfo {year} {2003}{\natexlab{a}})}\BibitemShut {NoStop}%
\bibitem [{\citenamefont {Berninger}\ \emph {et~al.}(2007)\citenamefont
  {Berninger}, \citenamefont {Stéfanov}, \citenamefont {Deachapunya},\ and\
  \citenamefont {Arndt}}]{Berninger2007}%
  \BibitemOpen
  \bibfield  {author} {\bibinfo {author} {\bibfnamefont {M.}~\bibnamefont
  {Berninger}}, \bibinfo {author} {\bibfnamefont {A.}~\bibnamefont {Stéfanov}},
  \bibinfo {author} {\bibfnamefont {S.}~\bibnamefont {Deachapunya}}, \ and\
  \bibinfo {author} {\bibfnamefont {M.}~\bibnamefont {Arndt}},\ }\href
  {\doibase 10.1103/PhysRevA.76.013607} {\bibfield  {journal} {\bibinfo
  {journal} {Phys. Rev. A}\ }\textbf {\bibinfo {volume} {76}},\ \bibinfo
  {pages} {013607} (\bibinfo {year} {2007})}\BibitemShut {NoStop}%
\bibitem [{\citenamefont {Brezger}\ \emph {et~al.}(2003)\citenamefont
  {Brezger}, \citenamefont {Arndt},\ and\ \citenamefont
  {Zeilinger}}]{Brezger2003}%
  \BibitemOpen
  \bibfield  {author} {\bibinfo {author} {\bibfnamefont {B.}~\bibnamefont
  {Brezger}}, \bibinfo {author} {\bibfnamefont {M.}~\bibnamefont {Arndt}}, \
  and\ \bibinfo {author} {\bibfnamefont {A.}~\bibnamefont {Zeilinger}},\ }\href
  {\doibase 10.1088/1464-4266/5/2/362} {\bibfield  {journal} {\bibinfo
  {journal} {J. Opt. B}\ }\textbf {\bibinfo {volume} {5}},\ \bibinfo {pages}
  {S82} (\bibinfo {year} {2003})}\BibitemShut {NoStop}%
\bibitem [{\citenamefont {Oberthaler}\ \emph {et~al.}(1996)\citenamefont
  {Oberthaler}, \citenamefont {Bernet}, \citenamefont {Rasel}, \citenamefont
  {Schmiedmayer},\ and\ \citenamefont {Zeilinger}}]{Oberthaler1996}%
  \BibitemOpen
  \bibfield  {author} {\bibinfo {author} {\bibfnamefont {M.~K.}\ \bibnamefont
  {Oberthaler}}, \bibinfo {author} {\bibfnamefont {S.}~\bibnamefont {Bernet}},
  \bibinfo {author} {\bibfnamefont {E.~M.}\ \bibnamefont {Rasel}}, \bibinfo
  {author} {\bibfnamefont {J.}~\bibnamefont {Schmiedmayer}}, \ and\ \bibinfo
  {author} {\bibfnamefont {A.}~\bibnamefont {Zeilinger}},\ }\href {\doibase
  10.1103/PhysRevA.54.3165} {\bibfield  {journal} {\bibinfo  {journal} {Phys.
  Rev.~A}\ }\textbf {\bibinfo {volume} {54}},\ \bibinfo {pages} {3165}
  (\bibinfo {year} {1996})}\BibitemShut {NoStop}%
\bibitem [{\citenamefont {Berman}(1997)}]{Berman1997}%
  \BibitemOpen
  \bibfield  {author} {\bibinfo {author} {\bibfnamefont {P.~R.}\ \bibnamefont
  {Berman}},\ }\href@noop {} {\emph {\bibinfo {title} {Atom Interferometry}}}\
  (\bibinfo  {publisher} {Acad. Press},\ \bibinfo {address} {New York},\
  \bibinfo {year} {1997})\BibitemShut {NoStop}%
\bibitem [{\citenamefont {Hornberger}\ \emph {et~al.}(2009)\citenamefont
  {Hornberger}, \citenamefont {Gerlich}, \citenamefont {Ulbricht},
  \citenamefont {Hackerm\"uller}, \citenamefont {Nimmrichter}, \citenamefont
  {Goldt}, \citenamefont {Boltalina},\ and\ \citenamefont
  {Arndt}}]{Hornberger2009}%
  \BibitemOpen
  \bibfield  {author} {\bibinfo {author} {\bibfnamefont {K.}~\bibnamefont
  {Hornberger}}, \bibinfo {author} {\bibfnamefont {S.}~\bibnamefont {Gerlich}},
  \bibinfo {author} {\bibfnamefont {H.}~\bibnamefont {Ulbricht}}, \bibinfo
  {author} {\bibfnamefont {L.}~\bibnamefont {Hackerm\"uller}}, \bibinfo
  {author} {\bibfnamefont {S.}~\bibnamefont {Nimmrichter}}, \bibinfo {author}
  {\bibfnamefont {I.}~\bibnamefont {Goldt}}, \bibinfo {author} {\bibfnamefont
  {O.}~\bibnamefont {Boltalina}}, \ and\ \bibinfo {author} {\bibfnamefont
  {M.}~\bibnamefont {Arndt}},\ }\href {\doibase 10.1088/1367-2630/11/4/043032}
  {\bibfield  {journal} {\bibinfo  {journal} {New J. Phys.}\ }\textbf {\bibinfo
  {volume} {11}},\ \bibinfo {pages} {043032} (\bibinfo {year}
  {2009})}\BibitemShut {NoStop}%
\bibitem [{\citenamefont {Schleich}(2001)}]{Schleich2001}%
  \BibitemOpen
  \bibfield  {author} {\bibinfo {author} {\bibfnamefont {W.~P.}\ \bibnamefont
  {Schleich}},\ }\href@noop {} {\emph {\bibinfo {title} {Quantum Optics in
  Phase Space}}}\ (\bibinfo  {publisher} {Wiley-VCH Verlag},\ \bibinfo
  {address} {Weinheim},\ \bibinfo {year} {2001})\BibitemShut {NoStop}%
\bibitem [{\citenamefont {Nimmrichter}\ and\ \citenamefont
  {Hornberger}(2008)}]{Nimmrichter2008a}%
  \BibitemOpen
  \bibfield  {author} {\bibinfo {author} {\bibfnamefont {S.}~\bibnamefont
  {Nimmrichter}}\ and\ \bibinfo {author} {\bibfnamefont {K.}~\bibnamefont
  {Hornberger}},\ }\href {\doibase 10.1103/PhysRevA.78.023612} {\bibfield
  {journal} {\bibinfo  {journal} {Phys. Rev. A}\ }\textbf {\bibinfo {volume}
  {78}},\ \bibinfo {pages} {023612} (\bibinfo {year} {2008})}\BibitemShut
  {NoStop}%
\bibitem [{\citenamefont {Hornberger}\ \emph {et~al.}(2004)\citenamefont
  {Hornberger}, \citenamefont {Sipe},\ and\ \citenamefont
  {Arndt}}]{Hornberger2004}%
  \BibitemOpen
  \bibfield  {author} {\bibinfo {author} {\bibfnamefont {K.}~\bibnamefont
  {Hornberger}}, \bibinfo {author} {\bibfnamefont {J.~E.}\ \bibnamefont
  {Sipe}}, \ and\ \bibinfo {author} {\bibfnamefont {M.}~\bibnamefont {Arndt}},\
  }\href {\doibase 10.1103/PhysRevA.70.053608} {\bibfield  {journal} {\bibinfo
  {journal} {Phys. Rev. A}\ }\textbf {\bibinfo {volume} {70}},\ \bibinfo
  {pages} {53608} (\bibinfo {year} {2004})}\BibitemShut {NoStop}%
\bibitem [{\citenamefont {Glauber}(1959)}]{Glauber1959}%
  \BibitemOpen
  \bibfield  {author} {\bibinfo {author} {\bibfnamefont {R.~J.}\ \bibnamefont
  {Glauber}},\ }\href@noop {} {\emph {\bibinfo {title} {High-energy collision
  theory}}},\ Lectures in Theoretical Physics, Vol. 1\ (\bibinfo  {publisher}
  {Wiley-Interscience},\ \bibinfo {address} {New York},\ \bibinfo {year}
  {1959})\BibitemShut {NoStop}%
\bibitem [{\citenamefont {Abramowitz}\ and\ \citenamefont
  {Stegun}(1965)}]{Abramowitz1965}%
  \BibitemOpen
  \bibfield  {author} {\bibinfo {author} {\bibfnamefont {M.}~\bibnamefont
  {Abramowitz}}\ and\ \bibinfo {author} {\bibfnamefont {I.}~\bibnamefont
  {Stegun}},\ }\href@noop {} {\emph {\bibinfo {title} {Handbook of Mathematical
  Functions}}}\ (\bibinfo  {publisher} {Dover Publications},\ \bibinfo
  {address} {New York},\ \bibinfo {year} {1965})\BibitemShut {NoStop}%
\bibitem [{\citenamefont {Nimmrichter}\ \emph
  {et~al.}(2011{\natexlab{b}})\citenamefont {Nimmrichter}, \citenamefont
  {Haslinger}, \citenamefont {Hornberger},\ and\ \citenamefont
  {Arndt}}]{Nimmrichter2011a}%
  \BibitemOpen
  \bibfield  {author} {\bibinfo {author} {\bibfnamefont {S.}~\bibnamefont
  {Nimmrichter}}, \bibinfo {author} {\bibfnamefont {P.}~\bibnamefont
  {Haslinger}}, \bibinfo {author} {\bibfnamefont {K.}~\bibnamefont
  {Hornberger}}, \ and\ \bibinfo {author} {\bibfnamefont {M.}~\bibnamefont
  {Arndt}},\ }\href {\doibase 10.1088/1367-2630/13/7/075002} {\bibfield
  {journal} {\bibinfo  {journal} {New. J. Phys.}\ }\textbf {\bibinfo {volume}
  {13}},\ \bibinfo {pages} {075002} (\bibinfo {year}
  {2011}{\natexlab{b}})}\BibitemShut {NoStop}%
\bibitem [{\citenamefont {Kreibig}\ and\ \citenamefont
  {Vollmer}(1995)}]{Kreibig1995}%
  \BibitemOpen
  \bibfield  {author} {\bibinfo {author} {\bibfnamefont {U.}~\bibnamefont
  {Kreibig}}\ and\ \bibinfo {author} {\bibfnamefont {M.}~\bibnamefont
  {Vollmer}},\ }\href@noop {} {\emph {\bibinfo {title} {Optical Properties of
  Metal Clusters}}}\ (\bibinfo  {publisher} {Springer},\ \bibinfo {address}
  {Berlin},\ \bibinfo {year} {1995})\BibitemShut {NoStop}%
\bibitem [{\citenamefont {Vacchini}(2007)}]{Vacchini2007b}%
  \BibitemOpen
  \bibfield  {author} {\bibinfo {author} {\bibfnamefont {B.}~\bibnamefont
  {Vacchini}},\ }\href {\doibase 10.1088/1751-8113/40/10/015} {\bibfield
  {journal} {\bibinfo  {journal} {Journal of Physics A: Mathematical and
  Theoretical}\ }\textbf {\bibinfo {volume} {40}},\ \bibinfo {pages} {2463}
  (\bibinfo {year} {2007})}\BibitemShut {NoStop}%
\bibitem [{\citenamefont {Hackermüller}\ \emph
  {et~al.}(2003{\natexlab{b}})\citenamefont {Hackermüller}, \citenamefont
  {Hornberger}, \citenamefont {Brezger}, \citenamefont {Zeilinger},\ and\
  \citenamefont {Arndt}}]{Hackermueller2003a}%
  \BibitemOpen
  \bibfield  {author} {\bibinfo {author} {\bibfnamefont {L.}~\bibnamefont
  {Hackermüller}}, \bibinfo {author} {\bibfnamefont {K.}~\bibnamefont
  {Hornberger}}, \bibinfo {author} {\bibfnamefont {B.}~\bibnamefont {Brezger}},
  \bibinfo {author} {\bibfnamefont {A.}~\bibnamefont {Zeilinger}}, \ and\
  \bibinfo {author} {\bibfnamefont {M.}~\bibnamefont {Arndt}},\ }\href
  {\doibase 10.1007/s00340-003-1312-6} {\bibfield  {journal} {\bibinfo
  {journal} {Appl. Phys. B}\ }\textbf {\bibinfo {volume} {77}},\ \bibinfo
  {pages} {781} (\bibinfo {year} {2003}{\natexlab{b}})}\BibitemShut {NoStop}%
\bibitem [{\citenamefont {Stibor}\ \emph {et~al.}(2005)\citenamefont {Stibor},
  \citenamefont {Hornberger}, \citenamefont {Hackermüller}, \citenamefont
  {Zeilinger},\ and\ \citenamefont {Arndt}}]{Stibor2005a}%
  \BibitemOpen
  \bibfield  {author} {\bibinfo {author} {\bibfnamefont {A.}~\bibnamefont
  {Stibor}}, \bibinfo {author} {\bibfnamefont {K.}~\bibnamefont {Hornberger}},
  \bibinfo {author} {\bibfnamefont {L.}~\bibnamefont {Hackermüller}}, \bibinfo
  {author} {\bibfnamefont {A.}~\bibnamefont {Zeilinger}}, \ and\ \bibinfo
  {author} {\bibfnamefont {M.}~\bibnamefont {Arndt}},\ }\href@noop {}
  {\bibfield  {journal} {\bibinfo  {journal} {Laser Physics}\ }\textbf
  {\bibinfo {volume} {15}},\ \bibinfo {pages} {10} (\bibinfo {year}
  {2005})}\BibitemShut {NoStop}%
\bibitem [{\citenamefont {Knight}\ \emph {et~al.}(1985)\citenamefont {Knight},
  \citenamefont {Clemenger}, \citenamefont {de~Heer},\ and\ \citenamefont
  {Saunders}}]{Knight1985}%
  \BibitemOpen
  \bibfield  {author} {\bibinfo {author} {\bibfnamefont {W.~D.}\ \bibnamefont
  {Knight}}, \bibinfo {author} {\bibfnamefont {K.}~\bibnamefont {Clemenger}},
  \bibinfo {author} {\bibfnamefont {W.~A.}\ \bibnamefont {de~Heer}}, \ and\
  \bibinfo {author} {\bibfnamefont {W.~A.}\ \bibnamefont {Saunders}},\ }\href
  {\doibase 10.1103/PhysRevB.31.2539} {\bibfield  {journal} {\bibinfo
  {journal} {Phys. Rev. B}\ }\textbf {\bibinfo {volume} {31}},\ \bibinfo
  {pages} {2539} (\bibinfo {year} {1985})}\BibitemShut {NoStop}%
\bibitem [{\citenamefont {de~Heer}(1993)}]{Heer1993}%
  \BibitemOpen
  \bibfield  {author} {\bibinfo {author} {\bibfnamefont {W.~A.}\ \bibnamefont
  {de~Heer}},\ }\href {\doibase 10.1103/RevModPhys.65.611} {\bibfield
  {journal} {\bibinfo  {journal} {Rev. Mod. Phys.}\ }\textbf {\bibinfo {volume}
  {65}},\ \bibinfo {pages} {611} (\bibinfo {year} {1993})}\BibitemShut
  {NoStop}%
\bibitem [{\citenamefont {Antoine}\ \emph {et~al.}(1999)\citenamefont
  {Antoine}, \citenamefont {Rayane}, \citenamefont {Allouche}, \citenamefont
  {Aubert-Frécon}, \citenamefont {Benichou}, \citenamefont {Dalby},
  \citenamefont {Dugourd}, \citenamefont {Broyer},\ and\ \citenamefont
  {Guet}}]{Antoine1999}%
  \BibitemOpen
  \bibfield  {author} {\bibinfo {author} {\bibfnamefont {R.}~\bibnamefont
  {Antoine}}, \bibinfo {author} {\bibfnamefont {D.}~\bibnamefont {Rayane}},
  \bibinfo {author} {\bibfnamefont {A.~R.}\ \bibnamefont {Allouche}}, \bibinfo
  {author} {\bibfnamefont {M.}~\bibnamefont {Aubert-Frécon}}, \bibinfo {author}
  {\bibfnamefont {E.}~\bibnamefont {Benichou}}, \bibinfo {author}
  {\bibfnamefont {F.~W.}\ \bibnamefont {Dalby}}, \bibinfo {author}
  {\bibfnamefont {P.}~\bibnamefont {Dugourd}}, \bibinfo {author} {\bibfnamefont
  {M.}~\bibnamefont {Broyer}}, \ and\ \bibinfo {author} {\bibfnamefont
  {C.}~\bibnamefont {Guet}},\ }\href {\doibase 10.1063/1.478455} {\bibfield
  {journal} {\bibinfo  {journal} {J. Chem. Phys.}\ }\textbf {\bibinfo {volume}
  {110 No. 12}},\ \bibinfo {pages} {5568 } (\bibinfo {year}
  {1999})}\BibitemShut {NoStop}%
\bibitem [{\citenamefont {Heer}\ and\ \citenamefont {Kresin}(2011)}]{Heer2011}%
  \BibitemOpen
  \bibfield  {author} {\bibinfo {author} {\bibfnamefont {W.~A.~D.}\
  \bibnamefont {Heer}}\ and\ \bibinfo {author} {\bibfnamefont {V.~V.}\
  \bibnamefont {Kresin}},\ }\enquote {\bibinfo {title} {Handbook of
  nanophysics: Clusters and fullerenes},}\ in\ \href@noop {} {\emph {\bibinfo
  {booktitle} {Handbook of Nanophysics}}},\ \bibinfo {editor} {edited by\
  \bibinfo {editor} {\bibfnamefont {K.~D.}\ \bibnamefont {Sattler}}}\ (\bibinfo
   {publisher} {Taylor \& Francis, CRC Press},\ \bibinfo {year} {2011})\ Chap.\
  \bibinfo {chapter} {Electric and magnetic dipole moments of free
  nanoclusters}, pp.\ \bibinfo {pages} {1--26}\BibitemShut {NoStop}%
\bibitem [{\citenamefont {Compagnon}\ \emph {et~al.}(2001)\citenamefont
  {Compagnon}, \citenamefont {Antoine}, \citenamefont {Rayane}, \citenamefont
  {Dugourd},\ and\ \citenamefont {Broyer}}]{Compagnon2001a}%
  \BibitemOpen
  \bibfield  {author} {\bibinfo {author} {\bibfnamefont {I.}~\bibnamefont
  {Compagnon}}, \bibinfo {author} {\bibfnamefont {R.}~\bibnamefont {Antoine}},
  \bibinfo {author} {\bibfnamefont {D.}~\bibnamefont {Rayane}}, \bibinfo
  {author} {\bibfnamefont {P.}~\bibnamefont {Dugourd}}, \ and\ \bibinfo
  {author} {\bibfnamefont {M.}~\bibnamefont {Broyer}},\ }\href {\doibase
  10.1007/s100530170131} {\bibfield  {journal} {\bibinfo  {journal} {Eur. Phys.
  J. D}\ }\textbf {\bibinfo {volume} {16}},\ \bibinfo {pages} {365 } (\bibinfo
  {year} {2001})}\BibitemShut {NoStop}%
\bibitem [{\citenamefont {Broyer}\ \emph {et~al.}(2007)\citenamefont {Broyer},
  \citenamefont {Antoine}, \citenamefont {Compagnon}, \citenamefont {Rayane},\
  and\ \citenamefont {Dugourd}}]{Broyer2007}%
  \BibitemOpen
  \bibfield  {author} {\bibinfo {author} {\bibfnamefont {M.}~\bibnamefont
  {Broyer}}, \bibinfo {author} {\bibfnamefont {R.}~\bibnamefont {Antoine}},
  \bibinfo {author} {\bibfnamefont {I.}~\bibnamefont {Compagnon}}, \bibinfo
  {author} {\bibfnamefont {D.}~\bibnamefont {Rayane}}, \ and\ \bibinfo {author}
  {\bibfnamefont {P.}~\bibnamefont {Dugourd}},\ }\href {\doibase
  10.1088/0031-8949/76/4/N05} {\bibfield  {journal} {\bibinfo  {journal} {Phys.
  Script.}\ }\textbf {\bibinfo {volume} {76}},\ \bibinfo {pages} {C135}
  (\bibinfo {year} {2007})}\BibitemShut {NoStop}%
\bibitem [{\citenamefont {Stefanov}\ \emph {et~al.}(2008)\citenamefont
  {Stefanov}, \citenamefont {Berninger},\ and\ \citenamefont
  {Arndt}}]{Stefanov2008}%
  \BibitemOpen
  \bibfield  {author} {\bibinfo {author} {\bibfnamefont {A.}~\bibnamefont
  {Stefanov}}, \bibinfo {author} {\bibfnamefont {M.}~\bibnamefont {Berninger}},
  \ and\ \bibinfo {author} {\bibfnamefont {M.}~\bibnamefont {Arndt}},\ }\href
  {\doibase 10.1088/0957-0233/19/5/055801} {\bibfield  {journal} {\bibinfo
  {journal} {Measurement Science \& Technology}\ }\textbf {\bibinfo {volume}
  {19}},\ \bibinfo {pages} {055801} (\bibinfo {year} {2008})}\BibitemShut
  {NoStop}%
\bibitem [{\citenamefont {Deachapunya}\ \emph {et~al.}(2008)\citenamefont
  {Deachapunya}, \citenamefont {Fagan}, \citenamefont {Major}, \citenamefont
  {Reiger}, \citenamefont {Ritsch}, \citenamefont {Stefanov}, \citenamefont
  {Ulbricht},\ and\ \citenamefont {Arndt}}]{Deachapunya2008}%
  \BibitemOpen
  \bibfield  {author} {\bibinfo {author} {\bibfnamefont {S.}~\bibnamefont
  {Deachapunya}}, \bibinfo {author} {\bibfnamefont {P.~J.}\ \bibnamefont
  {Fagan}}, \bibinfo {author} {\bibfnamefont {A.~G.}\ \bibnamefont {Major}},
  \bibinfo {author} {\bibfnamefont {E.}~\bibnamefont {Reiger}}, \bibinfo
  {author} {\bibfnamefont {H.}~\bibnamefont {Ritsch}}, \bibinfo {author}
  {\bibfnamefont {A.}~\bibnamefont {Stefanov}}, \bibinfo {author}
  {\bibfnamefont {H.}~\bibnamefont {Ulbricht}}, \ and\ \bibinfo {author}
  {\bibfnamefont {M.}~\bibnamefont {Arndt}},\ }\href {\doibase
  10.1140/epjd/e2007-00301-8} {\bibfield  {journal} {\bibinfo  {journal} {Eur.
  Phys. J. D}\ }\textbf {\bibinfo {volume} {46}},\ \bibinfo {pages} {307}
  (\bibinfo {year} {2008})}\BibitemShut {NoStop}%
\bibitem [{\citenamefont {Gerlich}\ \emph {et~al.}(2011)\citenamefont
  {Gerlich}, \citenamefont {Eibenberger}, \citenamefont {Tomandl},
  \citenamefont {Nimmrichter}, \citenamefont {Hornberger}, \citenamefont
  {Fagan}, \citenamefont {Tüxen}, \citenamefont {Mayor},\ and\ \citenamefont
  {Arndt}}]{Gerlich2011}%
  \BibitemOpen
  \bibfield  {author} {\bibinfo {author} {\bibfnamefont {S.}~\bibnamefont
  {Gerlich}}, \bibinfo {author} {\bibfnamefont {S.}~\bibnamefont
  {Eibenberger}}, \bibinfo {author} {\bibfnamefont {M.}~\bibnamefont
  {Tomandl}}, \bibinfo {author} {\bibfnamefont {S.}~\bibnamefont
  {Nimmrichter}}, \bibinfo {author} {\bibfnamefont {K.}~\bibnamefont
  {Hornberger}}, \bibinfo {author} {\bibfnamefont {P.}~\bibnamefont {Fagan}},
  \bibinfo {author} {\bibfnamefont {J.}~\bibnamefont {Tüxen}}, \bibinfo
  {author} {\bibfnamefont {M.}~\bibnamefont {Mayor}}, \ and\ \bibinfo {author}
  {\bibfnamefont {M.}~\bibnamefont {Arndt}},\ }\href {\doibase
  10.1038/ncomms1263} {\bibfield  {journal} {\bibinfo  {journal} {Nature
  Communications}\ }\textbf {\bibinfo {volume} {2}},\ \bibinfo {pages} {263}
  (\bibinfo {year} {2011})}\BibitemShut {NoStop}%
\bibitem [{\citenamefont {Eibenberger}\ \emph {et~al.}(2011)\citenamefont
  {Eibenberger}, \citenamefont {Gerlich}, \citenamefont {Arndt}, \citenamefont
  {Tüxen},\ and\ \citenamefont {Mayor}}]{Eibenberger2011}%
  \BibitemOpen
  \bibfield  {author} {\bibinfo {author} {\bibfnamefont {S.}~\bibnamefont
  {Eibenberger}}, \bibinfo {author} {\bibfnamefont {S.}~\bibnamefont
  {Gerlich}}, \bibinfo {author} {\bibfnamefont {M.}~\bibnamefont {Arndt}},
  \bibinfo {author} {\bibfnamefont {J.}~\bibnamefont {Tüxen}}, \ and\ \bibinfo
  {author} {\bibfnamefont {M.}~\bibnamefont {Mayor}},\ }\href {\doibase
  10.1088/1367-2630/13/4/043033} {\bibfield  {journal} {\bibinfo  {journal}
  {New J. Phys.}\ }\textbf {\bibinfo {volume} {13}},\ \bibinfo {pages} {43033}
  (\bibinfo {year} {2011})}\BibitemShut {NoStop}%
\bibitem [{\citenamefont {Tüxen}\ \emph {et~al.}(2011)\citenamefont {Tüxen},
  \citenamefont {Eibenberger}, \citenamefont {Gerlich}, \citenamefont {Arndt},\
  and\ \citenamefont {Mayor}}]{Tuexen2011}%
  \BibitemOpen
  \bibfield  {author} {\bibinfo {author} {\bibfnamefont {J.}~\bibnamefont
  {Tüxen}}, \bibinfo {author} {\bibfnamefont {S.}~\bibnamefont {Eibenberger}},
  \bibinfo {author} {\bibfnamefont {S.}~\bibnamefont {Gerlich}}, \bibinfo
  {author} {\bibfnamefont {M.}~\bibnamefont {Arndt}}, \ and\ \bibinfo {author}
  {\bibfnamefont {M.}~\bibnamefont {Mayor}},\ }\href {\doibase
  10.1002/ejoc.201100638} {\bibfield  {journal} {\bibinfo  {journal} {Eur. J.
  Org. Chem.}\ }\textbf {\bibinfo {volume} {25}},\ \bibinfo {pages} {4823}
  (\bibinfo {year} {2011})}\BibitemShut {NoStop}%
\bibitem [{\citenamefont {Bonin}\ and\ \citenamefont
  {Kresin}(1997)}]{Bonin1997}%
  \BibitemOpen
  \bibfield  {author} {\bibinfo {author} {\bibfnamefont {K.}~\bibnamefont
  {Bonin}}\ and\ \bibinfo {author} {\bibfnamefont {V.}~\bibnamefont {Kresin}},\
  }\href@noop {} {\emph {\bibinfo {title} {Electric-Dipole Polarizabilities of
  Atoms, Molecules and Clusters}}}\ (\bibinfo  {publisher} {World Scientific},\
  \bibinfo {year} {1997})\BibitemShut {NoStop}%
\bibitem [{\citenamefont {Vleck}(1965)}]{Vleck1965}%
  \BibitemOpen
  \bibfield  {author} {\bibinfo {author} {\bibfnamefont {J.~V.}\ \bibnamefont
  {Vleck}},\ }\href@noop {} {\emph {\bibinfo {title} {The theory of electric
  and magnetic susceptibilities}}}\ (\bibinfo  {publisher} {Oxford University
  Press London},\ \bibinfo {year} {1965})\BibitemShut {NoStop}%
\bibitem [{\citenamefont {Deachapunya}\ \emph {et~al.}(2007)\citenamefont
  {Deachapunya}, \citenamefont {Stefanov}, \citenamefont {Berninger},
  \citenamefont {Ulbricht}, \citenamefont {Reiger}, \citenamefont {Doltsinis},\
  and\ \citenamefont {Arndt}}]{Deachapunya2007}%
  \BibitemOpen
  \bibfield  {author} {\bibinfo {author} {\bibfnamefont {S.}~\bibnamefont
  {Deachapunya}}, \bibinfo {author} {\bibfnamefont {A.}~\bibnamefont
  {Stefanov}}, \bibinfo {author} {\bibfnamefont {M.}~\bibnamefont {Berninger}},
  \bibinfo {author} {\bibfnamefont {H.}~\bibnamefont {Ulbricht}}, \bibinfo
  {author} {\bibfnamefont {E.}~\bibnamefont {Reiger}}, \bibinfo {author}
  {\bibfnamefont {N.~L.}\ \bibnamefont {Doltsinis}}, \ and\ \bibinfo {author}
  {\bibfnamefont {M.}~\bibnamefont {Arndt}},\ }\href {\doibase
  10.1063/1.2721563} {\bibfield  {journal} {\bibinfo  {journal} {J. Chem.
  Phys.}\ }\textbf {\bibinfo {volume} {126}},\ \bibinfo {pages} {164304}
  (\bibinfo {year} {2007})}\BibitemShut {NoStop}%
\bibitem [{\citenamefont {Even}\ and\ \citenamefont {Lavie}(2012)}]{EVEN}%
  \BibitemOpen
  \bibfield  {author} {\bibinfo {author} {\bibfnamefont {U.}~\bibnamefont
  {Even}}\ and\ \bibinfo {author} {\bibfnamefont {N.}~\bibnamefont {Lavie}},\
  }\href@noop {} {\bibfield  {journal} {\bibinfo  {journal} {retrieved from
  http://sites.google.com/site/evenlavievalve/beam-properties}\ } (\bibinfo
  {year} {September 2012})}\BibitemShut {NoStop}%
\bibitem [{\citenamefont {Sansonetti}\ \emph {et~al.}(2001)\citenamefont
  {Sansonetti}, \citenamefont {Reader},\ and\ \citenamefont
  {Vogler}}]{Sansonetti2001}%
  \BibitemOpen
  \bibfield  {author} {\bibinfo {author} {\bibfnamefont {C.~J.}\ \bibnamefont
  {Sansonetti}}, \bibinfo {author} {\bibfnamefont {J.}~\bibnamefont {Reader}},
  \ and\ \bibinfo {author} {\bibfnamefont {K.}~\bibnamefont {Vogler}},\ }\href
  {\doibase 10.1364/AO.40.001974} {\bibfield  {journal} {\bibinfo  {journal}
  {Appl. Opt.}\ }\textbf {\bibinfo {volume} {40}},\ \bibinfo {pages} {1974}
  (\bibinfo {year} {2001})}\BibitemShut {NoStop}%
\bibitem [{\citenamefont {Rodewald}(2011)}]{Rodewald2011}%
  \BibitemOpen
  \bibfield  {author} {\bibinfo {author} {\bibfnamefont {J.}~\bibnamefont
  {Rodewald}},\ }\emph {\bibinfo {title} {Setup of an optical time-domain
  matter wave interferometer for heavy particles}},\ \href
  {http://ubdata.univie.ac.at/AC08823881} {Master's thesis},\ \bibinfo
  {school} {University of Vienna} (\bibinfo {year} {2011})\BibitemShut
  {NoStop}%
\bibitem [{\citenamefont {Hohensee}\ and\ \citenamefont
  {Müller}(2011)}]{Hohensee2011b}%
  \BibitemOpen
  \bibfield  {author} {\bibinfo {author} {\bibfnamefont {M.~A.}\ \bibnamefont
  {Hohensee}}\ and\ \bibinfo {author} {\bibfnamefont {H.}~\bibnamefont
  {Müller}},\ }\href {\doibase 10.1080/09500340.2011.606376} {\bibfield
  {journal} {\bibinfo  {journal} {J. Mod. Opt.}\ }\textbf {\bibinfo {volume}
  {58}},\ \bibinfo {pages} {2021} (\bibinfo {year} {2011})}\BibitemShut
  {NoStop}%
\bibitem [{\citenamefont {Dimopoulos}\ \emph {et~al.}(2007)\citenamefont
  {Dimopoulos}, \citenamefont {Graham}, \citenamefont {Hogan},\ and\
  \citenamefont {Kasevich}}]{Dimopoulos2007}%
  \BibitemOpen
  \bibfield  {author} {\bibinfo {author} {\bibfnamefont {S.}~\bibnamefont
  {Dimopoulos}}, \bibinfo {author} {\bibfnamefont {P.}~\bibnamefont {Graham}},
  \bibinfo {author} {\bibfnamefont {J.}~\bibnamefont {Hogan}}, \ and\ \bibinfo
  {author} {\bibfnamefont {M.}~\bibnamefont {Kasevich}},\ }\href {\doibase
  10.1103/PhysRevLett.98.111102} {\bibfield  {journal} {\bibinfo  {journal}
  {Phys. Rev. Lett.}\ }\textbf {\bibinfo {volume} {98}},\ \bibinfo {pages} {1}
  (\bibinfo {year} {2007})}\BibitemShut {NoStop}%
\bibitem [{\citenamefont {Hogan}\ \emph {et~al.}(2007)\citenamefont {Hogan},
  \citenamefont {Johnson},\ and\ \citenamefont {Kasevich}}]{Hogan2007a}%
  \BibitemOpen
  \bibfield  {author} {\bibinfo {author} {\bibfnamefont {J.}~\bibnamefont
  {Hogan}}, \bibinfo {author} {\bibfnamefont {D.~S.}\ \bibnamefont {Johnson}},
  \ and\ \bibinfo {author} {\bibfnamefont {M.}~\bibnamefont {Kasevich}}\
  }(\bibinfo  {publisher} {IOS Press, Oxford},\ \bibinfo {year} {2007})\ p.\
  \bibinfo {pages} {411}\BibitemShut {NoStop}%
\bibitem [{\citenamefont {Lan}\ \emph {et~al.}(2012)\citenamefont {Lan},
  \citenamefont {Kuan}, \citenamefont {Estey}, \citenamefont {Haslinger},\ and\
  \citenamefont {Müller}}]{Lan2012}%
  \BibitemOpen
  \bibfield  {author} {\bibinfo {author} {\bibfnamefont {S.-Y.}\ \bibnamefont
  {Lan}}, \bibinfo {author} {\bibfnamefont {P.-C.}\ \bibnamefont {Kuan}},
  \bibinfo {author} {\bibfnamefont {B.}~\bibnamefont {Estey}}, \bibinfo
  {author} {\bibfnamefont {P.}~\bibnamefont {Haslinger}}, \ and\ \bibinfo
  {author} {\bibfnamefont {H.}~\bibnamefont {Müller}},\ }\href {\doibase
  10.1103/PhysRevLett.108.090402} {\bibfield  {journal} {\bibinfo  {journal}
  {Phys. Rev. Lett.}\ }\textbf {\bibinfo {volume} {108}},\ \bibinfo {pages}
  {090402} (\bibinfo {year} {2012})}\BibitemShut {NoStop}%
\bibitem [{\citenamefont {Dickerson}\ \emph {et~al.}(2013)\citenamefont
  {Dickerson}, \citenamefont {Hogan}, \citenamefont {Sugarbaker}, \citenamefont
  {Johnson},\ and\ \citenamefont {Kasevich}}]{Dickerson2013}%
  \BibitemOpen
  \bibfield  {author} {\bibinfo {author} {\bibfnamefont {S.~M.}\ \bibnamefont
  {Dickerson}}, \bibinfo {author} {\bibfnamefont {J.~M.}\ \bibnamefont
  {Hogan}}, \bibinfo {author} {\bibfnamefont {A.}~\bibnamefont {Sugarbaker}},
  \bibinfo {author} {\bibfnamefont {D.~M.~S.}\ \bibnamefont {Johnson}}, \ and\
  \bibinfo {author} {\bibfnamefont {M.~A.}\ \bibnamefont {Kasevich}},\ }\href
  {\doibase 10.1103/PhysRevLett.111.083001} {\bibfield  {journal} {\bibinfo
  {journal} {Phys. Rev. Lett.}\ }\textbf {\bibinfo {volume} {111}},\ \bibinfo
  {pages} {083001} (\bibinfo {year} {2013})}\BibitemShut {NoStop}%
\bibitem [{\citenamefont {Teufel}\ \emph {et~al.}(2011)\citenamefont {Teufel},
  \citenamefont {Donner}, \citenamefont {Li}, \citenamefont {Harlow},
  \citenamefont {Allman}, \citenamefont {Cicak}, \citenamefont {Sirois},
  \citenamefont {Whittaker}, \citenamefont {Lehnert},\ and\ \citenamefont
  {Simmonds}}]{Teufel2011}%
  \BibitemOpen
  \bibfield  {author} {\bibinfo {author} {\bibfnamefont {J.~D.}\ \bibnamefont
  {Teufel}}, \bibinfo {author} {\bibfnamefont {T.}~\bibnamefont {Donner}},
  \bibinfo {author} {\bibfnamefont {D.}~\bibnamefont {Li}}, \bibinfo {author}
  {\bibfnamefont {J.~W.}\ \bibnamefont {Harlow}}, \bibinfo {author}
  {\bibfnamefont {M.~S.}\ \bibnamefont {Allman}}, \bibinfo {author}
  {\bibfnamefont {K.}~\bibnamefont {Cicak}}, \bibinfo {author} {\bibfnamefont
  {A.~J.}\ \bibnamefont {Sirois}}, \bibinfo {author} {\bibfnamefont {J.~D.}\
  \bibnamefont {Whittaker}}, \bibinfo {author} {\bibfnamefont {K.~W.}\
  \bibnamefont {Lehnert}}, \ and\ \bibinfo {author} {\bibfnamefont {R.~W.}\
  \bibnamefont {Simmonds}},\ }\href {\doibase 10.1038/nature10261} {\bibfield
  {journal} {\bibinfo  {journal} {Nature}\ }\textbf {\bibinfo {volume} {475}},\
  \bibinfo {pages} {359} (\bibinfo {year} {2011})}\BibitemShut {NoStop}%
\bibitem [{\citenamefont {Chan}\ \emph {et~al.}(2011)\citenamefont {Chan},
  \citenamefont {Mayer~Alegre}, \citenamefont {Safavi-Naeini}, \citenamefont
  {Hill}, \citenamefont {Krause}, \citenamefont {Groeblacher}, \citenamefont
  {Aspelmeyer},\ and\ \citenamefont {Painter}}]{Chan2011}%
  \BibitemOpen
  \bibfield  {author} {\bibinfo {author} {\bibfnamefont {J.}~\bibnamefont
  {Chan}}, \bibinfo {author} {\bibfnamefont {T.~P.}\ \bibnamefont
  {Mayer~Alegre}}, \bibinfo {author} {\bibfnamefont {A.~H.}\ \bibnamefont
  {Safavi-Naeini}}, \bibinfo {author} {\bibfnamefont {J.~T.}\ \bibnamefont
  {Hill}}, \bibinfo {author} {\bibfnamefont {A.}~\bibnamefont {Krause}},
  \bibinfo {author} {\bibfnamefont {S.}~\bibnamefont {Groeblacher}}, \bibinfo
  {author} {\bibfnamefont {M.}~\bibnamefont {Aspelmeyer}}, \ and\ \bibinfo
  {author} {\bibfnamefont {O.}~\bibnamefont {Painter}},\ }\href {\doibase
  10.1038/nature10461} {\bibfield  {journal} {\bibinfo  {journal} {Nature}\
  }\textbf {\bibinfo {volume} {478}},\ \bibinfo {pages} {89} (\bibinfo {year}
  {2011})}\BibitemShut {NoStop}%
\bibitem [{\citenamefont {Zawisky}\ \emph {et~al.}(2002)\citenamefont
  {Zawisky}, \citenamefont {Baron}, \citenamefont {Loidl},\ and\ \citenamefont
  {Rauch}}]{Zawisky2002}%
  \BibitemOpen
  \bibfield  {author} {\bibinfo {author} {\bibfnamefont {M.}~\bibnamefont
  {Zawisky}}, \bibinfo {author} {\bibfnamefont {M.}~\bibnamefont {Baron}},
  \bibinfo {author} {\bibfnamefont {R.}~\bibnamefont {Loidl}}, \ and\ \bibinfo
  {author} {\bibfnamefont {H.}~\bibnamefont {Rauch}},\ }\href {\doibase
  10.1016/S0168-9002(01)01253-0} {\bibfield  {journal} {\bibinfo  {journal}
  {Nucl. Instr. and Meth. in Phys. Res. A}\ }\textbf {\bibinfo {volume}
  {481}},\ \bibinfo {pages} {406} (\bibinfo {year} {2002})}\BibitemShut
  {NoStop}%
\bibitem [{\citenamefont {Chiow}\ \emph {et~al.}(2011)\citenamefont {Chiow},
  \citenamefont {Kovachy}, \citenamefont {Chien},\ and\ \citenamefont
  {Kasevich}}]{Chiow2011}%
  \BibitemOpen
  \bibfield  {author} {\bibinfo {author} {\bibfnamefont {S.}~\bibnamefont
  {Chiow}}, \bibinfo {author} {\bibfnamefont {T.}~\bibnamefont {Kovachy}},
  \bibinfo {author} {\bibfnamefont {H.}~\bibnamefont {Chien}}, \ and\ \bibinfo
  {author} {\bibfnamefont {M.}~\bibnamefont {Kasevich}},\ }\href {\doibase
  10.1103/PhysRevLett.107.130403} {\bibfield  {journal} {\bibinfo  {journal}
  {Phys. Rev. Lett.}\ }\textbf {\bibinfo {volume} {107}},\ \bibinfo {pages}
  {130403} (\bibinfo {year} {2011})}\BibitemShut {NoStop}%
\bibitem [{\citenamefont {Friedman}\ \emph {et~al.}(2000)\citenamefont
  {Friedman}, \citenamefont {Patel}, \citenamefont {Chen}, \citenamefont
  {Tolpygo},\ and\ \citenamefont {Lukens}}]{Friedman2000}%
  \BibitemOpen
  \bibfield  {author} {\bibinfo {author} {\bibfnamefont {J.~R.}\ \bibnamefont
  {Friedman}}, \bibinfo {author} {\bibfnamefont {V.}~\bibnamefont {Patel}},
  \bibinfo {author} {\bibfnamefont {W.}~\bibnamefont {Chen}}, \bibinfo {author}
  {\bibfnamefont {S.~K.}\ \bibnamefont {Tolpygo}}, \ and\ \bibinfo {author}
  {\bibfnamefont {J.~E.}\ \bibnamefont {Lukens}},\ }\href {\doibase
  10.1038/35017505} {\bibfield  {journal} {\bibinfo  {journal} {Nature}\
  }\textbf {\bibinfo {volume} {406}},\ \bibinfo {pages} {43} (\bibinfo {year}
  {2000})}\BibitemShut {NoStop}%
\bibitem [{\citenamefont {der Wal}\ \emph {et~al.}(2000)\citenamefont {der
  Wal}, \citenamefont {Haar}, \citenamefont {Wilhelm}, \citenamefont
  {Schouten}, \citenamefont {Harmans}, \citenamefont {Orlando}, \citenamefont
  {Lloyd},\ and\ \citenamefont {Mooij}}]{Wal2000}%
  \BibitemOpen
  \bibfield  {author} {\bibinfo {author} {\bibfnamefont {C.~H.~V.}\
  \bibnamefont {der Wal}}, \bibinfo {author} {\bibfnamefont {A.~C. J.~T.}\
  \bibnamefont {Haar}}, \bibinfo {author} {\bibfnamefont {F.~K.}\ \bibnamefont
  {Wilhelm}}, \bibinfo {author} {\bibfnamefont {R.~N.}\ \bibnamefont
  {Schouten}}, \bibinfo {author} {\bibfnamefont {C.~J. P.~M.}\ \bibnamefont
  {Harmans}}, \bibinfo {author} {\bibfnamefont {T.~P.}\ \bibnamefont
  {Orlando}}, \bibinfo {author} {\bibfnamefont {S.}~\bibnamefont {Lloyd}}, \
  and\ \bibinfo {author} {\bibfnamefont {J.~E.}\ \bibnamefont {Mooij}},\ }\href
  {\doibase 10.1126/science.290.5492.773} {\bibfield  {journal} {\bibinfo
  {journal} {Science}\ }\textbf {\bibinfo {volume} {290}},\ \bibinfo {pages}
  {773} (\bibinfo {year} {2000})}\BibitemShut {NoStop}%
\bibitem [{\citenamefont {Korsbakken}\ \emph {et~al.}(2010)\citenamefont
  {Korsbakken}, \citenamefont {Wilhelm},\ and\ \citenamefont
  {Whaley}}]{Korsbakken2010}%
  \BibitemOpen
  \bibfield  {author} {\bibinfo {author} {\bibfnamefont {J.~I.}\ \bibnamefont
  {Korsbakken}}, \bibinfo {author} {\bibfnamefont {F.~K.}\ \bibnamefont
  {Wilhelm}}, \ and\ \bibinfo {author} {\bibfnamefont {K.~B.}\ \bibnamefont
  {Whaley}},\ }\href {\doibase 10.1209/0295-5075/89/30003} {\bibfield
  {journal} {\bibinfo  {journal} {EPL}\ }\textbf {\bibinfo {volume} {89}},\
  \bibinfo {pages} {30003} (\bibinfo {year} {2010})}\BibitemShut {NoStop}%
\bibitem [{\citenamefont {Nimmrichter}\ and\ \citenamefont
  {Hornberger}(2013)}]{Nimmrichter2013}%
  \BibitemOpen
  \bibfield  {author} {\bibinfo {author} {\bibfnamefont {S.}~\bibnamefont
  {Nimmrichter}}\ and\ \bibinfo {author} {\bibfnamefont {K.}~\bibnamefont
  {Hornberger}},\ }\href {\doibase 10.1103/PhysRevLett.110.160403} {\bibfield
  {journal} {\bibinfo  {journal} {Phys. Rev. Lett.}\ }\textbf {\bibinfo
  {volume} {110}},\ \bibinfo {pages} {160403} (\bibinfo {year}
  {2013})}\BibitemShut {NoStop}%
\bibitem [{\citenamefont {Marshall}\ \emph {et~al.}(2003)\citenamefont
  {Marshall}, \citenamefont {Simon}, \citenamefont {Penrose},\ and\
  \citenamefont {Bouwmeester}}]{Marshall2003}%
  \BibitemOpen
  \bibfield  {author} {\bibinfo {author} {\bibfnamefont {W.}~\bibnamefont
  {Marshall}}, \bibinfo {author} {\bibfnamefont {C.}~\bibnamefont {Simon}},
  \bibinfo {author} {\bibfnamefont {R.}~\bibnamefont {Penrose}}, \ and\
  \bibinfo {author} {\bibfnamefont {D.}~\bibnamefont {Bouwmeester}},\ }\href
  {\doibase 10.1103/PhysRevLett.91.130401} {\bibfield  {journal} {\bibinfo
  {journal} {Phys. Rev. Lett.}\ }\textbf {\bibinfo {volume} {91}},\ \bibinfo
  {pages} {130401} (\bibinfo {year} {2003})}\BibitemShut {NoStop}%
\bibitem [{\citenamefont {Romero-Isart}\ \emph {et~al.}(2011)\citenamefont
  {Romero-Isart}, \citenamefont {Pflanzer}, \citenamefont {Blaser},
  \citenamefont {Kaltenbaek}, \citenamefont {Kiesel}, \citenamefont
  {Aspelmeyer},\ and\ \citenamefont {Cirac}}]{Romero-Isart2011b}%
  \BibitemOpen
  \bibfield  {author} {\bibinfo {author} {\bibfnamefont {O.}~\bibnamefont
  {Romero-Isart}}, \bibinfo {author} {\bibfnamefont {A.~C.}\ \bibnamefont
  {Pflanzer}}, \bibinfo {author} {\bibfnamefont {F.}~\bibnamefont {Blaser}},
  \bibinfo {author} {\bibfnamefont {R.}~\bibnamefont {Kaltenbaek}}, \bibinfo
  {author} {\bibfnamefont {N.}~\bibnamefont {Kiesel}}, \bibinfo {author}
  {\bibfnamefont {M.}~\bibnamefont {Aspelmeyer}}, \ and\ \bibinfo {author}
  {\bibfnamefont {J.~I.}\ \bibnamefont {Cirac}},\ }\href {\doibase
  10.1103/PhysRevLett.107.020405} {\bibfield  {journal} {\bibinfo  {journal}
  {Phys. Rev. Lett.}\ }\textbf {\bibinfo {volume} {107}},\ \bibinfo {pages}
  {20405} (\bibinfo {year} {2011})}\BibitemShut {NoStop}%
\end{thebibliography}%

\end{document}